%%%%%%%%%%%%%%%%%%%%%%%%%%%%%%%%%%%%%%%%%%%%%%%%%%%%%%%%%%%%%%%%%%%%%%%%%%
\input epsf
\input lanlmac
%\draftmode
%\baselineskip =20pt plus 1 pt minus 1 pt
%%%%%%%%%%%%%%%%%%%%
% personal macros  %
%%%%%%%%%%%%%%%%%%%%
%\font\tay=eusb10  
%\font\tentit=cmmib10  
%\font\ninetit=cmmib9
%\font\seventit=cmmib7 
%\font\fivetit=cmmib5  
%\newfam\titfam
%\textfont\titfam=\tentit
%\scriptfont\titfam=\seventit
%\scriptscriptfont\titfam=\fivetit
%\def\tit{\fam\titfam\tentit}
%
\def\CA{{\cal A}} \def\CB{{\cal B}} \def\CC{{\cal C}} \def\CD{{\cal D}}
 \def\CF{{\cal F}} \def\CG{{\cal G}} \def\CH{{\cal H}}
\def\CI{{\cal I}}   
\def\CM{{\cal M}} \def\CN{{\cal N}} \def\CO{{\cal O}} \def\CP{{\cal P}}
 \def\CR{{\cal R}} \def\CS{{\cal S}} \def\CT{{\cal T}}
 \def\CV{{\cal V}}  
 \def\CZ{{\cal Z}}
\font\sevenbf=cmbx7 \font\fivebf=cmbx5 \font\threebf=cmbx5 scaled 800
\newfam\sbffam\def\sbf{\fam\sbffam\sevenbf}\textfont\sbffam=\sevenbf
\scriptfont\sbffam=\fivebf\scriptscriptfont\sbffam=\threebf
\def\rvec{{\bf \vec r}} 
\def\rvecR{{{\bf\vec r}_{\sbf R}\null}}
\def\rveccm{{{\bf\vec r}_{\rm cm}}}
\def\rvecRcm{{\left.{\rvecR}\right._{\rm cm}}}
\def\rangleR{{\rangle}_{\sbf R}} 
\def\CHR{{\cal H}^{\sbf R}} 
\def\kBT{{\rm k}_{\rm B}T}
\def\bR{{b_{\sbf R}}}
\def\ZR{\CZ_{\sbf R}}
\def\kvec{{\bf \vec k}}

\def\qvec{{\bf \vec q}}
\def\mvec{{\bf \vec m}}
\def\kbf{{\bf k}}
\def\qbf{{\bf q}}
\def\rbf{{\bf r}}
\def\Cbf{{\bf C}}

\font\twelverm=cmr10 scaled \magstep1
\font\twelvessb=cmssbx10 scaled \magstep1
%\font\twelvessb=wncyss10 scaled \magstep1
\def\npr{\hbox{\twelverm :}}
\def\ii{{\rm i}}
\def\ee{{\rm e}}
\def\sprod{\mathop{\Pi}}

\def\ssum{\mathop{\Sigma}}

\def\RR{\relax{\rm I\kern-.18em R}}
\def\NN{\relax{\rm I\kern-.18em N}}
\def\IOne{\relax{\rm 1\kern-.30em I}}
\def\Rr{\relax{\ninerm I\kern-.22em \ninerm R}}
\def\Nn{\relax{\ninerm I\kern-.22em \ninerm N}}
\def\setminusp{\hbox{$/ \kern -3pt {}_p$}}
\def\ssetminusp{\hbox{$\scriptstyle / \kern -2pt {}_p$}}
\def\lesssim{\hbox{\raise.4ex \hbox{$<$} \kern-1.1em \lower.8ex \hbox{$\sim$}}}
\def\buildlim#1\under#2{\mathrel{\mathop{\kern0pt #2}\limits_{#1}}}
%%%%%%%%%%%%%%%%%%%%%%%%%%%%%%%%%%%%%%%%%%%%%%%%%%%%%%%%%%%%%%%%%%%%%%%%%%%
\gdef\href#1#2{{#2}}
%\Title{SPhT/97-001\ \ cond-mat/9702136}{
\Title{T/97-001}{
\vbox{
\vskip -10.0truecm
\leftline{\epsfxsize=7truecm\epsfysize=0.cm\epsfbox{logo-spht.eps}}
 %\vskip -2truecm
\vskip 1.5truecm
 \centerline{{\twelvessb Renormalization Theory for the}}
 \centerline{{\twelvessb Self-Avoiding Polymerized Membranes}}   
}}
\centerline{Fran\c cois
%\href{file:/home/wasa2/david/FD/fd.jpeg}
{David}\footnote{$^\dagger$}{Member of CNRS},
%Bertrand Duplantier{$^?$}{}
Bertrand Duplantier
and Emmanuel
%\href{file:/home/wasa2/david/FD/eg.jpeg}
{Guitter}}
%\bigskip\centerline{Service de Physique Th\'eorique\footnote{$^\star$}{
\bigskip\centerline{ 
C.E.A. Saclay,
%\href{http://amoco.saclay.cea.fr/index.html}
{Service de Physique Th\'eorique}
%\footnote{$^\star$}{Laboratoire de la Direction des Sciences de la Mati\`ere du Commissariat \`a l'Energie Atomique}
%, CEA-Saclay
}
%\centerline{C.E. Saclay}
\centerline{F-91191 Gif-sur-Yvette Cedex, FRANCE}
\vskip 1.truecm
\centerline{\bf Abstract}{
\ninerm
\textfont0=\ninerm
\font\ninemit=cmmi9
\font\sevenmit=cmmi7
\textfont1=\ninemit
\scriptfont0=\sevenrm
\scriptfont1=\sevenmit
\baselineskip=11pt 
\bigskip
%\centerline{
%The search for a theory of self-avoiding tethered surfaces is over.
%}

We prove the renormalizability of the generalized Edwards model for
self-avoiding polymerized membranes.
This is done by use of
a short distance multilocal operator product expansion, which extends the
methods of local field theories to a large class of models with
non-local singular interactions.
This ensures the existence of scaling laws for crumpled self-avoiding membranes,
and validates the direct renormalization method used for polymers and membranes.
This also provides a framework for explicit perturbative calculations.
We discuss hyperscaling relations for the configuration exponent and contact
exponents.
We finally consider membranes with long range interactions and at the
$\Theta$-point.
\bigskip
\par
}
\vskip .3in
\Date{PACS numbers: 05.20.-y, 11.10.Gh, 11.17.+y}
%\Date{\vbox{\hbox{PACS numbers: 05.20.-y, 11.10.Gh, 11.17.+y}
%            \hbox{Submitted for publication to}}} 
%for preliminary versions, specify \draftmode at some point 
\hfuzz 1.pt
\vfill\eject
%\bye

\listtoc
\writetoc
\vfill\eject

\nref\rJerus{{\sl Statistical Mechanics of Membranes and
Surfaces}, Proceedings of the Fifth Jerusalem Winter School for Theoretical
Physics (1987), D. R. Nelson, T. Piran and S. Weinberg Eds., World Scientific,
Singapore (1989).}

\nref\rHouches{{\sl Fluctuating Geometries in Statistical Mechanics and Field
Theory}, F. David, P. Ginsparg and J. Zinn-Justin eds., Les Houches Summer
School Session LXLL (Elsevier Science, 1996).}

\nref\rNelPel{D.~R.~Nelson and L.~Peliti, J. de Physique {\bf 48}
(1987) 1085.}

\nref\rPelLei{L.~Peliti and S. Leibler, Phys. Rev. Lett. {\bf 54} (1985)
1690.}

\nref\rKaKaNe{Y. Kantor, M. Kardar and D.~R.~Nelson, Phys. Rev. Lett.
{\bf 57} (1986) 791; Phys. Rev. {\bf A 35} (1987) 3056.}

\nref\rPaKaNeI{M.~Paczuski, M. Kardar and D.~R.~Nelson, Phys. Rev. Lett.
{\bf 60} (1988) 2638.}

\nref\rDaGu{F.~David and E.~Guitter, Europhys. Lett. {\bf 5} (1988) 709.}

\nref\rArLuI{J.~A. Aronovitz and T.~C.~Lubensky, Phys. Rev. Lett. {\bf 60}
(1988) 2634.}

\nref\rPaKaI{M.~Paczuski and M.~Kardar, Phys. Rev. {\bf A 39} (1989) 6086.}

\nref\rKanNe{Y. Kantor and D.~R.~Nelson, Phys. Rev. Lett. {\bf 58} (1987)
2774; Phys. Rev. A {\bf 36} (1987) 4020.}

\nref\rEdwards{S.~F.~Edwards, Proc. Phys. Soc. Lond. 85 (1965) 613.}

\nref\rKaNe{M. Kardar and D. R. Nelson, Phys. Rev. Lett. {\bf 58}
(1987) 1289, 2280 (E); Phys. Rev. {\bf A 38} (1988) 966.}

\nref\rArLu{J. A. Aronovitz and T. C. Lubensky, Europhys. Lett. {\bf 4}
(1987) 395.}

\nref\rFix{M. Fixman, J. Chem. Phys. 23 (1955) 1656.}

\nref\rDesClo{J.~des~Cloizeaux, J. de Physique {\bf 42} (1981) 635.}

\nref\rDesCloJan{J. des Cloizeaux and G. Jannink, {\sl Polymers in
solution, their modeling and structure}, Clarendon Press, Oxford, (1990).}

\nref\rHwa{T. Hwa, { Phys. Rev.} {\bf A 41} (1990) 1751.}

\nref\rBDone{B.~Duplantier, Phys. Rev. Lett. {\bf 58} (1987) 2733.}

\nref\rDeGennes{P.-G. de Gennes, Phys. Lett. A 38 (1972) 339.}

\nref\rDombGreen{C. Domb and M.S. Green, eds., Phase Transitions and Critical
Phenomena, Vol. 6 (Academic Press, New York, 1976).}

\nref\rZinn{J. Zinn-Justin, {\sl Quantum Field Theory and Critical Phenomena}
(Clarendon Press, Oxford, 1989).}
\nref\rBenMah{M. Benhamou and G, Mahoux, J. de Physique 47 (1986) 556.}
\nref\rDuplII{B. Duplantier, J. de Physique 47 (1986) 559.}
\nref\rDuHwKa{ B.~Duplantier, T.~Hwa and M.~Kardar, Phys. Rev. Lett.
{\bf 64} (1990) 2022.}

\nref\rIIM{F. David. B. Duplantier and E. Guitter, Phys. Rev. Lett. 
{\bf 70} (1993) 2205; Nucl. Phys. {\bf B394} (1993) 555.}

\nref\rDaDuGuI{F. David, B. Duplantier and E. Guitter, Phys. Rev. Lett. {\bf 72}
(1994) 311.}

\nref\rDaWiI{F. David and K. J. Wiese, Phys. Rev. Lett. {\bf 76} (1996) 4564.}
\nref\rDaWiII{F. David and K. J. Wiese, Saclay Preprint T96/089, 
cond-mat/9608022, to appear in Nucl. Phys. B.}
\nref\rDaWiIII{K.J. Wiese and F. David, {Nucl. Phys.} {\bf B 450} (1995) 495.}

\nref\rBD{B. Duplantier, Phys. Rev. Lett. {\bf 62} (1989) 2337.}
\nref\rLasLip{See also M. L\"assig and R. Lipowsky, Phys. Rev. Lett. {\bf 70}
(1993) 1131.}

\nref\rBlum{L.~M.~Blumenthal, {\sl Theory and application of distance geometry},
Clarendon Press, Oxford, 1953.}

\nref\rElitzur{S. Elitzur, Nucl. Phys. B 212 (1983) 501.}
\nref\rDavidIR{F. David. Commun. Math. Phys. 81 (1981) 149.}

\nref\rSchoenberg{I.J. Schoenberg, Ann. Math. 38 (1937) 787.}

\nref\rBerLam{M. Berg\`ere and Y. M. P. Lam, J. Math. Phys. 17 (1976) 1546.}

\nref\rBerDav{M. Berg\`ere and F. David, Ann. Phys. 142 (1982) 416.} 

\nref\rBDW{B. De Witt, {\sl The Spacetime
Approach to Quantum Field Theory} in {\sl Relativity, Groups and Topology II},
Les Houches session XL (1983), B. De Witt and R. Stora Eds. , North-Holland
(1984)}

\nref\rBDtwo{B. Duplantier, {\sl Statistical Mechanics of Self-avoiding Crumpled
Manifolds}, in \rJerus .}

\nref\rGuiPal{E. Guitter and J. Palmeri, Phys. Rev. A 45 (1992) 734.}
\nref\rGoul{M. Goulian, J. de Physique II (1991) 1327.}
\nref\rLeDous{P. Le Doussal, J. Phys. A 25 (1992) L469.}
\nref\rDeGeTri{P.-G. de Gennes, J. de Physique 36 (1975) L-55.}

\newsec {Introduction}
Polymerized or tethered membranes and their statistical properties are a rich
and interesting subject
[\xref\rJerus,\xref\rHouches] , still under investigations.
These objects can be viewed as a simple two dimensional generalization of one
dimensional polymers \rNelPel .
However such membranes exhibit a larger variety of behavior than
polymers, due to the relevance of bending rigidity for two dimensional
objects (as can be seen by simple dimensional analysis) 
[\xref\rNelPel,\xref\rPelLei] .
In particular they might undergo a crumpling transition, separating a
high temperature (low rigidity) crumpled phase from a low temperature
(high rigidity) flat phase with anomalous elasticity
[\xref\rKaKaNe-\xref\rKanNe]  %\rPaKaNeI\rDaGu\rArLuI\rPaKaI\rKaNe 
.
Also different behaviors are obtained if the membrane is fluid, i.e. without
shear modulus.

In this paper we consider only the case of tethered (polymerized) membranes,
with internal in-plane elasticity, and we focus on the effect of self-avoidance
constraints on the crumpled phase.
For that purpose, a generalization of the celebrated Edwards model
\rEdwards\ 
for self-avoiding polymers has been introduced several years ago in
[\xref\rKaNe,\xref\rArLu] .
This model incorporates the internal elastic properties of the membrane
(in the crumpled phase) by a simple Gaussian entropic term, and self-avoidance
through a local contact interaction.
It allows for a systematic perturbative expansion in the steric interaction,
and for a renormalization group treatment similar to the direct renormalization
used for polymers within the Edwards model.
At zero order, the crumpled membrane is described by a free Gaussian term,
leading to a size exponent $\nu$ given by
\eqn\eNu{
\nu\ =\ {2-D\over 2} \ \ ,
}
where $D$ is the internal dimension of the object ($D=1$ for polymers,
$D=2$ for membranes).
Dimensional analysis shows that the canonical dimension for the steric
interaction is 
\eqn\eEps{
\epsilon\ =\ 2D-{2-D\over 2}d
\ \ ,}
where $d$ is the dimension of the bulk space in which the membrane is
embedded ($d=3$ for most physical situations).
Taking $\epsilon=0$ sets the upper critical dimension $d_c$
\eqn\eUCd{
d_c(D)\ =\ {4D\over 2-D}
}
below which self-avoidance is relevant (in the usual renormalization group
sense).
For polymers $d_c(1)=4$ and a systematic $\varepsilon=4-d$ expansion can be
performed, using perturbative renormalization group techniques
%\rFix\rDesClo\rDesCloJan.
[\xref\rFix-\xref\rDesCloJan] .
However for membranes $d_c(2)=+\infty$, and such a perturbative expansion cannot
be performed directly at $D=2$.
Still the model can be considered at a first stage for $D$-dimensional
membranes with internal dimension $0<D<2$, and renormalization calculations
have been performed at first order 
[\xref\rKaNe,\xref\rArLu,\xref\rHwa,\xref\rBDone]
, by expanding around the critical
dimension both with respect to $D$ and $d$. 
Thus starting from a $D=D_0<2$ and $d=d_c(D_0)$, one can reach the physical
line $D=2$, $d<\infty$ describing two dimensional membranes.

The calculations within this model are parallel to those of the direct 
renormalization for polymers
[\xref\rDesClo,\xref\rDesCloJan] .
The consistency of these calculations for polymers (in other words the
renormalizability of the Edwards model) is ensured by the famous equivalence
\rDeGennes\ 
between the Edwards model and a local quantum field theory (QFT) in the $d$-dimensional
embedding space: the O($n$) model in the limit $n\to 0$,
and by mapping the results of renormalization group theory for local QFT
[\xref\rDombGreen,\xref\rZinn]\ onto the Edwards model
[\xref\rBenMah,\xref\rDuplII] . 
Such an equivalence breaks down as soon as $D\ne 1$.
In order to validate the previous calculations for membranes, an important
issue is therefore the renormalizability of the generalized
Edwards model for arbitrary $D$.
This is not a simple problem.
Indeed,
although local in the embedding space, the interaction is non-local in the
internal $D$-dimensional space, since contacts may involve points arbitrarily
far apart along the membrane.
Thus the Edwards model can be viewed as 
a non-local field theory in internal space.
For general non-local theories no theory of renormalization exists.

A first step in this direction was made in \rDuHwKa , where it is
shown that the theory
is renormalizable at first order in perturbation.
In this paper we establish the renormalizability of this theory to all orders 
in perturbation theory.
This is done by using previous results by the authors for a simpler
model \rIIM , together with a new short-distance multi-local operator product
expansion (MOPE), which extends methods of local field theories to a large
class of models with non-local interactions, to which the generalized
Edwards model belongs.
A summary of these results has already been published in \rDaDuGuI , and they
are presented here in much greater details, with some applications.
Our results validate the direct renormalization scheme used before. 
They also ensure the existence of renormalization group equations leading
to scaling laws for self-avoiding membranes in the crumpled phase.
They finally provide a general and practical framework for explicit
calculations, which have been used in particular in 
[\xref\rDaWiI,\xref\rDaWiII] where explicit 
${\cal O}(\epsilon^2)$ estimates
for the size exponent $\nu$ for self-avoiding membranes have been obtained,
and in \rDaWiIII\ where the behavior of tricritical membranes at the
$\Theta$-point is studied more thoroughly.

The paper is organized as follows.
In Section~2 we recall the definition of the model and its perturbative
expansion.
This expansion admits a diagrammatic representation in terms of "dipoles"
interacting via a Coulomb potential in internal $D$-dimensional space.
A proper definition of the model in non-integer dimension $D$ is obtained
by use of "distance geometry", following a previous  work \rIIM .
Finally, we address the problem of infra-red (I.R.) divergences, first showing
that observables which are invariant under global translations in embedding
space are I.R. finite at zero order.

In Section~3 we focus on short distances, or ultra-violet (U.V.), divergences,
by first identifying those configurations which are singular, and
by showing that they always correspond to a limit of short distances between 
some end-points of the dipoles.
Around these configurations, we derive a multilocal operator product
expansion (MOPE), involving general multilocal operators, which fully
encodes the behavior of the amplitudes for the model at short distances.
This MOPE is then used to classify the U.V. singular configurations by
power counting. 
We show that these U.V. divergences are proportional to the insertion of
multi-local operators.
At the upper critical dimension, i.e. at $\epsilon=0$, the only
dangerous operators are those present in the original Edwards model,
which strongly suggests that the model is renormalized onto itself.

Section~4 is devoted to the proof of renormalizability of the model:
U.V. divergences can be subtracted and the theory can be made U.V. finite at 
$\epsilon=0$ by adding appropriate counterterms to the Hamiltonian of
the Edwards model.
We focus on the simple case of infinite membranes with internal flat geometry.
We first show that the model is U.V. finite for $\epsilon>0$
(super-renormalizable case) (Section~4.1),
as well as I.R. finite when $\epsilon \to 0_+$, as long as translationally
invariant observables are concerned (Section~4.2).
We then analyse the U.V. divergences at $\epsilon=0$, making repeated use of 
the formalism previously developed in \rIIM\ for the model of a
free membrane interacting with a single point [\xref\rBD,\xref\rLasLip] .
Indeed, this formalism, developed for local but singular interactions, can
be adapted to the class of singular multilocal interactions that we consider
here.
In Section~4.3, single U.V. singularities (superficial divergences) are
classified in terms of dipole configurations which are called diagrams,
following \rIIM .
A subtraction prescription for these superficial divergences is then 
introduced.
In Section~4.4 the nested structure of multiple U.V. divergences and the
required
subtractions are presented on simple examples at second order.
Iterating this analysis naturally leads to a formulation of renormalization
in terms
of nested families of successive divergent subdiagrams, which are analogous to
Zimmermann's forests of renormalization theory in perturbative local
QFT.
A forest formula for a subtraction operator is given in Section~4.5, and is
shown to correspond to the introduction of multilocal counterterms in the
Hamiltonian of the model.
Section~4.6 is the core of the proof.
We show that the renormalized observables are U.V. finite at $\epsilon=0$,
by using a strategy analogous to that of \rIIM .
The distance integration domain is decomposed in sectors, analogous to Hepp's
sectors in perturbative local QFT, and forests of subtractions are reorganized
with respect to each sector in order to ensure U.V. finiteness.
The argument is quite technical, and relies on the techniques developed in
\rIIM , but is presented here in a consistent way.
Finally in Section~4.7 we discuss some additional points, in particular we
show that renormalization does not affect the I.R. finiteness of invariant
observables at $\epsilon=0$.

The rest of the paper presents applications of this renormalization formalism.
In Section~5 we derive the renormalization group equations for the model, and
we show how the scaling laws for the self-avoiding membrane model are derived
for $\epsilon>0$.
As an application, we perform renormalization at first order in perturbation
theory, using our formalism, and we show how the $\CO(\epsilon)$ result of
[\xref\rKaNe,\xref\rArLu]\ for the size exponent $\nu$ is recovered.

Section~6 is devoted to the important case of finite membranes.
Indeed, the direct renormalization method requires the study of finite 
membranes (or polymers), in particular since it is the internal size $L$
of the membrane which fixes the renormalization scale.
In Sections~6.1-4 
we show in details how our formalism can be extended and applied to the case of
finite (open or closed) membranes,  possibly with non-flat internal geometry
(i.e. with frozen Gaussian curvature), and to the case of several interacting
membranes.
We show in particular in Section~6.5
how to derive finite size scaling laws for finite
membranes, and we discuss the hyperscaling relations 
\rBDone\ between the
size exponent $\nu$ and the configuration exponent $\gamma$.
In Section~6.6
we discuss in details the relation between the direct renormalization
method and our renormalization formalism, and we
prove that the scaling assumptions
which are the starting point of direct renormalization are indeed correct.
Finally in Section~6.7 we discuss the contact exponents which describe the
scaling laws for contacts between different elements of the membrane.
In particular we show how to take into account properly the edge effects
in the calculation of edge contact exponents.

Finally Section~7 is devoted to two other models of interacting membranes,
which are also amenable to our formalism.
The first case is flexible tethered membranes with long range repulsive
interactions, for instance uniformly charged membrane with non-screened
Coulomb interactions.
We show that for long-range interactions our formalism implies that there is
a renormalization of the internal elasticity, but no renormalization of the 
interaction (charge), and that this leads to the fact (already well known for
polymers)
that the size exponent $\nu$ is exactly that given by a Gaussian variational
method.
The second case is that of membranes with competing attractive and repulsive
short range interactions, corresponding to the generalization of polymers
at the $\Theta$-point (separating the swollen phase from the collapsed phase).
While for polymers it is well know that a repulsive three-body interaction is
relevant to describe the scaling properties of the $\Theta$-point (at least in
the $\varepsilon=4-d$ expansion), we show that for membranes there is a
competition between this three-body interaction and a modified two-body interaction. 
We then perform analytically the first order calculation of the renormalization
group flow for this new interaction.
We refer to \rDaWiIII\ for a subsequent and
more complete analysis of this problem and a treatment of the
crossover between the two interactions.

Section~8 is the conclusion and discusses still open problems.
Some more technical points are discussed in the appendices.

%\vfill\break
\newsec {The Perturbative Expansion}
\subsec {Perturbation Theory Diagrammatics}
In this section we define the model by giving the formal rules of its
perturbative expansion.
We start from the Edwards Hamiltonian, generalized to the case of a
manifold with arbitrary internal dimension $D$.
Each point on the manifold is labeled by a $D$-dimensional vector 
$x\in\RR^D$
%($x=\{x_i\}$, $i=1,D$)%
.
The position of the points $x$ in the embedding $d$-dimensional Euclidean space
defines a vector field
\foot{In this paper, bold quantities like $\rvec=\{{\bf r}^\mu\}$,
$\mu=1,\ldots ,d$ refer to the external
$d$-dimensional space, and non-bold quantities like
$x=\{x^\alpha\}$, $\alpha=1,\ldots ,D$ to the internal $D$-dimensional space.}
$\rvec(x)\in\RR^d$.
The continuum Hamiltonian reads
\eqn\eEdwards{
%\beta\,{\cal H}
{\raise.2ex\hbox{$\CH[\rvec]$}/\raise -.2ex\hbox{${\rm k}_{\rm B}T$}}
\ =\ {1\over 2}\,\int d^Dx\,\big(\nabla_{\! x}\rvec (x)\big)^2+
{b\over 2}\int d^Dx\int d^Dx'\ \delta^{d}\big(\rvec (x)-\rvec (x')\big)
\ .}
The first Gaussian term 
$(\nabla_{\! x}\rvec (x))^2=\sum_{\alpha=1}^D(\partial\rvec(x)/\partial
x^\alpha)^2$
describes the local elastic energy.
This term alone describes for $b=0$ a ``phantom" polymerized membrane with
no self-avoidance, and is mostly of entropic origin.
The second term is a two-body short range $\delta$-potential which describes
weak self-avoidance with excluded volume parameter $b>0$.
This term is both singular (it is a singular function -- in fact the Dirac
 distribution in $\RR^d$ -- of the field $\rvec$)
and non-local in the internal space $\RR^D$.
In this respect this theory is quite different from usual local field theories.
With our choice of units, the Hamiltonian \eEdwards\ is dimensionless.
The Hamiltonian \eEdwards\ is invariant under global translations in
embedding space
$\rvec(x)\to\rvec(x)+\rvec_0$.
In \eEdwards\ the integrals run over $\RR^D$, which means that we consider
an infinite membrane.
The case of finite membranes will be discussed in Sec. 6.

The partition function is defined as a sum over all membrane configurations
$\rvec(x)$.
More precisely, in this continuum formulation it is given by the
functional integral:
\eqn\ePartFunc{
\CZ_b\ =\ \int \CD[\rvec(x)]\,\exp\left(
-{\raise.2ex\hbox{$\CH[\rvec]$}/\raise -.2ex\hbox{${\rm k}_{\rm B}T$}}
\right)
\ .
}
%We normalize the measure $\CD[\rvec(x)]$ so that the partition function
%for the free Gaussian membrane (with $b=0$) is $\CZ_0=1$.
It has a perturbative expansion in $b$, formally given by
expanding the exponential of the contact interaction
\eqn\eZPertExp{
\CZ_b\ =\ \CZ_0\ \sum_{N=0}^\infty\,{(-b/2)^N\over N!}\,
\int\sprod_{i=1}^{2N}{d^Dx_i}
\bigl\langle
\sprod_{a=1}^{N}\delta^{d}(\rvec(x_{2a})-\rvec(x_{2a-1}))
\bigr\rangle_0
\ ,
}
where $\CZ_0$ is the partition function of the Gaussian manifold,
and
$\langle\cdots\rangle_0$ denotes the average with respect to the
Gaussian manifold ($b=0$):
\eqn\evevo{
\langle(\cdots)\rangle_0\ =\ {1\over\CZ_0}\,
\int \CD[\rvec(x)]\,\exp\left(-{1\over 2}\int d^Dx\,
(\nabla_x\rvec(x))^2\right)(\cdots)
\ .
}

Physical observables are provided by average values of operators, which
must be invariant under global translations.
Using Fourier representation,
local operators can always be generated by the exponential operators
(or vertex operators), of the form
\eqn\eVertOp{
V_{\qvec}(z)\ =\ {\rm e}^{\ii\qvec\cdot\rvec(z)}
\ .
}
In perturbation theory the field $\rvec(x)$ will be treated as a massless free 
field and the momenta $\qvec$ will appear as the
``charges" associated with the translations in $\RR^d$.
Translationally invariant operators are then provided by ``neutral" products
of such local operators,
\eqn\eNeutObs{
O_{\qvec_1,\cdots,\qvec_P}(z_1,\cdots,z_P)\ =\ \prod_{l=1}^{P}\,
V_{\qvec_l}(z_l)
\qquad,\qquad \qvec_{\rm total}\ =\ \sum_{l=1}^{P}\,\qvec_l\ =\ {\vec{\bf 0}}
\ .
}
The perturbative expansion for these observables is simply
\eqn\eCorrFirst{
\bigl\langle \sprod_{l=1}^{P}{\rm e}^{\ii \qvec_l\cdot\rvec(z_l)}\bigr\rangle
\ =\ {1\over \CZ_b}\,\sum_{N=0}^\infty\,{(-b/2)^N\over N!}\,
\int\sprod_{i=1}^{2N}{d^Dx_i}
\bigl\langle
\sprod_{l=1}^{P}{\rm e}^{\ii \qvec_l\cdot\rvec(z_l)}
\sprod_{a=1}^{N}\delta^{d}(\rvec(x_{2a})-\rvec(x_{2a-1}))
\bigr\rangle_0
\ .
}
Each $\delta$ function in \eZPertExp\ and \eCorrFirst\ 
can itself be written in terms of two exponential operators as
\eqn\eExpRep{
\delta^{d}(\rvec(x_{2})-\rvec(x_{1}))\ =\ 
\int{d^d\kvec_{1}d^d\kvec_{2}\over (2\pi)^d}
\,\delta^d(\kvec_{1}+\kvec_{2})
\,{\rm e}^{\ii \kvec_{1}\cdot \rvec(x_{1})}
\,{\rm e}^{\ii \kvec_{2}\cdot \rvec(x_{2})}
\ .
}
Viewing again the momenta $\kvec_1$, $\kvec_2$ as charges assigned to the
points $x_1$, $x_2$, the bi-local operator \eExpRep\ corresponds to a dipole,
with charges $\kvec_1=\kvec$, $\kvec_2=-\kvec$, integrated over its internal
charge $\kvec$. 
We depict graphically each such dipole as on 
\fig\fDipole{\hyperref\hDipole{A dipole}}%
.
\topinsert
\centerline{\epsfxsize=3.truecm\epsfbox{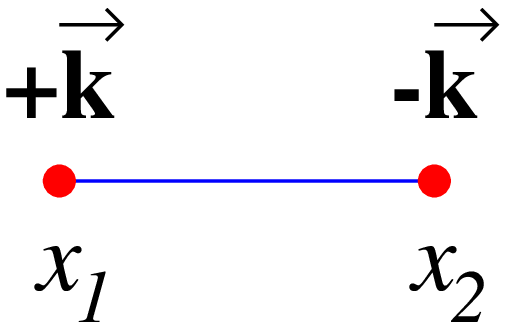}}
\centerline{\hyperdef\hDipole{userfigure}{uDipole}{\fDipole}}
%\centerline{\fDipole}
\endinsert

Similarly, the product of bi-local operators in \eZPertExp\ and \eCorrFirst\ 
can be written as an ensemble of $N$ dipoles, that is as the  product of
$2N$ vertex operators with $N$ ``dipolar constraints"
%integrated over their $2N$ internal charges $\kvec_a$,
\eqn\eDipConst{
\CC_a\{\kvec_i\}\ =\ (2\pi)^d\delta^d(\kvec_{2a-1}+\kvec_{2a})
\ ,
}
then integrated over all internal charges $\kvec_i$:
\eqn\eProdExp{
\sprod_{a=1}^{N}\delta^{d}(\rvec(x_{2a})-\rvec(x_{2a-1}))=
\int\sprod_{i=1}^{2N}{d^d\kvec_i\over (2\pi)^d}
\sprod_{a=1}^{N}\CC_a\{\kvec_i \}
\sprod_{i=1}^{2N}{\rm e}^{\ii \kvec_i\cdot \rvec(x_i)}
}
Products of such bi-local operators and of external vertex operators,
as in \eCorrFirst , are depicted by diagrams such as that of 
\fig\fDiagr{\hyperref\hDiagr{A Diagram}}
\topinsert
\centerline{\epsfxsize=7.5truecm\epsfbox{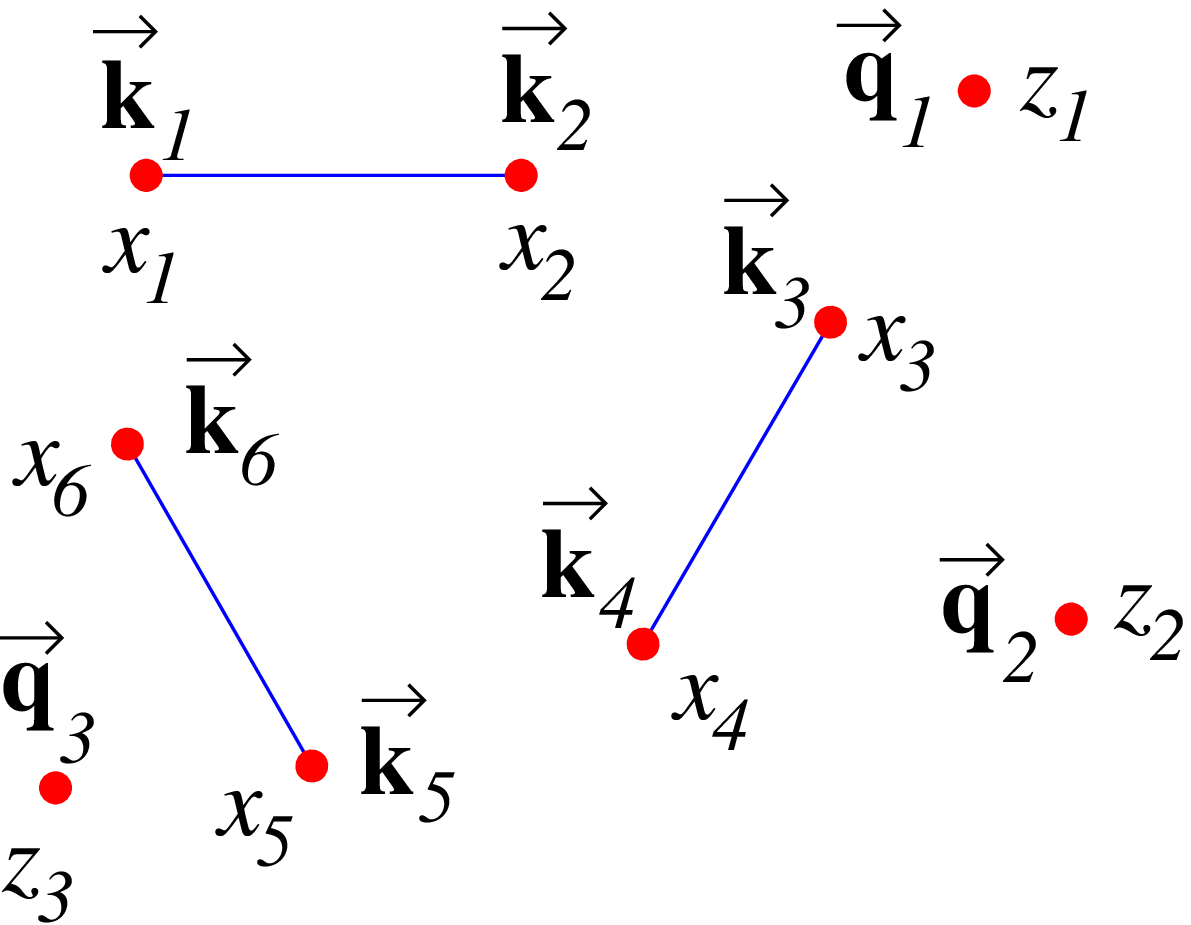}}
\centerline{\hyperdef\hDiagr{userfigure}{uDiagr}{\fDiagr}}
%\centerline{\fDiagr}
\endinsert

The Gaussian average in \eZPertExp\ is easily performed.
We use the identity
\eqn\eGausAv{
\bigl\langle \sprod\limits_{i}{\rm e}^{\ii\kvec_i\cdot\rvec(x_i)}\bigr\rangle_0
\ =
\ \exp\Big({-{1\over 2}\sum\limits_{i,j}\kvec_i\cdot\kvec_j 
\langle{\bf r}(x_i){\bf r}(x_j)\rangle_0}\Big)
\ ,
}
where ${\bf r}$ is any single component of $\rvec$.
Thanks to the neutrality condition $\sum_i\kvec_i=0$, we can rewrite it as
%The massless propagator $\langle{\bf r}(x_i){\bf r}(x_j)\rangle_0$
%is I.R. divergent for $D<2$, but thanks to the neutrality condition
%$\sum_i\kvec_i=0$, we can rewrite the l.h.s. of \eGausAv\ as
\eqn\eGausBis{
\bigl\langle \sprod\limits_{i}{\rm e}^{\ii\kvec_i\cdot\rvec(x_i)}\bigr\rangle_0
\ =\ 
\exp\Big({-{1\over 2}\sum\limits_{i,j}\kvec_i\cdot\kvec_j 
G(x_i-x_j)}
%{-1\over 2}\langle({\bf r}(x_i)-{\bf r}(x_j))^2\rangle_0}
\Big)
\ ,
}
with the translationally invariant two-point function
\eqn\eMlssProp{
G(x_i,x_j)\ =\ 
-\,{1\over 2}\,\langle \big({\bf r}(x_i)-{\bf r}(x_j)\big)^2\rangle_0\ =\ 
-\,{|x_i-x_j|^{2-D}\over (2-D)S_{D}}
\ .
}
%is the massless propagator in $D$ dimensions for 
%any single component ${\bf r}$ of $\rvec$,
%with 
Here 
\eqn\eVolSph{
S_{D}\ =\ {2\,\pi^{D/2}\over\Gamma(D/2)}
}
is the volume of the unit sphere in $D$ dimensions.
In \eMlssProp\ and \eVolSph\ the standard analytic continuation in the
dimension $D$ to $0<D<2$ is performed.

%Strictly speaking the massless propagator \eMlssProp\ in real space is free of
%infra-red divergences only for $D>2$.
%For $D\le 2$ a divergent term proportional to $m^{D-2}$, with $m$ some 
%infra-red mass regulator, appears when $m\to 0$.
%In the following we shall use \eMlssProp\ in the range $0<D<2$, which means
%that the I.R. divergence is subtracted by a finite part prescription.
%This appears to be justified for translationally invariant observables,
%thanks to the neutrality condition in \eNeutObs .
%We shall discuss these problems of infra-red divergences in Section 2.3.

Integration over the momenta $\kvec_i$ then gives for the $N$'th
term of the perturbative expansion for the partition function $\CZ_b$
\eZPertExp\ the ``manifold integral"
\eqn\eZManInt{
%\hbox{\it some factor}\,
(2\pi)^{-Nd/2}\,
\int \sprod_{i=1}^{2N} d^D x_i\,\Delta\{x_i\}^{-{d\over 2}}\,
}
with $\Delta\{x_i\}$ the determinant associated with the quadratic form 
(on $\RR^{2N}$) $Q\{\kbf_i\}=\ssum\limits_{i,j=1}^{2N} \kbf_i \kbf_jG(x_i,x_j)$ 
restricted to the $N$-dimensional vector space defined by the
$N$ neutrality constraints $\CC_a\{\kbf_i\}$, $\kbf_{2a}+\kbf_{2a-1}=0$.
$\Delta\{x_i\}$ is given explicitly by the determinant of the
$N\times N$ matrix $\Delta_{ab}$
(with row and columns labeled by the dipoles indices $a,b=1,\cdots ,N$)
\eqn\eDelExpl{
\Delta\ =\ \det\big(\Delta_{ab}\big)\quad,\quad
\Delta_{ab}\ =\ 
G(x_{2a-1},x_{2b-1})+G(x_{2a},x_{2b})-G(x_{2a-1},x_{2b})-G(x_{2a},x_{2b-1})
}

Similarly, the $N$'th term in the perturbative expansion of the $P$-point
observable \eCorrFirst\ is
\eqn\eVManInt{
%\hbox{\it some factor}\,
(2\pi)^{-Nd/2}\,
\int \sprod_{i=1}^{2N} d^D x_i\,\Delta\{x_i\}^{-{d\over 2}}\,
%{\rm e}^{-{\scriptscriptstyle{1\over 2}}\sssum\limits_{l,m=1}^{P}
%\qvec_l\cdot\qvec_m\, {{\scriptstyle \N^{lm}}
\exp\left(-\,{1\over 2}\ssum\limits_{l,m=1}^{P}
\qvec_l\cdot\qvec_m\, {N^{lm}\over\Delta}\right)
}
$N^{lm}$ is the $(lm)$ minor of the $(P+N)\times(P+N)$
matrix 
\eqn\ePplusNMatrix{
%\left[
%\matrix{
%%
%%\overbrace{G_{lm}}^P&\overbrace{\Gamma_{lb}}^N\cr\Gamma_{am}&\Delta_{ab}\cr
%%
%(G_{lm})_{1\le l,m\le P}&(\Gamma_{lb})_{1\le l\le P\atop 1\le b\le N}\cr
%(\Gamma_{am})_{1\le a\le N\atop 1\le m\le P}&(\Delta_{ab})_{1\le a,b\le N}\cr
%}
%\right]
\left[
\matrix{
%G_{lm}
G(z_l,z_m)
&
%\Gamma_{lb}
G(z_l,x_{2b-1})-G(z_l,x_{2b})
\cr
%\Gamma_{am}
G(x_{2a-1},z_m)-G(x_{2a},z_m)
&
\Delta_{ab}
\cr
}
\right]_{1\le l,m\le P\atop 1\le a,b \le N}
\ .
}
%with
%\eqn\eGlm{
%G_{lm}\ =\ G(z_l,z_m)
%\quad,\quad
%\Gamma_{am}\ =\ \Gamma_{ma}\ =\ 
%G(x_{2a-1},z_m)\,-\,G(x_{2a},z_m)
%}

%\vfill\break
\subsec {The Distance Measure}

In the integrals \eZManInt\ and \eVManInt\ the dimension $d$ of external
space appears only as a power, and can be considered as a continuous parameter
\foot{Strictly speaking the observables \eVManInt\ are then defined as functions
of the invariants $\qvec_l\cdot\qvec_m$.}
$0<d<\infty$.
For the reasons explained above, we also need to treat the internal dimension
$D$ as a continuous variable $0<D<2$.
This can be done, thanks to the formalism developed in \rIIM .
The integrand in \eZManInt\ 
is a well defined function
of the relative Euclidean distances between the internal points
$d_{ij}=|x_i-x_j|$, thanks to the Euclidean invariance of the model
in internal $D$-dimensional space.
The integral over the $M=2N$ points $x_i$ in $\RR^D$ can be rewritten as
an integral over the squared distances 
\eqn\esqrdist{
a_{ij}\ = \ d_{ij}^2\ =\ (x_i-x_j)^2
}
times a global internal translation factor ${\cal V}=\int_{\RR^D}d^Dx_1$
(the volume of the membrane):
\eqn\eDistMeas{
\int_{\RR^D}\prod_{i=1}^{M=2N} d^Dx_i\ =\ {\cal V}\,\int_{{\cal A}_M}
\prod_{\scriptscriptstyle 1\le i<j\le M} \!\!da_{ij}\,\mu_M^D[a_{ij}]
}
with the measure on the integration domain $\CA_M$ (defined below)
\eqn\eMuMeas{
\mu_M^D[a_{ij}]\ = 2^{-{M(M-1)\over2}}\,S_D\,S_{D-1}\cdots S_{D-M+2}\,
\Big(P_M[a_{ij}]\Big)^{(D-M)/2}
\ ,
}
where $P_M[a_{ij}]$ is the Cayley-Menger determinant
\rBlum
\eqn\eCayMen{
P_M[a_{ij}]\ =\ {(-1)^M\over 2^{M-1}}\,
\left|
\matrix{0&1&1&\ldots&1\cr 1&0&a_{12}&\ldots&a_{1M}\cr
1&a_{21}&0&\ldots&a_{2M}\cr
\vdots&\vdots&\vdots&\ddots&\vdots\cr
1&a_{M1}&a_{M2}&\ldots&0\cr
}\right|
}
The integration domain ${\cal A}_M$ is the subset of the variables
$a_{ij}$ which can be
realized as actual squared distances between $M$ points in $\RR^{M-1}$.
It is given by the $M-1$ constraints $P_K[a]\ge 0$, $2\le K\le M$
\foot{
For instance in the case of three points ($M=3$),
the two non-trivial constraints
$P_2\ge 0$ and $P_3\ge 0$ imply that the squared distance variables $a_{ij}$
are positive, and satisfy the triangular inequalities
$|\sqrt{a_{12}}-\sqrt{a_{23}}|\le\sqrt{a_{13}}\le\sqrt{a_{12}}+\sqrt{a_{23}}$.
}%
.
Geometrically the Cayley-Menger determinant is simply the squared volume of the
$(M-1)$-dimensional parallelepiped generated by the vertices $x_1$, $\ldots$,
$x_M$.
Originally the correspondence \eDistMeas\ is valid only for integer dimension
$D\ge M-1$.
It can then be analytically continued for non-integer dimension $0<D<\infty$.
For $D>M-2$ the measure \eMuMeas\ remains integrable. 
For $D\le M-2$, \eMuMeas\ becomes singular due to the vanishing of
the sphere volumes $S_{D-K}$ at $D=K-L$, $L\in\NN$, i.e. for integer $D\le M-2$, and to the
divergences of $P_M[a]^{-(D-M)/2}$ when the determinant $P_M[a]$ vanishes.
It has been shown in \rIIM\ that these divergences occur on the boundary of
$\CA_M$, and can be dealt with for non-integer $D$ by a finite part
integration prescription.
Thus $\mu_M^D[a]$ becomes for $D\le M-2$ a well-defined distribution
with support on ${\cal A}_M$.
For integer $D\le M-2$ this distribution concentrates on the boundary of
$\CA_M$, and more specifically on those submanifolds which describes
distances which can be realized as actual distances between $M$ points in
$\RR^D$.
%for $D\le M-2$, which can be constructed
%explicitly by analytic continuation in $D$ and finite part prescriptions
%to deal with the divergences which occur when $P_M[a]$ vanishes.

For correlation functions like \eCorrFirst\ one has to perform the integral
\eVManInt .
The integrand is now a function of the squared distances
%$a_{ij}=(x_i-x_j)^2$, $a_{il}=a_{li}=(x_i-z_l)^2$ and $a_{lm}=(z_l-z_m)^2$
$a_{ij}$, $a_{il}$ and $a_{lm}$
between the $M=2N$
internal points $x_i$ and the $P$ external points $z_l$.
The integral over the internal points is rewritten as an integral over
the variables $a_{ij}$ and $a_{il}$, the distances between the
external points, that is the $a_{lm}$, being kept fixed:
\eqn\eMeasExt{
\int_{\RR^D}\prod_{i=1}^{M} d^Dx_i\ =\ \int_{{\cal A}_{M,P}}
\prod_{\scriptscriptstyle 1\le i<j\le M} \!\!\!da_{ij}
\prod_{1\le i\le M\atop 1\le l\le P}\!\!da_{il}
\ \mu_{M,P}^D[a_{ij},a_{il},a_{lm}]
\ ,
}
with the measure depending now on the Cayley-Menger determinant
$P_P[a_{lm}]$ relative to the $P$ external points, and on the
Cayley-Menger determinant
%$P_{M+P}\left[\left(\matrix{a_{ij}&a_{il}\cr a_{mj}&a_{lm}\cr}\right)\right]$
$P_{M+P}[a_{ij},a_{il},a_{lm}]$
relative to the $M+P$ external and internal points:
\eqn\eMeasExt{
\mu_{M,P}^D[a_{ij},a_{il},a_{lm}]\ =\ 2^{-{M(M+2P-1)\over 2}}\,S_{D-P+1}\ldots
S_{D-P-M+2}\,{\Big(
P_{M+P}[a_{ij},a_{il},a_{lm}]
\Big)^{D-M-P\over 2}
\over
\Big(
P_P[a_{lm}]\Big)^{D-P\over 2}
}
\ .
}
$\CA_{M,P}\equiv\CA_{M,P}(a_{lm})$ is now the subset of the variables
($a_{ij}$,$a_{il}$) which can be realized as
actual distances between $M$ points and between these $M$ points and the
$P$ fixed external points in $\RR^{M+P-1}$.
For generic $a_{lm}$ in $\CA_P$, this measure is integrable for
$D>M+P-1$, and becomes a distribution for $D\le M+P-1$.
Alternative parametrizations of the measures \eMuMeas\ and \eMeasExt\ are
given in \rIIM .

%\vfill\break
\subsec {Infra-Red Problems}

The perturbative expansion of the model has potential infra-red divergences
for $D\le 2$.
In fact, infra-red divergences arise already for the free theory ($b=0$)
if we deal with observables which are not translationally invariant.
Indeed, the free massless propagator which appears in \eGausAv\ is
infra-red singular for $D\le 2$.
It can be defined properly by introducing an infra-red regulator (represented by
a length scale $L$), for instance by considering  a finite membrane
\foot{This requires a proper treatment of the global zero mode, see
\rIIM .}
with internal linear size of order $L$, or by adding a mass term
$\int m^2 \rvec ^2$ in the Hamiltonian \eEdwards , with $m\propto 1/L$,
corresponding to a confining potential in external space.
With such regulators, the free propagator is finite and behaves,
for $L\gg |x_i-x_j|$, as
\eqn\eTruePr{
\langle {\bf r}(x_i){\bf r}(x_j)\rangle_0\ {\buildlim L\to\infty\under =} \ 
{L^{2-D}\over (2-D)S_D}\,-\,{|x_i-x_j|^{2-D}\over (2-D)S_{D}}\,+\,%\ldots
{\cal O}(L^{-D})
\ .
}
For translationally invariant observables, we have seen that
the neutrality conditions over the external momenta and the neutrality
constraints $\CC_a$ imply that in \eGausAv\ the infra-red divergent propagator
can be replaced by the translationally invariant two-point function
$G(x_i,x_j)$ given by \eMlssProp , which is infra-red finite.
In fact $G(x_i,x_j)$ can be considered as the infra-red finite part of
the free propagator \eTruePr , defined for instance by letting $L\to\infty$ for
$D>2$, and then by analytic continuation to $D<2$.
This is a first example of cancellation of infra-red divergences.

Another source of infra-red divergences comes from the integration over
internal points $x_i$ in \eZManInt\ and \eVManInt , when some of these points
go to infinity.
However, we expect these divergences to disappear when considering average
values like \eCorrFirst , by a compensation of these divergences in the
numerator, given by integrals like \eVManInt , and those in the denominator
$\CZ_b$, given by integrals like \eZManInt .
At variance with the cancellations of I.R. divergences in the integrand, which
occur for $0<D<2$ and for any $0<d<\infty$,
these cancellations should
occur only in the vicinity of the upper critical dimension
$d_c(D)=4D/(2-D)$ where the model is renormalizable.
For instance, one can check by an explicit calculation that at first order in
$b$, the invariant observables \eCorrFirst\ are I.R. finite as long as
$\epsilon=2D-(2-D)d/2<D$.
Similarly, the free energy density $f=-\ln(\CZ_b)/\CV$ ($\CV$ being the
volume of the membrane) is I.R. finite at order $b$, provided that
$\epsilon<D$.
At this stage, we can only make the following conjecture:
\medskip
{\it The term of order $b^n$ of the perturbative expansion of invariant
observables is I.R. finite for $0<D<2$ and $\epsilon<D/n$}.
\medskip
This statement is analogous to the Elitzur theorem
[\xref\rElitzur,\xref\rDavidIR] , which states that invariant
observables are IR finite in two dimensional non-linear sigma models.
We shall come back to this point later.

%\vfill\break
\newsec {Ultra-Violet Singularities and the Multilocal Operator Product
Expansion}
\subsec {UV Singular Configurations}
%- Generalized Sch\"onberg Theorem %or whatever Name is has
%- Coulombic Formulation}
A first condition 
for the manifold integrals \eZManInt\ and \eVManInt\ to be well defined is that
the determinant $\Delta\{x_i\}$ must be nonnegative.
%Moreover, the integrand becomes singular if this determinant vanishes.
This can be shown by methods used to prove the so-called
Sch\"onberg theorem \rSchoenberg , that we now discuss, using a convenient electrostatic
formulation.

Let us consider the quadratic form
\eqn\eQuadF{
Q\{{\bf k}_i\}\ =\ \sum_{i,j=1}^{M} {\bf k}_i{\bf k}_j\,G(x_i,x_j)
%\quad,\quad
%G(x_i,x_j)\ =\ -\,{\Gamma(D/2)\,\pi^{-D/2}\over 2(2-D)}\,|x_i-x_j|^(2-D)
%Q\{{\bf k}_i\}\ =\ -\,\sum_{i,j=1}^{2N} {\bf k}_i{\bf k}_j\,|x_i-x_j|^{2\nu}
}
restricted to the $(M-1)$-dimensional subspace $E\subset\RR^M$
defined by the global neutrality constraint
\eqn\eGlNeu{
\sum_{i=1}^{M}{\bf k}_i\ =\ 0
\ .
}
Here $G(x_i,x_j)$ is given by \eMlssProp , so that $Q\{{\bf k_i}\}$
can be viewed as the electrostatic energy of a globally neutral system of
$M$ charges ${\bf k}_i$ in $\RR^D$.
We still take $0<D<2$, and make use of distance geometry to view
$Q$ as a function of the squared distances $a_{ij}=(x_i-x_j)^2$, taken  
in the domain ${\cal A}_{M}$, which can be realized as actual
squared distances between $M$ points $x_i$ in $\RR^{M'}$, with $M'\ge M-1$
integer.

\medskip
\noindent{\it Sch\"onberg Theorem:}

For $0<D<2$, the quadratic form  $Q\{{\bf k}_i\}$ is positive on $E$.
Moreover, $Q\{\bf k_i\}=0$ if and only if the local ``charge density" (defined
in $\RR^{M'}$, $M'\ge M-1$, by $\rho(x)=\sum_{i=1}^{M}{\bf k}_i\delta(x-x_i)$)
is zero everywhere.

\medskip
\noindent{\bf Proof}

We simply use the Fourier transform representation of the kernel
$-|x_i-x_j|^{2\nu}$ in $\RR^{M'}$, valid for $0<\nu<1$,
\eqn\eFTG{
-\,|x_i-x_j|^{2\nu}\ =\ C(M',\nu)\,\int {d^{M'}\! q\over (2\pi)^{M'}}
\,|q|^{-2\nu-M'}\, \left({\rm e}^{{\rm i}q(x_i-x_j)}-1\right)
\ ,
}
with $C(M',\nu)$ a positive constant (for $0<\nu<1$)
\eqn\eFTCst{
C(M',\nu)\ =\ 4^{\nu+{M'\over 2}}\,\pi^{{M'\over 2}}\,
{\nu\Gamma(\nu+{M'\over 2})\over\Gamma(1-\nu)}
\ .
}
Taking $\nu=(2-D)/2$ and using the neutrality constraint
$\sum_i{\bf k}_i=0$ we get:
\eqn\eFTQ{
Q\{{\bf k}_i\}\ =\ {C(M',\nu)\over (2-D)S_D}\,
\int {d^{M'}\! q\over (2\pi)^{M'}} \,|q|^{-2\nu-M'}\, 
\left|\sum_{i=1}^{M}{\bf k_i}\,{\rm e}^{{\rm i}qx_i}\right|^2
\ .
}
Since the kernel $|q|^{-2\nu-M'}$ is positive, the positivity of $Q$ follows.
Moreover,
$Q=0$ if and only if the Fourier transform of the charge density $\rho(x)$
\eqn\eFTrho{
\hat\rho(q)\ =\ \sum_{i=1}^{M}{\bf k_i}\,{\rm e}^{{\rm i}qx_i}
}
is zero, i.e. if $\rho$ vanishes everywhere.

\medskip
We now consider the quadratic form $Q$ associated to the $M=2N$ points $x_i$
in \eZManInt .
Since $Q\{{\bf k}_i\}$ is positive on the $(M-1)$-dimensional subspace
$E$, it is also
positive on the $N$-dimensional subspace $F\subset E$ defined by the $N$ dipole
neutrality constraints $\CC_a\{{\bf k}_i\}$
\eqn\eFVecSp{
F\ =\ \left\{\{{\bf k}_i\}\subset\RR^{2N}:\ 
{\bf k}_{2a-1}-{\bf k}_{2a}=0,\ a=1,\ldots N\right\}
}
This implies that $\Delta\{x_i\}=\det_F(Q)\ge 0$ in ${\cal A}_{2N}$.
Similarly, the quadratic form
\eqn\eQbar{
%{\bar Q}\{\qvec_l\}\ =\ \sum_{l,m=1}^P\qvec_l\cdot\qvec_m\,{N^{lm}\over\Delta}
{\bar Q}\{\qbf_l\}\ =\ \sum_{l,m=1}^P\qbf_l\cdot\qbf_m\,{N^{lm}\over\Delta}
}
in the P-point integral \eVManInt , which depends on the $P$ external charges 
$\qbf_l$, such that $\sum_{l=1}^P\qbf_l=0$, is given by the minimum over all
possible internal charges $\{\kbf_i\}\in F$ of the global quadratic form
$Q\{\qbf_l,\kbf_i\}$. Therefore $\bar Q\{\qbf_l\}\ge 0$.

Still the integrals \eZManInt\ and \eVManInt\ may diverge if $\Delta\{x_i\}=0$,
i.e. 
if and only if the quadratic form $Q$ has a non-zero isotropic
vector $\{\kbf_i\}\in F$, such that
\eqn\eIsoVec{
Q\{\kbf_i\}\ =\ 0\quad,\quad\{\kbf_i\}\neq\{0,\ldots ,0\}
\ .
}
The Sch\"onberg theorem allows us to identify precisely when this happens:
$\Delta\{x_i\}=0$ if and only if the $2N$ dipoles have their positions such
that the charge density $\rho(x)$ can be kept zero everywhere,
while some dipoles have (appropriate) non-zero charges.
If all the points $i$ occupy distinct positions in $\RR^D$, this is
clearly impossible.
Thus $\Delta\{x_i\}=0$ first requires that some points must coincide.
More precisely:
\item{(1)} To any singular configuration is associated a partition of the $2N$
points into ``atoms" ${\CP}$, such that the points $i$ which
belong to the same atom ${\CP}$ share the same position $x_\CP$
(in the distance space this means $i,j\in\CP\Rightarrow a_{ij}=0$).
Some atoms $\CP$ may consist of only one point $\{i\}$.
Since the points $i$ are end-points of the dipoles $a$, the atoms $\CP$
are assembled into ``molecules",
defined as connected sets of atoms attached by dipoles
(see
\fig\fmolecule{\hyperref\hmolecule{A singular configuration} is
obtained by contracting some points and by forming ``molecules".}%
).
A singular configuration is thus associated to a grouping of the points into
molecules.
\par\noindent
Moreover, the condition of zero charge density means that each atom $\CP$
must be neutral:
\eqn\eNeuAt{
\sum_{i\in\CP}\kbf_i\ =\ 0
\ ,
}
while at least one dipole carries a non-zero charge.
Clearly this is possible if and only if the non-zero dipoles belong to a closed
loop in the molecule.
Thus $\Delta\{x_i\}=0$ requires in addition that:
\item{(2)} One of the molecules contains at least one closed loop of
dipoles.

\topinsert
%\centerline{\epsfxsize=15.truecm\epsfbox{coloredmolecules.eps}}
\centerline{\epsfbox{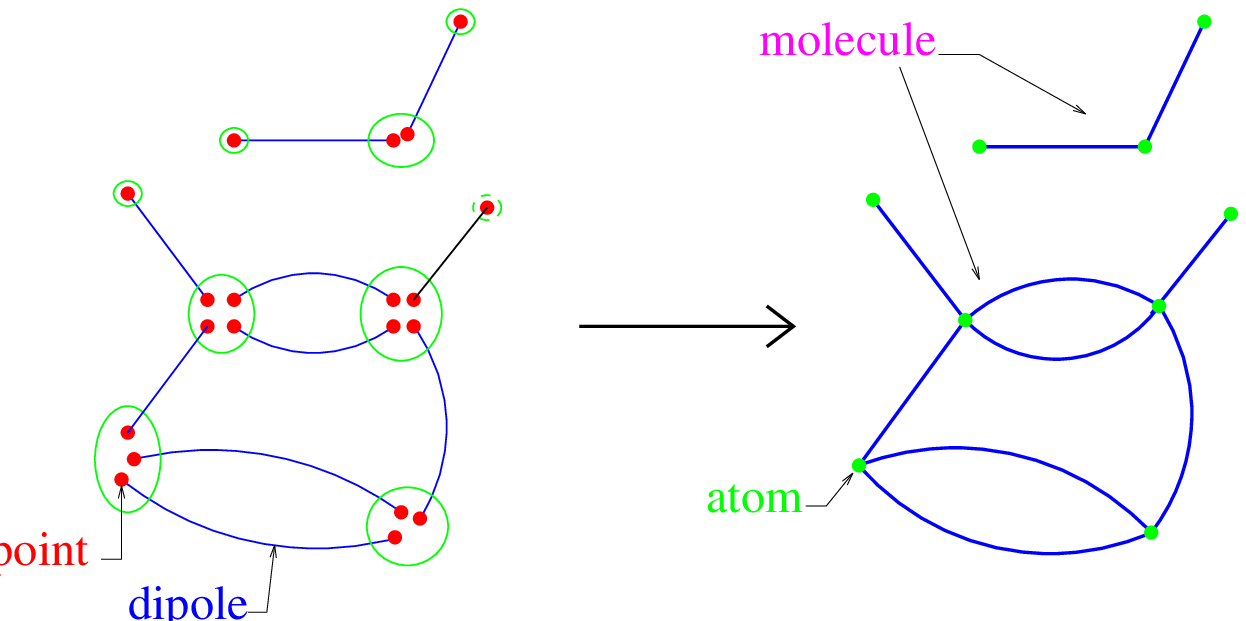}}
\centerline{\hyperdef\hmolecule{userfigure}{umolecule}{\fmolecule}}
\endinsert

\noindent
In fact, for a given molecule,
%the non-zero components of the isotropic vector, i.e.
the non-zero charges ${\bf k}_i$
are carried by the one-particle irreducible
(1PI) components of the molecule, that is by the connected
subsets which remain connected when any line (dipole) is removed.
%{\sl As we shall show below, this implies that UV singular configurations are
%given by collections of disjoint 1PI molecules, and that each 1PI molecule
%gives an independent singularity.}

%\vfill\break
\subsec {The Multilocal Operator Product Expansion (MOPE)}

We now analyse the short distance singularity associated with a given singular
configuration, i.e.  a molecule characterized by a set of atoms $\CP$ with
positions $x_\CP$, and by a set of dipolar links $a$ between these atoms.
%If the singular configuration consists of more than one molecule, the
%following treatment can be applied independently for each molecule.

For this purpose, we consider the points $i$ which belong to a given atom
$\CP$ and see how the integrands $\Delta^{-d/2}$ and
$\Delta^{-d/2}{\rm e}^{-{1\over 2}{\bar Q}}$ behave when the points $i$ collapse
inside $\CP$ ($x_i\to x_\CP$, $i\in\CP$).
The singularity can be analyzed in distance space, by letting the squared
distances $a_{ij}$ go to zero for $i$ and $j$ belonging to the same
atoms.
A more convenient and physical way to study this behavior is to
formulate the problem in terms of the ``physical" positions $x_i$ of the
points $i$ and $x_\CP$ of the atoms $\CP$.
Let
\eqn\eyi{
y_i\ =\ x_i-x_\CP\quad,\quad {\rm for}\ i\in\CP
}
be the relative position of point $i$ with respect to its atom $\CP$.

The short distance singularity of $\Delta^{-d/2}$ is analyzed by performing
a small $y_i$ expansion of the product of bilocal operators
\eqn\eProdExp{
\sprod_{a=1}^{N}\delta^{d}(\rvec(x_{2a})-\rvec(x_{2a-1}))=
\int\sprod_{i=1}^{2N}{d^d\kvec_i\over (2\pi)^d}
\sprod_{a=1}^{N}\CC_a\{\kvec_i \}
\sprod_{i=1}^{2N}{\rm e}^{\ii \kvec_i\cdot \rvec(x_i)}
}
in the Gaussian theory.
This is done as follows:
\item{(1)}
We regroup the vertex operators in each atom $\CP$.
For each atom we factorize the product of vertex operators into
its expectation value times its ``normal product"
\eqn\eNorProd{
\sprod_{i\in\CP}\ee^{\ii\kvec_i\cdot\rvec(x_i)}\ =\ 
\langle\sprod_{i\in\CP}\ee^{\ii\kvec_i\cdot\rvec(x_i)}\rangle_0\,
\npr \sprod_{i\in\CP}\ee^{\ii\kvec_i\cdot\rvec(x_i)}\npr
\ .
}
This can be considered as the definition of the normal product $\npr\quad\npr$.
At the diagrammatic level, this means that when evaluating the expectation
value of the l.h.s. of \eNorProd\ times the other ``external" operators, using
Wick's theorem (which leads to \eGausAv)
one can always factorize out the contribution involving pairs of
points $(i,j)$ in the atom $\CP$, which is nothing but the average value:
\eqn\eAvVal{
\ee^{-{1\over 2}\ssum\limits_{i,j\in\CP}\kvec_i\cdot\kvec_i\langle{\bf r}(x_i)
{\bf r}(x_j)\rangle_0}\ =\ \langle\sprod_{i\in\CP}\ee^{\ii\kvec\cdot\rvec(x_i)}
\rangle_0
\ .
}
What remains is the contribution involving propagators between a point $i$ in
the atom $\CP$ and the other points not in $\CP$.
This is precisely the definition of the normal product, which consists in
omitting internal contractions between points inside the $\npr\quad\npr$.

\item{(2)} Now the normal product relative to the atom $\CP$ is a smooth
function of the positions $x_i$ in a neighborhood of $x_\CP$, since the
short distance singularities at $x_i=x_j$ have been factorized out.
Therefore we can perform a Taylor expansion of this normal product in the $y_i$,
whose coefficients are normal ordered local operators located at $x_\CP$,
and which is of the form
\eqn\eTayExp{\eqalign{
\npr \sprod_{i\in\CP}\ee^{\ii\kvec_i\cdot\rvec(x_i)}\npr
\ & =\ 
\sprod_{i\in\CP}\ee^{y_i\cdot{\partial\over\partial x_i}}
\npr\sprod_{i\in\CP}\ee^{\ii\kvec_i\cdot\rvec(x_i)}\npr\Big|_{x_i=x_\CP}
\cr
&=\ \sum_{A}C^A\{y_i,\kvec_i\}\,\npr A(x_\CP)\ee^{\ii(\ssum\limits_{i\in\CP}
\kvec_i)\cdot \rvec(x_\CP)}\npr
\ .\cr
}}
The local operators $A(x_\CP)$ are monomials in the derivatives of the
field $\rvec$ at the point $x_\CP$.
The corresponding coefficients $C^A\{y_i,\kvec_i\}$ are monomials in
the $y_i$ and $\kvec_i$ variables, for $i\in\CP$.
For instance, the first terms of this Taylor expansion are given explicitly
by:
\eqn\eTExplic{\eqalign{
\npr \sprod_{i\in\CP}\ee^{\ii\kvec_i\cdot\rvec(x_i)}\npr\ 
& =\ \npr\ee^{\ii (\ssum\limits_{i\in \CP}\kvec_i)\cdot\rvec(x_\CP)}\npr
\,+\,\ii\ssum_{i\in\CP}y_i\kbf^\mu_i\,\npr{\nabla_{\! x_i}}
\rbf^\mu(x_\CP)\ee^{\ii (\ssum\limits_{i\in \CP}\kvec_i)\cdot\rvec(x_\CP)}\npr
\cr
&
%\hskip3em
-\, {1\over 2}\ssum_{i,j\in\CP}y_i y_j\kbf^\mu_i\kbf^\nu_j\,\npr
{\nabla_{\! x_i}}\rbf^\mu(x_\CP)
{\nabla_{\! x_j}}\rbf^\mu(x_\CP)
\ee^{\ii (\ssum\limits_{i\in \CP}\kvec_i)\cdot\rvec(x_\CP)}\npr
\cr
& \hskip3em +\,
{\ii\over 2}\ssum_{i\in\CP}y_i y_i\kbf^\mu_i\,\npr
{\nabla_{\! x_i}\!\nabla_{\! x_i}}\rbf^\mu(x_\CP)
\ee^{\ii (\ssum\limits_{i\in \CP}\kvec_i)\cdot\rvec(x_\CP)}\npr
\,+\,\cdots
\cr
}}
where the summation over the $D$ components $y_i^\alpha$ of the $y_i$'s are
implicit:
$y_i\nabla_{\! x_i}=\ssum_{\alpha=1}^D y_i^\alpha \partial/\partial x_i^\alpha$,
$y_i y_i \nabla_{\! x_i}\nabla_{\! x_i}=
\ssum_{\alpha,\beta=1}^D y_i^\alpha y_i^\beta
\partial^2/\partial x_i^\alpha\partial x_i^\beta$, $\ldots$

\item{(3)} From translation invariance in internal space $\RR^D$, the average
value \eAvVal\ relative to the points $i\in\CP$ is the same when evaluated in
terms of relative positions, i.e. $\langle \rbf(y_i)\rbf(y_j)\rangle_0$.

\item{(4)} Thanks to the global neutrality condition on the $\kvec_i$ in
\eProdExp , we can replace all propagators by their infra-red
finite part \eMlssProp .
This means on the one hand that the correlation function
$\langle \rbf(y_i)\rbf(y_j)\rangle_0$
of point (3) above is taken to be $G(y_i,y_j)$, building the factor
\eqn\eAVbis{
\ee^{-{1\over 2}\ssum\limits_{i,j\in\CP}\kvec_i\cdot\kvec_i G(y_i,y_j)}
}
in \eAvVal .
On the other hand, the operators in \eTayExp\ are also defined through
the normal order prescription, which uses the I.R. finite propagator \eMlssProp.
This means that when evaluating correlation functions of products of such
operators, located at positions $x_\CP$, $x_{\CP'}$, $\ldots$
one uses the propagators $G(x_\CP,x_{\CP'})$ and their derivatives.

\item{(5)}
We now gather the contributions relative to the atoms $\CP$ and links $a$
of a given
%connected one-particle irreducible (1PI) component $\CM$ of the molecule. 
connected molecule $\CM$.
%If the molecule has more than one 1PI component, or if the singular
%configuration consists in more than one molecule, the
%following treatment can be applied independently for each 1PI component $\CM$.
If the singular configuration consists in more than one molecule, the
following treatment can be applied independently for each molecule $\CM$.
For each atom $\CP$ we introduce its total charge $\kvec_\CP=\sum_{i\in\CP}
\kvec_i$.
This is done by inserting
$1=\int d^d\kvec_\CP\,\delta^d(\kvec_\CP-\ssum_{i\in\CP}\kvec_i)$ in \eProdExp ,
which becomes an integral over the $\kvec_i$ and the $\kvec_\CP$.
The constraints $\CC_a$ for the dipolar links $a$ in the component $\CM$
ensure that $\sum_{\CP\in\CM}\kvec_\CP=0$.
We therefore rewrite
\eqn\ePrConst{
%\sprod_{\CP\in\CM}\delta^d(\kvec_\CP-\ssum_{i\in\CP}\kvec_i)\,
\sprod_{a\in\CM}\CC_a\{\kvec_i\}\ =\ 
%\sprod_{\CP\in\CM}\delta^d(\kvec_\CP-\ssum_{i\in\CP}\kvec_i)\,
\mathop{{\sprod}'}_{a\in\CM}\CC_a\{\kvec_i\}\,(2\pi)^d\,
\delta^d(\ssum_{\CP\in\CM}\kvec_\CP)
\ ,
} 
where ${\sprod}'$ means that one of the constraints has been omitted
arbitrarily (the result does not depend on the link $a$ which is omitted).

\item{(6)} We now Taylor expand each $\delta^d(\kvec_\CP-\ssum_{i\in\CP}\kvec_i)$
in powers of $\kvec_\CP$:
\eqn\eTEkp{
\delta^d(\kvec_\CP-\ssum_{i\in\CP}\kvec_i)\ =\ \sum_{\mvec}
{(\kvec_\CP)^{\mvec} \over \mvec !}
\left(\nabla_{\!\kvec}\right)^{\mvec}
\delta^d(-\ssum_{i\in\CP}\kvec_i)
}
where the sum runs of $d$-uples $\mvec$ of non-negative integers
%$\mvec=\{m_\mu,\ \mu=1,\cdots ,d\}$%
$\mvec=\{m_\mu\}_{\mu=1,\cdots ,d}$%
, with the standard notation:
\eqn\eStNot{
(\kvec)^\mvec\ =\ \sprod_{\mu=1}^d \left(\kbf^\mu\right)^{m_\mu}
\qquad
\left(\nabla_{\!\kvec}\right)^\mvec\ =\ \sprod_{\mu=1}^d\left({\partial
\over\partial\kbf^\mu}\right)^{m_\mu}
\qquad
\mvec !\ =\ {(m_1+\cdots+m_d)!\over m_1!\cdots m_d!}
}

\item{(7)}
At this stage the product of bilocal operators \eProdExp\ has been expanded
as a sum of terms of the form
\eqn\eSumTerm{
\Cbf^{\{A_\CP,\mvec_\CP\}}\{y_i\}\cdot\Phi_{\{A_\CP,\mvec_\CP\}}\{x_\CP\}
\ ,
}
where
\eqn\eCMOPE{\eqalign{
\Cbf^{\{A_\CP,\mvec_\CP\}}\{y_i\}&\ \equiv\ \int\sprod_{i\in\CM}
{d^d\kvec_i\over (2\pi)^d}\,
\mathop{{\sprod}'}_{a\in\CM}\CC_a\{\kvec_i\}\times\cr
&\hskip -3.em\times\sprod_{\CP\in\CM}
\Bigg\{C^{A_\CP}\{y_i,\kvec_i\}_{i\in\CP}
\,{1\over \mvec_\CP!}\,
(\nabla_\kvec)^{\mvec_\CP}\delta^d(-\!\ssum_{i\in\CP}\kvec_i)\,
\ee^{-{1\over 2}\!\ssum\limits_{i,j\in\CP}\!\kvec_i\cdot\kvec_j G(y_i,y_j)}
\Bigg\}\cr}
}
and
\eqn\ePhiMOPE{
\Phi_{\{A_\CP,\mvec_\CP\}}\{x_\CP\}\equiv
\int\!\!\sprod_{\CP\in\CM}d^d\kvec_\CP\,(2\pi)^d\,
\delta^d(\!\ssum_{\CP\in\CM}\!\kvec_\CP)\sprod_{\CP\in\CM}
\Bigg\{(\kvec_\CP)^{\mvec_\CP}
\npr A_\CP(x_\CP)\,\ee^{\ii\kvec_\CP\cdot\rvec(x_\CP)}\npr\Bigg\}
}
The sum extends over all collections of $\{A_\CP,\mvec_\CP\}$ relative to
the atoms $\CP$ of the molecule $\CM$.
Once integrated over the $\kvec_i$, \eCMOPE\ builds a function of
the relative positions $y_i$ inside the atoms.
Since $G(y_i,y_j)$ is a pure power of $|y_i-y_j|$, this function is
moreover homogeneous with respect to the $y_i$.
The integral over the $\kvec_\CP$ in \ePhiMOPE\ can be performed by
performing a Fourier transform of the $\delta$-function, thus introducing
a variable $\rvec$:
\eqn\ePhiExpl{
\Phi_{\{A_\CP,\mvec_\CP\}}\{x_\CP\}\ =\ 
\int d^d\rvec\,\sprod_{\CP\in\CM}
\Bigg\{\npr A_\CP(x_\CP)\,
(\ii\nabla_{\!\rvec})^{\mvec_\CP}\delta^d(\rvec-\rvec(x_\CP)) \npr\Bigg\}
\ .
}
Eq. \ePhiExpl\ describes generically a multilocal operator, involving
contact interaction (at point $\rvec$) in external space between the points
$x_\CP$, associated with the different atoms $\CP$ of the molecule $\CM$, with
local operators $A_\CP(x_\CP)$ at each point $x_\CP$.
As can be seen in \eTayExp , these local operators, which must be
translationally invariant, involve only derivatives of $\rvec(x)$.

\medskip
\noindent{\it Multilocal Operator Product Expansion (MOPE):}
\medskip
The result of points (1)--(7) above is the following multilocal operator
product expansion property:
When approaching the singular configuration characterized by the connected
molecule $\CM$, the product of bilocal operators corresponding to the
dipoles $a$ of the molecule can be expanded as a series of multilocal
operators $\Phi$ 
\eqn\eTheMOPE{
\sprod_{a\in\CM}\delta^{d}(\rvec(x_{2a})-\rvec(x_{2a-1}))\ =\ 
\sum_{\{A_\CP,\mvec_\CP\}\atop\CP\in\CM}
\Cbf^{\{A_\CP,\mvec_\CP\}}\{y_i\}\cdot\Phi_{\{A_\CP,\mvec_\CP\}}\{x_\CP\}
}
The $\Phi$'s are multilocal $|\CM|$-body contact operators, where
$|\CM|$ is the number of atoms in $\CM$.
%They are functions of the positions $x_\CP$ of these atoms only.
They depend only on the positions $x_\CP$ of these atoms $\CP$.
%The coefficients $\Cbf$ of the MOPE depend on the full structure of the
%molecule $\CM$, and are homogeneous functions of the relative coordinates
%$\{y_i=x_i-x_\CP\}_{i\in\CP}$
%inside the atoms $\CP$ of the positions $x_i$ of the bilocal operators
%in the l.h.s. of \eTheMOPE .
The coefficients $\Cbf$ of the MOPE depend on full structure of the
molecule $\CM$, and on the relative coordinates $\{y_i=x_i-x_\CP\}_{i\in\CP}$
of the end-points
inside each atom $\CP$ of $\CM$ separately.
%of end points $i$ of the dipoles.
They do not depend on the relative coordinates $\{x_i-x_j\}$ of end-points
belonging to different atoms ($i\in\CP$, $j\in\CP'$, $\CP\neq\CP'$).
Finally the coefficients $\Cbf$ are homogeneous functions of the $y_i$'s, with
a degree of homogeneity ${\rm deg}[\Cbf]$ which depends on the molecule
$\CM$ and on the $\{\CA_\CP,\mvec_\CP\}$.

\medskip
A direct application of this MOPE is the short distance expansion for
the integrands $\Delta^{-d/2}$ and $\Delta^{-d/2}\ee^{-{1\over 2}\bar Q}$ in
\eZManInt\ and \eVManInt .
For instance, we have
\eqn\eDelMOPE{\eqalign{
(2\pi)^{-Nd/2}\Delta\{x_i\}^{-d/2}\ &=\ 
\langle\sprod_{a=1}^N\delta^{d}(\rvec(x_{2a})-\rvec(x_{2a-1}))\rangle_0\cr
=\ \sum_{\{A_\CP,\mvec_\CP\}\atop\CP\in\CM}&
\Cbf^{\{A_\CP,\mvec_\CP\}}\{y_i\}\cdot
\langle\sprod_{a\notin\CM}\delta^{d}(\rvec(x_{2a})-\rvec(x_{2a-1}))
\Phi_{\{A_\CP,\mvec_\CP\}}\{x_\CP\}\rangle_0
}}
where the expectation value of a product of bilocal operators times a
multilocal operator in the r.h.s of \eDelMOPE\ can be evaluated
by the same rules as in Section~2.1, using the Fourier representation of
\ePhiMOPE\ for $\Phi$.
It is a function of the positions $\{x_i\}_{i\notin\CM}$ and of the
$\{x_\CP\}_{\CP\in\CM}$.
Using the Sch\"onberg theorem, one can show that, like $\Delta^{-d/2}$,
this function is well defined and non-singular if the $\{x_i\}_{i\notin\CM}$
and the $\{x_\CP\}_{\CP\in\CM}$ do not coincide.

%\vfill\break
\subsec {Power-Counting}

We are now in a position to analyze the singular configurations and the
associated divergences.
Given a connected molecule $\CM$, we contract the points $i$
towards the center of the atoms $\CP$ by rescaling the $y_i=x_i-x_\CP$
by a {\it global} factor $\lambda$.
The behavior of the integrand in \eZManInt\ and \eVManInt\ when $\lambda\to 0$
is given by the MOPE.
Since $\Cbf^{\{A_\CP,\mvec_\CP\}}\{y_i\}$ is a
homogeneous function of the $\{y_i\}$, each term of the MOPE gives a single
power of $\lambda$.
From \eCMOPE and using the fact that $\kbf$ scales as $y^{-\nu}$,
the degree of $\Cbf^{\{A_\CP,\mvec_\CP\}}\{y_i\}$ is given by
\eqn\eDegC{
{\rm deg}[\Cbf^{\{A_\CP,\mvec_\CP\}}]\ =\ 
d\nu\cdot(\#_{{\rm links}\atop a\in\CM}-1)-d\nu\cdot
\#_{{\rm points}\atop i\in\CM}
+\ssum_{{\rm atoms}\atop\CP\in\CM}\left(|\mvec_\CP|\nu - {\rm dim}[\CA_\CP]
+d\nu\right)
}
with, using \eStNot
\eqn\eModMP{
|\mvec|=\ssum_{\mu=1}^d m_\mu
\ .
}
To obtain the corresponding {\it degree of convergence} $\omega$,
we have to integrate over the positions for all the points $i$ but one
inside each atom $\CP$, since the position $x_\CP$ of each atom is fixed in the
contraction process.
This degree of convergence is thus given by
\eqn\eOmega{
\omega_\CM^{\{A_\CP,\mvec_\CP\}}\ =\ {\rm deg}[\Cbf^{\{A_\CP,\mvec_\CP\}}]
+D(\#_{{\rm points}\atop i\in\CM}-\#_{{\rm atoms}\atop\CP\in\CM})
}
Using the fact that 
$\#_{{\rm points}}=2\,\#_{{\rm links}}$
and defining the {\it number of internal loops} of the connected
molecule $\CM$ as
\eqn\eNLoop{
\#_{{\rm loops}}=\#_{{\rm links}}-\#_{{\rm atoms}}+1
}
one obtains
\eqn\eOMEGA{
\omega_\CM^{\{A_\CP,\mvec_\CP\}}\ =\ 
D\,(\#_{{\rm atoms}}-2)+\epsilon\,(\#_{{\rm loops}})
+\ssum_{{\rm atoms}}\left(|\mvec_\CP|\,\nu - {\rm dim}[\CA_\CP]\right)  
}
where we recall that $\epsilon$ and $\nu$ are given by
\eEps\ and \eNu\ respectively.

The last sum in \eOMEGA , relative to the atoms $\CP$,
is always positive, since the local operators
$\CA_\CP$ generated in the MOPE are monomials in (multiple) $x$-derivatives 
of $\rvec$, and have negative dimension (for $D<2$).
Indeed,
\eqn\eDimA{
{\rm dim}[\CA]\ =\ -\#_{x{\rm -derivatives}}+\nu\,\#_{\rvec}\ \le\ 
(\nu-1)\#_{\rvec}\,=\,-{D\over 2}\#_{\rvec}\ \le\ 0
}
This sum is nothing but the dimension of the ``dressing" by
$\{\CA_\CP,\mvec_\CP\}$ of the molecule $\CM$.
The smallest degree $\omega_\CM$ is obtained by setting all the $\CA_\CP=1$
(the identity operator) and all the $\mvec_\CP=\vec 0$.
This defines the {\it superficial degree of convergence} of the molecule
$\CM$:
\eqn\eSOmega{
\omega_\CM\ =\ D\,(\#_{\rm atoms}-2)\,+\,\epsilon\,(\#_{\rm loops})
}

\medskip
Only the molecules $\CM$ such that
\eqn\eNegOm{
\omega_\CM\ \le\ 0
}
give short-distance U.V. divergences in the integrals \eZManInt\ and \eVManInt .
From the MOPE, we expect that the divergences given by a coefficient
$\Cbf^{\{A_\CP,\mvec_\CP\}}$ will be proportional
to the insertion of the corresponding multi-local operator
$\Phi_{\{A_\CP,\mvec_\CP\}}\{x_\CP\}$.
Moreover, it is easy to check that the degree of convergence
$\omega_\CM^{\{A_\CP,\mvec_\CP\}}$ is related to the canonical dimension of
the multi-local operator $\Phi_{\{A_\CP,\mvec_\CP\}}\{x_\CP\}$, integrated over the whole space, by
\eqn\eDegOme{
\omega_\CM^{\{A_\CP,\mvec_\CP\}}\ =\ -\,
{\rm dim}\Bigg[\underbrace{\int\cdots\int}_{\#_{\rm atoms}\atop\CP\in\CM}
\Phi_{\{A_\CP,\mvec_\CP\}}\Bigg]\,+\,\epsilon\,
\#_{{\rm links}\atop a\in\CM}
}
as expected from dimensional analysis.

Depending on the sign of $\epsilon$ there are three cases:
\item{(1)} $\epsilon>0$: only the molecules such that $\#_{\rm atoms}=1$
and with a number of loops $\#_{\rm loops}$ small enough diverge
($\#_{\rm loops}\le D(2-\#_{\rm atoms})/\epsilon$).
There is a finite number of such divergent molecules, and the
model will be {\it super-renormalizable}.
\item{(2)} $\epsilon=0$: only the molecules such that $\#_{\rm atoms}=1$ or $2$,
but with $\#_{\rm loops}$ arbitrary, are divergent.
There is an infinite number of divergent molecules, but we will show that they 
correspond to a {\it finite} number of multi-local operators.
The model will be {\it strictly-renormalizable}.
Such divergent configurations are depicted on 
\fig\fdivdiag{General structure of \hyperref\hdivdiag{UV divergent diagrams}
with $N=1$ (a) and $N=2$  (b) atoms.}
.
\item{(3)} $\epsilon<0$: all the molecules with a large enough $\#_{\rm loops}$
diverge.
There is an infinite number of divergent molecules, which
corresponds to an infinite number of multi-local operators (in fact all of
them).
The theory will be {\it non-renormalizable}.

%\fig\fdivdiag{General structure of \hyperref\hdivdiag{UV divergent diagrams}
%with $N=1$ and $N=2$ atoms.}
\topinsert
\centerline{\epsfbox{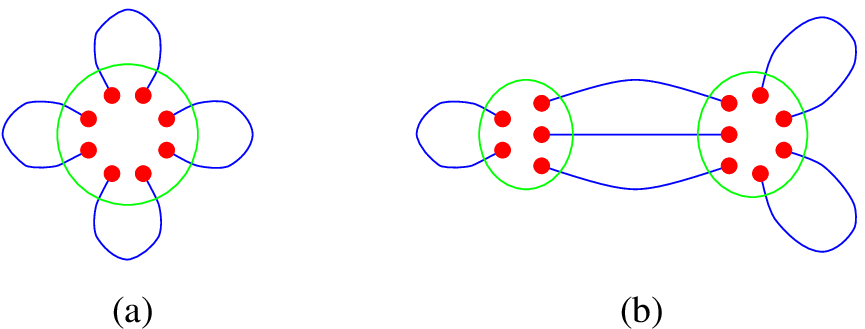}}
\centerline{\hyperdef\hdivdiag{userfigure}{udivdiag}{\fdivdiag}}
%\centerline{\fdivdiag}
\endinsert

%\vfill\break
\newsec {Renormalization}
We are now in a position to analyse the UV divergences of the theory at
$\epsilon=0$ and to show that the theory is renormalizable, that is that
the UV divergences can be subtracted and that the theory can be made
finite, in perturbation theory, by adding appropriate counterterms to the
Hamiltonian.
We shall follow the general strategy of the so called dimensional regularization
scheme, as used in local field theories and 
for the model of \rIIM : The theory is first defined for
$\epsilon>0$ and small, and $\epsilon$ is considered as a regularization
parameter.
Then we construct the renormalized theory by adequate subtractions and 
by taking the limit $\epsilon\to 0$.

\subsec {The theory for $\epsilon>0$}
When $\epsilon>0$, the only divergent configurations are the molecules
$\CM$ made of one atom and of $N$ dipoles, with $N\le D/\epsilon$.
According to the results of the previous section, the corresponding
singularities are given by the MOPE coefficients ${\bf C}^{A}$ when the $N$
dipoles are contracted toward
a single local operator $A(x)$ (we can here assimilate $A(x)$ with the
``1-local operator"
${\bf\Phi}_{\{A,\mvec=\vec 0\}}\{x\}$ defined by \ePhiExpl , and
``1-local operators" with $\mvec\ne\vec 0$ are automatically equal to zero).
The corresponding degree of convergence $\omega_\CM^A$ of each term is given by
\eOMEGA, and is 
\eqn\omegonept{
\omega_\CM^A\ =\ -D+\epsilon N-{\rm dim}[A]
}
We see that only the identity operator $A_0={\bf 1}$ with
${\rm dim}[{\bf 1}]=0$ gives a divergence.
Less relevant operators give $\omega_\CM^A>0$, for instance the least
irrelevant one is $(\nabla\rvec)^2$ which has dimension
${\rm dim}[(\nabla\rvec)^2]=D$, so that $\omega=\epsilon N>0$.

Since for $\epsilon>0$ there is a finite number of divergent molecules, and
since the associated divergences in the manifold integrals are proportional to
the unity operator, this suggests that:
\item{(1)} Those divergences are cancelled by adding to the bare Hamiltonian
\eEdwards\ a counterterm $\Delta\CH/\kBT$ independent of $\rvec$ and of the
form
\eqn\CTvol{
\Delta\CH/\kBT\ =\ \Delta(b;\epsilon)\int d^Dx 1
}
\item{(2)} As a consequence, these short distance divergences should be
absent from the manifold integrals when one computes in the interacting theory
expectation values of physical observables $O$ as defined for instance by
\eNeutObs .
Indeed the divergences, being independent of the configuration field $\rvec$,
factorize out of the functional integral $\int \CD[\rvec]$ and should
cancel between the numerator in \eCorrFirst\ and the normalization denominator
factor $\CZ_b$.
This means that perturbation theory for physical observables in the infinite
membrane model is ultra-violet finite as long as $\epsilon>0$.
\par

Let us give the outline of the proof.
For simplicity we denote by $\phi\{x,y\}$ the two-point operator 
\eqn\edefPhi{
\phi\{x,y\}\ =\ \delta^d(\rvec(x)-\rvec(y))\ .
}
and we consider the $N$-th term of the perturbative series expansion of the
expectation value of an observable $O_{\{\qvec_i\}}(\{z_i\})$, as defined by
\eNeutObs .
This term is proportional to the integral of the connected expectation value
(in the free theory) of $O$ times the product of $N$ $\phi$ operators
\eqn\conMInt{
O^{(N)}\{z_i\}\ =\ 
\int\sprod_{i=1}^{2N}d^Dx_i\,
\langle O_{\{\qvec_i\}}(\{z_i\}) \sprod_{a=1}^N\phi\{x_{2a-1},x_{2a}\}
\rangle_0^{\rm conn}
}
The subscript $\langle\cdots\rangle_0$ refers to the expectation value in
the free theory with $b=0$.
The connected expectation value (e. v.)
$\langle\cdots\rangle^{\rm conn}$ of a product of $N$
operators is defined here by inverting the well known relation
\eqn\econpart{
\langle A_1\cdots A_N\rangle\ =\ \sum_{{\rm partitions}\atop
{\rm of}\ \{1,\cdots,N\}}\,\prod_{{\rm subsets\ of}\atop{\rm the\ partition}}
\langle\sprod_{i\in{\rm subset}}A_i\rangle^{\rm conn}
}
which gives
\eqn\enoncon{
\langle A_1\cdots A_N\rangle^{\rm conn}\ =\ \sum_{{\rm partitions}\atop
{\rm of}\ \{1,\cdots,N\}}\,
w({\rm partition})\,
%(-1)^{|{\rm part}|-1}\,(|{\rm part}|-1)!\,
\prod_{{\rm subsets\ of}\atop{\rm the\ partition}}
\langle\sprod_{i\in{\rm subset}}A_i\rangle
}
with\eqn\efactpart{
w({\rm partition})\ =\ (-1)^{|{\rm part}|-1}\,(|{\rm part}|-1)!\,
}
where $|{\rm part}|$ is the number of distinct subsets (elements) of the
partition.

With these notations, in \conMInt\ the observable $O$ has to be considered as
a single multilocal operator, although it is a product of local
vertex operators. 
In other words, the partitions must not split the operator $O$.

Potential short distance divergences occur in the integral \conMInt\ when
the mutual distances between a subset of the $N$ dipoles become small.
For simplicity let us assume that these are the first $M$ dipole
$\phi\{x_{2a-1},x_{2a}\}$, $1\le a\le M\le N$, and let us contract these dipoles
towards the origin $0$.
If we rescale the $x_i\to \lambda x_i$, $1\le i\le 2M$, the MOPE ensures us
that when $\lambda\to 0$ the non-connected e.v. behaves as
\eqn\eintgrd{
%\langle\sprod_{a=1}^N\phi\rangle_0
\langle O\,\underbrace{\phi\cdots\phi}_M\,\underbrace{\phi\cdots\phi}_{N-M}
\rangle_0
\ =\ \lambda^{-M\,d\nu}
\,C_{\underbrace{\phi\cdots\phi}_M}^{\bf 1}\,
\langle O\,{\bf 1} \,\underbrace{\phi\cdots\phi}_{N-M}\rangle_0
\,+\,\CO(\lambda^{-M\,d\nu\,+\,D})
}
where $C_{\phi\cdots\phi}^{\bf 1}$ is the MOPE
coefficient for the contraction of the first $M$ $\phi$ operators toward
the identity operator ${\bf 1}$ and is an homogeneous function of
$\{x_1,x_2,\cdots, x_{2M-1},x_{2M}\}$ with degree $-Md\nu$.
The $\CO(\lambda^{-M\,d\nu\,+\,D})$ terms are associated with the insertion of
less relevant operators.
We can use this result, together with \enoncon\ to evaluate the short-distance
behavior of the connected e. v. , by first decomposing it into non-connected
e.v. of subsets of operators, then using the MOPE and finally re-expressing the
result in terms of connected e.v.
The final result is quite simple.
To evaluate the small $\lambda$ behavior of
$\langle O\phi\cdots\phi \rangle_0^{\rm conn}$
we have to consider the {\it restricted} partitions of the $N+1$ operators
$\{O,\phi,\cdots ,\phi\}$, such that each subset of the
partition contains at least one of the $M$ contracted $\phi$ operators,
then use naively the MOPE for each subset, and then take the connected e.v.
for the resulting product of operators relative to each subset.
The final result is of the form
\eqn\econgrd{\eqalign{
\langle O\,\underbrace{\phi\cdots\phi}_M\,
\underbrace{\phi\cdots\phi}_{N-M}
\rangle_0^{\rm conn}\ =\ \lambda^{-M\,d\nu}\,
\sum_{{\rm restricted}\atop{\rm partitions}}
&
w({\rm partition})\ 
C^{\bf 1}_{\underbrace{\phi\cdots\phi}_{M_1}}
\langle O\,{\bf 1}\,
\underbrace{\phi\cdots\phi}_{N_1-M_1}\rangle^{\rm conn}_0\,\times
\cr
&\times\,
\prod_{{{\rm other\ subsets}\atop{\rm not\ containing\ }O}}
C_{\underbrace{\phi\cdots\phi}_{M_i}}^{\bf 1}
\langle {\bf 1} \,
\underbrace{\phi\cdots\phi}_{N_i-M_i}\rangle_0^{\rm conn}
}
}
In \econgrd\ we have separated the first subset of the partition, which
contains $O$, from the other subsets, which do not contain $O$.
Now we simply use the fact that in the contribution of each restricted
partition all the $\langle\cdots\rangle^{\rm conn}$ contain the
${\bf 1}$ operator, and that there is at least one of the subsets of the
partition which contains also some other non-contracted operator, for instance
the observable $O$.
Now, using for instance \enoncon\ it is trivial to show that any
connected e.v. of products of operators which contains the identity
operator ${\bf 1}$ vanishes identically, except of course the single e.v. of
the identity operator $\langle {\bf 1}\rangle^{\rm conn}=1$.
This ensures us that each restricted partition in the r.h.s. of \econgrd\ 
gives zero and that in fact
\eqn\econzero{
\langle O\,\underbrace{\phi\cdots\phi}_M\,\underbrace{\phi\cdots\phi}_{N-M}
\rangle_0^{\rm conn}\ =\ \CO(\lambda^{-M\,d\nu\,+\,D})
}
Now, when integrating over the positions of the $2M$ endpoints of the contracted
dipoles {\it but one} (which represents the position of the inserted operator),
we obtain that the degree of convergence of the configuration is in fact
\eqn\edegcon{
\omega_\CM^{\rm conn}\ =\ 
(2M-1)D-Md\nu+D\ =\ M\epsilon\ >\ 0 \qquad{\rm if}\qquad \epsilon\ >\ 0
}
This ensures that for $\epsilon>0$ superficial UV divergences are absent from
the e.v. of physical observables.

This result is a strong indication that the perturbation theory for e. v.
of physical observables is UV finite for $\epsilon>0$, but is not a 
rigorous proof.
We would need in fact to perform a much more detailed analysis of the
convergence of the manifold integrals when the distances between ``nested"
subsets of points are contracted toward zero in a hierarchical way.
This ``nest analysis" will not be presented here, but follows the analysis
presented in the next subsections for the U.V. convergence.
%In fact an outline of the nest analysis will be given in the next subsections
%for the fully renormalized theory at $\epsilon=0$.
%This analysis leads (as a simple application) to the proof of the UV convergence
%of the unrenormalized theory for $\epsilon>0$.

%\vfill\break
\subsec {Cancellation of Infra-red Divergences}
Before discussing the full renormalization of the model at $\epsilon=0$,
we must check that  perturbation theory for the infinite membrane makes sense
when $\epsilon\to 0_+$, and does not suffer from infra-red divergences.
As already discussed in subsec.~2.3., we expect strong I.R. divergences to occur
for non-vanishing $\epsilon>0$, but we made the conjecture that at a fixed
$N$-th order in perturbation theory, translationally invariant observables
are I.R. finite as long as $\epsilon<D/N$ (and in the domain $0<D<2$ where
perturbation theory is a priori well defined).
The arguments developed in the previous subsection to deal with
U.V. divergences for $\epsilon>0$ can be adapted to deal with I.R. divergences
for $\epsilon\le 0$, as we now show.

Let us go back to the manifold integral \conMInt\ which gives the $N$-th
term of the perturbative series of the e.v. of the observable $O$
\eqn\conmint{
O^{(N)}\{z_i\}\ =\ 
\int\sprod_{i=1}^{2N}d^Dx_i\,
\langle O_{\{\qvec_i\}}(\{z_i\}) \sprod_{a=1}^N\phi\{x_{2a-1},x_{2a}\}
\rangle_0^{\rm conn}
}
with
\eqn\einvobs{
O_{\{\qvec_i\}}(\{z_i\})\ =\ \ee^{\ii\ssum\limits_{i=1}^P\qvec_i\rvec(z_i)}
}
and the additional constraint
\eqn\enetcond{
\ssum_{i=1}^P\qvec_i\ =\ \vec {\bf 0}
}
Potential I.R. divergences occur in the integral \conmint\ when some of the 
dipole end-points positions $x_i$ become large.
We can analyse these I.R. singularities following the general method
of [\xref\rDavidIR,\xref\rBerDav]
.
The most divergent I.R. singularities are given by configurations when
a subset of the $N$ dipoles goes to infinity.
For simplicity, let us consider the case when these are the last $M$ dipoles
$\phi\{x_{2a-1},x_{2a}\}$, $N-M+1\le a\le N$, with $1\le M\le N$.
We shall prove that when we rescale these dipoles positions by
\eqn\eIRscal{
x_i\,\to\,\rho x_i\qquad;\qquad i\ge 2(N-M)+1
}
the integrand in \conmint\ scales when $\rho\to\infty$ as
\eqn\eIRInt{
\langle O_{\{\qvec_i\}}(\{z_i\})
\sprod_{a=1}^{N-M}\phi\{x_{2a-1},x_{2a}\}
\sprod_{a=N-M+1}^N\phi\{\rho x_{2a-1},\rho x_{2a}\}
\rangle_0^{\rm conn}
\ =\ \CO(\rho^{-M\,d\nu-D})
}
This is sufficient to show that the I.R. degree of divergence of this
configuration is
\eqn\eIRdeg{
M\,(2D\,-\,d\nu)\,-\,D\ =\ M\,\epsilon\,-\,D
}

To show \eIRInt , we first use the fact that the integrand is an homogeneous
function  of the $q_i$'s, the $z_i$'s and the $x_j$'s, with degree $-Nd\nu$
under the rescaling
\eqn\glresc{
\qvec_i\,\to\,\lambda^{-\nu}\qvec_i\qquad,\qquad
z_i\,\to\,\lambda z_i\qquad;\qquad
x_j\,\to\,\lambda x_j\qquad;\qquad 
}
This implies that the I.R. rescaling of the last $M$ dipoles is equivalent
to the U.V. rescaling of the observable $O$ and of the $N-M$ first dipoles
toward the origin
\eqn\eUVInt{\eqalign{
\langle & O_{\{\qvec_i\}}(\{z_i\})
\sprod_{a=1}^{N-M}\phi\{x_{2a-1},x_{2a}\}
\sprod_{a=N-M+1}^N\phi\{\rho x_{2a-1},\rho x_{2a}\}
\rangle_0^{\rm conn}
\cr
&\ =\ 
\rho^{-N\,d\nu}\,
\langle O_{\{\rho^\nu\qvec_i\}}(\{\rho^{-1}z_i\})
\sprod_{a=1}^{N-M}\phi\{\rho^{-1}x_{2a-1},\rho^{-1}x_{2a}\}
\sprod_{a=N-M+1}^N\phi\{x_{2a-1},x_{2a}\}
\rangle_0^{\rm conn}
}
}
To study the large $\rho$ behavior of the r.h.s. of \eUVInt\ amounts to
look at
the small distance behaviour of $O$ times the product of $N-M$ dipoles,
when the distances are contracted by a factor $\lambda=\rho^{-1}\to 0$,
{\it and} when the ``external momenta" $\qvec_i$ are simultaneously
expanded by a factor $\lambda^{-\nu}=\rho^\nu$.
This limit is different from the short distance behavior of products of
operators studied in subsection 3.2, where the $O$ operator was not involved.
However, the general method used there to establish the MOPE structure can be
easily extended to our purpose. 
Indeed, the only difference is that now we have to study the short distance
behavior of the product
\eqn\enewprod{
O_{\{\qvec_i\}}(\{z_i\})\sprod_{a=1}^{N-M}\phi\{x_{2a-1},x_{2a}\}
}
under the rescaling
\eqn\enewresc{
\qvec_i\to\lambda^{-\nu}\qvec_i\quad,\quad
z_i\to\lambda z_i\quad,\quad x_j\to\lambda x_j\quad,\quad\lambda\to 0
}
Since $O$ is a product of exponential operators, similar to those
appearing in the Fourier representation \eProdExp\ of the dipole operators
$\phi$, and since the rescaling \enewresc\ for the external momenta $\qvec_i$
is the same than the rescaling done for the internal momenta $\kvec_j$ in the
MOPE derivation, repeating step by step the arguments of subsection 3.2,
we obtain that the product of operators \enewprod\ has also a short distance 
MOPE expansion, which is of the form
\eqn\enewMOPE{
O_{\{\qvec_i\}}(\{z_i\})\sprod_{a=1}^{N-M}\phi\{x_{2a-1},x_{2a}\}
\ =\ \sum_A {\bf C}_{O\underbrace{\phi\cdots\phi}_{N-M}}^A(\{\qvec_i\},\{z_i\}
,\{x_j\})\,\cdot\,A(0)
}
where the sum runs over all local operators $A$ of the form
\eqn\enewA{
A\ =\ {\rm monomial}(\nabla \rvec,\nabla^2\rvec,\cdots)
\ee^{\ii(\ssum\limits_i\qvec_i)\rvec}
}
and where the MOPE coefficient
${\bf C}_{O\phi\cdots\phi}^A(\{\qvec_i\},\{z_i\},\{x_j\})$ is a homogeneous
function under the rescaling \enewresc , with degree
\eqn\edegC{
{\rm degree}[{\bf C}_{O\phi\cdots\phi}^A]\ =\ -(N-M)\nu d -{\rm dim}[A]
}
As before this result comes from simple power counting, since ${\rm dim}[O]=0$
and ${\rm dim}[\phi]=-\nu d$.
The ${\bf C}_{O\phi\cdots\phi}^A$ coefficients can be calculated explicitly,
but we do not need that at this stage.
The most singular operator $A$ is 
\eqn\emostA{
V_\qvec\ =\ \ee^{\ii\qvec\rvec}\qquad;\qquad\qvec=\ssum_i\qvec_i
}

Putting together \eUVInt\ and \enewMOPE , we obtain the I.R. behavior of the
non-connected e.v.
\eqn\eIRnc{
\langle O\,\phi\cdots\phi \underbrace{\phi\cdots\phi}_M\rangle_0\ =\ 
\rho^{-M\,d\nu}\,{\bf C}_{O\underbrace{\phi\cdots\phi}_{N-M}}^{V\qvec}\,
\langle V_\qvec\,\underbrace{\phi\cdots\phi}_M\rangle_0
\ +\ \CO(\rho^{-M\,d\nu-D})
}
We must now use \enoncon\ and repeat the arguments of subsection 4.1 to
obtain from \eIRnc\ the I.R. behavior of the corresponding
{\it connected} e.v.
We obtain a result similar to \econgrd :
we have a restricted sum over the partitions of $O$ and the $N$ $\phi$
operators, with the condition that each subset of the partition must
contain at least one of the operators which are not I.R. expanded, that is
$O$ or one of the first $N-M$ $\phi$.
Inside each subset of the partition, we perform the U.V. contraction of
$O$ (if it is in the subset), and of the first $\phi$, and we obtain
through the MOPE an operator $A$, and the connected
partition function $\langle A\phi\cdots\phi\rangle_0^{\rm conn}$.
If the subset contains $O$, the most singular $A$ is $V_\qvec$,
if the subset does not contains $O$, the most singular $A$ is the identity
operator ${\bf 1}$.
We thus obtain an expansion of the form
\eqn\eIRcon{\eqalign{
\langle O\,\phi\cdots\phi \underbrace{\phi\cdots\phi}_M\rangle_0^{\rm conn}
\ =\ &
\rho^{-M\,d\nu}\sum_{{\rm restricted}\atop{\rm partitions}}
w({\rm partition})\ 
{\bf C}_{O\phi\cdots\phi}^{V_\qvec}
\langle V_\qvec\phi\cdots\phi\rangle_0^{\rm conn}
\times\cr
&\times
\prod_{{{\rm other\ subsets}\atop{\rm not\ containing\ }O}} {\bf C}_{\phi\cdots\phi}^{\bf 1}
\langle {\bf 1}\phi\cdots\phi\rangle_0^{\rm conn}
\,+\,\CO(\rho^{-M\,\nu d-D})
}}
Now, when integrating over the positions of the $M$ last dipoles, we see that
I.R. divergences occur only at $\epsilon=0$ from the $\CO(\rho^{-M\nu d})$
terms in \eIRcon.
As long as there is at least one subset in the partition without $O$ and
with one non-contracted $\phi$, the partition contributes with a
$\langle{\bf 1}\phi\cdots\phi\rangle_0^{\rm conn}=0$.
Thus the only dangerous partitions are those such that {\it all} the $M$
non-contracted $\phi$ are in the same subset as $O$.
These dangerous partitions contribute by a factor
$\langle V_\qvec\phi\cdots\phi\rangle_0^{\rm conn}$.

It is at this last step that the invariance of the observable $O$ 
under global translations of the membrane plays a crucial role.
Indeed, this means that the total charge $\qvec=\sum_i\qvec_i=\vec{\bf 0}$.
But from \emostA\ we then have
\eqn\eAzeroV{
V_{\qvec=\vec{\bf 0}}\ =\ {\bf 1}
}
and therefore the contribution of the dangerous partitions is proportional to
\eqn\ezeroIR{
\langle V_{\qvec=\vec{\bf 0}}\underbrace{\phi\cdots\phi}_M\rangle_0^{\rm conn}
\ =\ 
\langle {\bf 1}\underbrace{\phi\cdots\phi}_M\rangle_0^{\rm conn}
\ =\ 0
}
and vanishes, since $M>0$.

This establishes \eIRInt, and shows that ``superficial" I.R. divergences are
absent from the manifold integral \conmint\ when
$\epsilon\le D/N$, as long as translationally invariant observables (with
zero external charge) are considered.
A complete analysis of the I.R. structure of the manifold integrals,
taking into account nested I.R. configurations, should be required to
prove rigorously the I.R. finiteness of perturbation theory for
$\epsilon\le 0$.
We shall not give this analysis here, but refer to the general results
presented in the next subsections.

%\vfill\break
\subsec {Superficial U.V. Divergences at $\epsilon=0$}
We thus have shown that perturbation theory for invariant observables is
UV finite for $\epsilon>0$, and I.R. finite for $\epsilon\le 0$.
We now discuss in more details the structure of the short distance
singularities at $\epsilon=0$, which have been characterized in subsect. 3.3.
We then define the U.V. subtraction operator for a single singular
configuration (divergent molecule). 
This is the first, but essential, ingredient in the construction of the
full renormalization operator and in the proof of the renormalizability
for the model.

Power counting and the MOPE show that at $\epsilon=0$, there are two kinds of
U.V. singular configurations:
molecules with one atom, with U.V. degree of
convergence $\omega=-D$, as depicted on \fdivdiag~(a), and molecules with
two atoms, with U.V. degree $\omega=0$, as depicted on \fdivdiag~(b).
According to the MOPE, the first case is expected to give UV singularities
(when the distances within the single atom become small) proportional to
the insertion of local operators, while the second case will give
singularities (when the distances within the two atoms independently become
small) proportional to the insertion of bi-local operators.
We shall show that by adding adequate counterterms to the
manifold integrands, proportional to insertions of these local or bi-local
operators, we improve the short-distance behavior of these integrands, without
spoiling the large distance I.R. convergence.

In the following we shall denote the integrand in the manifold integral
\conMInt\ and \conmint\ by
\eqn\eIntgrd{
\langle O_{\{\qvec_i\}}(\{z_i\})\sprod_{a=1}^N\phi\{x_{2a-1},x_{2a}\})
\rangle_0^{\rm conn}
%\ =\ I_{\{\qvec_i\}}^{\rm conn}(\{z_i\},\{x_j\})
\ =\ I^{\rm conn}(x_j)
}
and we consider it as a function of the $2N$ dipole end-points positions
$\{x_j\}$ only.
This is justified since the $P$ external points $z_i$ and external momenta
$\qvec_i$ are kept fixed, and since it appears that no additional
U.V. singularities occur when some $x_j$'s are close to a $z_i$.
The arguments that we shall present below apply equally to the
{\it non-connected} e.v.
%$\langle O\phi\cdots\phi\rangle_0=I$.
\eqn\eIntnc{
\langle O_{\{\qvec_i\}}(\{z_i\})\sprod_{a=1}^N\phi\{x_{2a-1},x_{2a}\})
\rangle_0
%\ =\ I_{\{\qvec_i\}}^{\rm conn}(\{z_i\},\{x_j\})
\ =\ I(x_j)
}
which appear in \eCorrFirst, as far as U.V. problems are concerned
(but which are I.R. divergent).
We shall rely on the methods and on the notations used in
\rIIM, where the renormalization of the model of a random manifold interacting
with a single impurity has been considered.

\medskip
\noindent
$\bullet$ One atom divergences:

First let us consider the singularity associated to a molecule made
of one atom.
This atom, denoted by $\CP$, is made of the $2P$ end-points of the $P$ dipoles
that will be contracted to a single point.
For simplicity we denote 
\eqn\edefP{
|\CP|\,=\,2P\,=\,\hbox{number of points in $\CP$}
}
Let us consider more closely the behavior of the integrand $I(x_j)$ when this
contraction is performed.
For that purpose, we chose a point $w$ inside $\CP$, that we call the
{\it root} of the
atom $\CP$ (at that stage the choice of $w$ is completely arbitrary),
and we rescale the distances inside the atom $\CP$ by a factor $\lambda$
\eqn\escalP{
x_i\,\to\,x_i(\lambda)=\left\{
\eqalign{&x_w+\lambda(x_i-x_w)\quad{\rm if}\quad i\in\CP\cr
&x_i\quad{\rm if}\quad i\not\in\CP}\right.
}
Under this rescaling, the integrand $I(x_j)$ has its small $\lambda$
expansion given by the MOPE, and it is of the form
\eqn\eIExp{
I(x_j(\lambda))\ =\ \sum_{\sigma}\,\lambda^\sigma\,\CT^{(\sigma)} I(x_j)
}
The term of order $\lambda^\sigma$, $\CT^{(\sigma)}I$, is given by the
insertions of 
local operators $A={\rm monomial}(\nabla\rvec, \nabla^2\rvec,\cdots)$
with dimension such that
\eqn\eSigDim{
\sigma\ =\ -Pd\nu-{\rm dim}[A]
\ =\ -Pd\nu+p-q\nu\qquad p\ge q>0\ {\rm integers}
}
and for the non-connected integrand $I(x_j)$ it is explicitly given by
\eqn\eExpCTI{
\CT^{(\sigma)}I(x_j)\ =\ \sum_{A:\ {\rm dim}[A]=\atop -Pd\nu-\sigma}
C_{\underbrace{\phi\cdots\phi}_P}^A(x_j-x_w)\,\langle A(x_w)\phi\cdots\phi
\rangle_0
}
Only the first terms in the expansion \eIExp, such that
\eqn\eSigDiv{
\sigma\le \sigma_{\rm max}(\CP)\ =\ -D(|\CP|-1)
}
give an U.V. divergence when the $|\CP|$ points in $\CP$ are contracted towards
$x_w$. 
We can subtract this divergence by subtracting from $I(x_j)$ these
U.V. divergent terms, thus obtaining 
\eqn\euvsubI{
I(x_j)-\sum_{\sigma\le\sigma_{\rm max}}\CT^{(\sigma)}I(x_j)
}
but we must be careful with I.R. divergences.
The terms with $\sigma<-D(|\CP|-1)$ are I.R. convergent, since when all the
$x_j$ in $\CP$ {\it but} the root $x_w$ go to infinity the corresponding
integral $\int\prod\limits_{j \ne w}dx_j \CT^{(\sigma)}I$ scale as
$\int d\lambda/\lambda\,\lambda^{\sigma-\sigma_{\rm max}}$.
%$\lambda^{D(|\CP|-1)+\sigma}$.
The term with $\sigma=-D(|\CP|-1)$ is I.R. divergent when $\epsilon=0$, since
the corresponding integral scales as $\int d\lambda/\lambda$.
We must introduce an I.R. regulator, in order to perform this subtraction
as small distances (in order to subtract the U.V. divergence), but not at
large distances (in order to keep the I.R. convergence of the original
integral).
A simple I.R. regulator consists in subtracting the $\CT^{(\sigma_{\rm max})}I$ term
only if the distances between all the $|\CP|$ points inside the atom $\CP$
are smaller than a fixed distance $L$.
We write the characteristic function
\def\Chi{\raise .5ex\hbox{$\chi$}}
\eqn\eChiP{
\Chi_\CP(x_j)\ =\ \prod_{i,j\in\CP}\theta(L-|x_i-x_j|)
}
and we thus define the subtraction operation as
\eqn\euvirSub{
\CT_\CP I\ =\ \sum_{\sigma<\sigma_{\rm max}}\CT^{(\sigma)}I\,+\,\CT^{(\sigma_{\rm max})} I\cdot
\Chi_\CP
}
With these definitions, it is now clear that the subtracted integrand
\eqn\eSubInt{
I\,-\,\CT_\CP I
}
behaves, under the rescaling \escalP, as 
\eqn\eSubScl{
%\big(I\,-\,\CT_\CP I\big)(x_j(\lambda))\ =\ o(\lambda^{-(|\CP|-1)D})\quad
%{\rm when}\ \lambda\to 0\ {\rm and}\ \infty
\big(I\,-\,\CT_\CP I\big)(x_j(\lambda))\ =\ 
\cases{\CO(\lambda^{\sigma_{\rm max}-\delta_+})\ \ \delta_+>0\ \ 
\lambda\to +\infty\cr
\CO(\lambda^{\sigma_{\rm max}+\delta_-})\ \ \delta_->0\ \ \lambda\to 0\cr}
}
This ensures both U.V. and I.R. convergence with respect to the integration
over the $x_j$'s in the atom $\CP$, the position of the root $x_w$,
that is of the atom $\CP$ being kept fixed.

\medskip
\noindent
$\bullet$ Two atom divergences:

We now consider the divergence given by a molecule made of two atoms, such
as those depicted on \fdivdiag~(b).
Let us denote $\CC_1$ and $\CC_2$ these two atoms, which are disconnected
subsets of respectively $|\CC_1|$ and $|\CC_2|$ points (note that we may have
$|\CC_1|\ne|\CC_2|$).
We shall call the collection of these two disconnected atoms
\eqn\eDiagTwo{
\CP\ =\ \{\CC_1,\CC_2\}
}
a {\it diagram}, and we call $\CC_1$ and $\CC_2$ its {\it connected components}.
We now consider the behaviour of the integrand $I(x_j)$ when both atoms
are contracted simultaneously.
For that purpose, we chose a root $x_{w_1}$ inside $\CC_1$ and a root 
$x_{w_2}$ inside $\CC_2$, and rescale the distances inside $\CC_1$ and
$\CC_2$ by the factor $\lambda$
\eqn\esclPP{
x_i\,\to\,x_i(\lambda)\,=\ \left\{\eqalign{
&x_{w_1}+\lambda(x_i-x_{w_1})\quad{\rm if}\quad i\in\CC_1\cr
&x_{w_2}+\lambda(x_i-x_{w_2})\quad{\rm if}\quad i\in\CC_2\cr
&x_i\qquad{\rm otherwise}
}\right.
}
Under this rescaling, the integrand $I(x_j)$ has a small $\lambda$ expansion
of the form \eIExp, which is still given by the MOPE, but now in terms of the
insertion of bi-local operators
$\Phi(x_{w_1},x_{w_2})$
with the powers of $\lambda$ given by
$\sigma=-(|\CC_1|+|\CC_2|)d\nu/2-{\rm dim}[\Phi]$.
Again we denote by $\CT^{(\sigma)}I$ the coefficient of the power
$\lambda^\sigma$ in \eIExp, and we subtract all the U.V. divergent terms
such that
\eqn\eSgDvPP{
\sigma\le\sigma_{\rm max}=-D(|\CC_1|-1+|\CC_2|-1)
}
In fact, power counting shows that for these diagrams with two atoms,
there is only one divergent term with $\sigma=\sigma_{\rm max}$, but we may 
discuss as well the general case where there are several divergent terms.
Again, the subtraction of the term with $\sigma=\sigma_{\rm max}$ gives I.R.
divergencies, and we must modify this last subtraction through a I.R. regulator.
We write the characteristic function for the diagram $\CP$ as the
the product of the characteristic functions of its components
\eqn\eCHiPP{
\Chi_\CP(x_j)\ =\ \Chi_{\CC_1}(x_j)\,\Chi_{\CC_2}(x_j)
}
and define the subtraction operation $\CT_\CP$
for the diagram $\CP$ corresponding to
the divergent molecule made of two atoms $\CC_1$ and $\CC_2$ by the same
formula \euvirSub\ as for one atom.
With these definitions, the subtracted integrand behaves under the
rescaling as
\eqn\eSbScPP{
I-\CT_\CP I\ =\ o(\lambda^{-(|\CC_1|+|\CC_2|-2)D})
}
This ensures both U.V. and I.R. convergence with respect to the integration
over the $x_j$'s inside the two components $\CC_1$ and $\CC_2$ of $\CP$.

\medskip\noindent$\bullet$ Sum over roots

We have seen that we need to choose a root, i.e. a contraction point, inside
the connected components of the divergent diagrams.
However we expect that the final result for the subtraction operation is
independent of the choice of root.
For this reason, and in order to have a symmetric formulation, it is better
to sum over possible roots, with an adequate coefficient.
For the simple situation discussed here (subtraction of superficial
divergences), the simplest choice is to choose the same weight for all possible
roots, as done in \rIIM .
For the diagram of \fdivdiag~(a) with one atom $\CP$ and one root $w\in\CP$,
we call the pair $\{\CP,w\}$ a {\it rooted diagram}
\eqn\eRoCoDi{
\CP_w\ =\ \{\CP,w\}
}
and we assign to each rooting a weight
\eqn\eWeOn{
W(\CP_w)\ =\ {1\over |\CP|}
}
Similarly, the diagram with two atoms $\CC_1$ and $\CC_2$ and two roots
$w_1$ and $w_2$ is called a rooted diagram with two components
\eqn\eRoTwDi{
\CP_w\ =\ \Big\{\{\CC_1,w_1\},\{\CC_2,w_2\}\Big\}\ =\ 
\{{\CC_1}_{w_1},{\CC_2}_{w_2}\}
}
with weight
\eqn\eWeTw{
W(\CP_w)\ =\ {1\over |\CC_1||\CC_2|}
}
These rooted diagrams are depicted on 
\fig\frootdiag{Rooted \hyperref\hrootdiag{UV divergent diagrams}
with $N=1$ and $N=2$ connected components (atoms).}
\topinsert
\centerline{\epsfbox{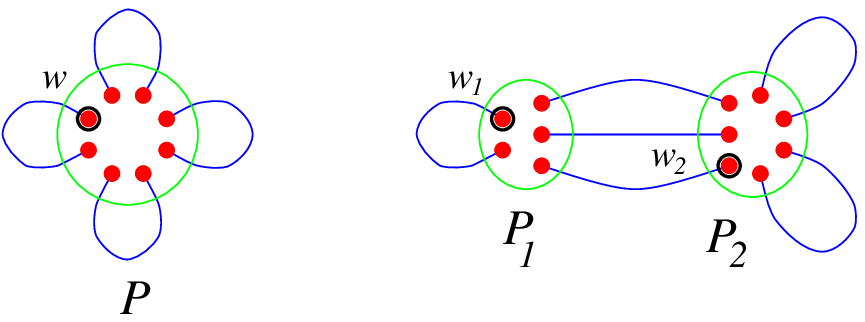}}
\centerline{\hyperdef\hroot{userfigure}{urootdiag}{\frootdiag}}
\endinsert
With these notations, since we have by definition
\eqn\eSumW{
\sum_{{\rm roots}\ w}W(\CP_w)\ =\ 1
}
we can replace in \eSubScl\ and \eSbScPP\ 
the subtraction operator $\CT_\CP$, which depends in fact
explicitly on the rooting $w$, by
\eqn\eRoTay{
\CT_{\CP}\ \to\ \sum_{{\rm roots}\ w}W(\CP_w)\,\CT_{\CP_w}
}

%\subsec{Full renormalization: expression in terms of forests of divergent diagrams}
\subsec{U.V. divergences and subdivergences: the example of the 2nd order.}

We can construct now a general
renormalization operation which makes perturbation theory finite.
As in standard local field theories, the renormalization procedure is defined
by an iterative procedure.

First we subtract the divergent diagrams which appear at the first order
in perturbation theory.
These are the diagrams which form 1-loop connected molecules, as depicted on
\fig\fOldiag{The diagrams giving \hyperref\hOldiag{divergences at one loop}}%
.
We subtract these 1-loop superficial divergences by applying the subtraction
operation $\CT_\CP$ to all families $\CF^{(1)}$ of disjoint 1-loop divergent
diagrams
$\CF^{(1)}=\{\CP_1,\ldots,\CP_l\}$.
We use the notations of \rIIM\ and call two diagrams {\it disjoint} if 
all their connected components (atoms) are mutually disjoint.
%, that is if 
%the intersection of the two diagrams $\CP_1$ and $\CP_2$,
%$\CP_1\wedge\CP_2=\emptyset$.
\topinsert
%\centerline{\epsfxsize=10.truecm\epsfbox{c1loopdiagram.eps}}
\centerline{\epsfbox{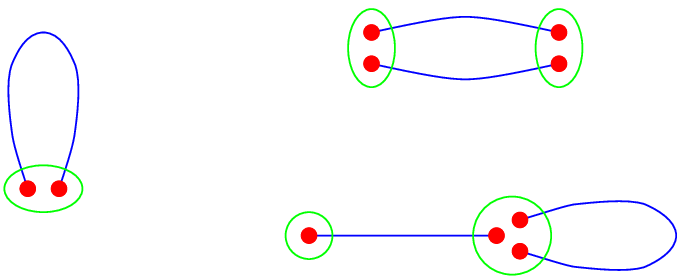}}
\centerline{\hyperdef\hOldiag{userfigure}{uOldiag}{\fOldiag}}
\endinsert

%If we consider that the empty set $\emptyset$ as such a family and
%associate to it the ``identity subtraction operator" $1$ (which does not
%change the integrand),
We can perform these subtractions by applying
to the integrand $I$ the subtraction operation
\eqn\eRIloop{
{\bf R}^{(1)}\ =\ 1\,+\,\sum_{\CF^{(1)}}\prod_{\CP\in\CF^{(1)}}
\left(-\CT_\CP\right)I
}
We note that with our definition \euvirSub\ of the operators $\CT_\CP$,
they commute for disjoint diagrams $\CP_1$ and $\CP_2$.
This subtraction is sufficient to  make perturbation theory U.V. finite
at first order (we do not give the proof at this stage).
It is simple to see that
this subtraction operation ${\bf R}^{(1)}$
amounts to add counterterms $\delta\CH^{(1)}$
to the bare Hamiltonian $\CH$,
proportional to the insertion of the operators ${\bf 1}$, 
$(\nabla\rvec)^2$ and $\phi$.

We now want to subtract superficial UV divergences which appear at second order,
once 1-loop divergences have been subtracted, i.e. for the theory defined by
the Hamiltonian $\CH+\delta\CH^{(1)}$.
We thus have to subtract 2-loops divergent diagrams, which are depicted on
\fig\fTldiag{The diagrams giving \hyperref\hTldiag{divergences at two loops}}
(some diagrams, similar to the third diagram of \fOldiag , are in fact not
divergent and are omitted for simplicity).
These subtractions are performed by including in \eRIloop\ families
of disjoint 1-or-2-loop divergent diagrams.
\midinsert
%\centerline{\epsfxsize=10.truecm\epsfbox{c2loopdiagram.eps}}
\centerline{\epsfbox{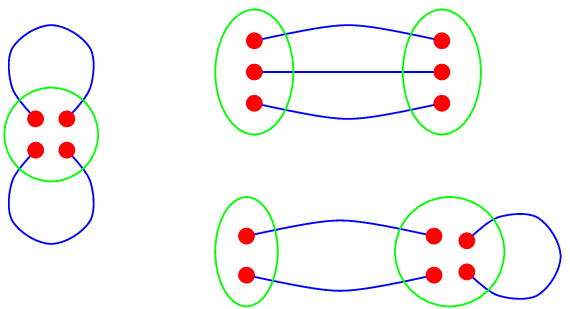}}
\centerline{\hyperdef\hTldiag{userfigure}{uTldiag}{\fTldiag}}
\endinsert

But we have also to subtract the new 1-loop divergent diagrams which appear
because of the additional term $\delta\CH^{(1)}$.
These diagrams are depicted on 
\fig\fOIdiag{The 1-loop diagrams containing insertions of 1-loop counterterms
and giving also \hyperref\hOldiag{divergences at two loops},
with the corresponding nested 2-loop divergent diagrams.
The cross \raise-.5ex\hbox{\epsfbox{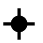}}
denotes the insertion of a local $1$ or
$(\nabla\rvec)^2$ operator.}%
, and contain insertions of the operators
${\bf 1}$, $(\nabla\rvec)^2$ and $\phi$, times the corresponding 1-loop
counterterms, which are associated with 1-loop divergent diagrams.
``Expanding" these contractions into the corresponding diagrams, we can
rewrite the additional subtractions in terms of 2-loop divergent diagrams
containing 1-loop divergent sub-diagrams,
as depicted on \fOIdiag .
\midinsert
%\centerline{\epsfxsize=10.truecm\epsfbox{c11loopdiagram.eps}}
\centerline{\epsfbox{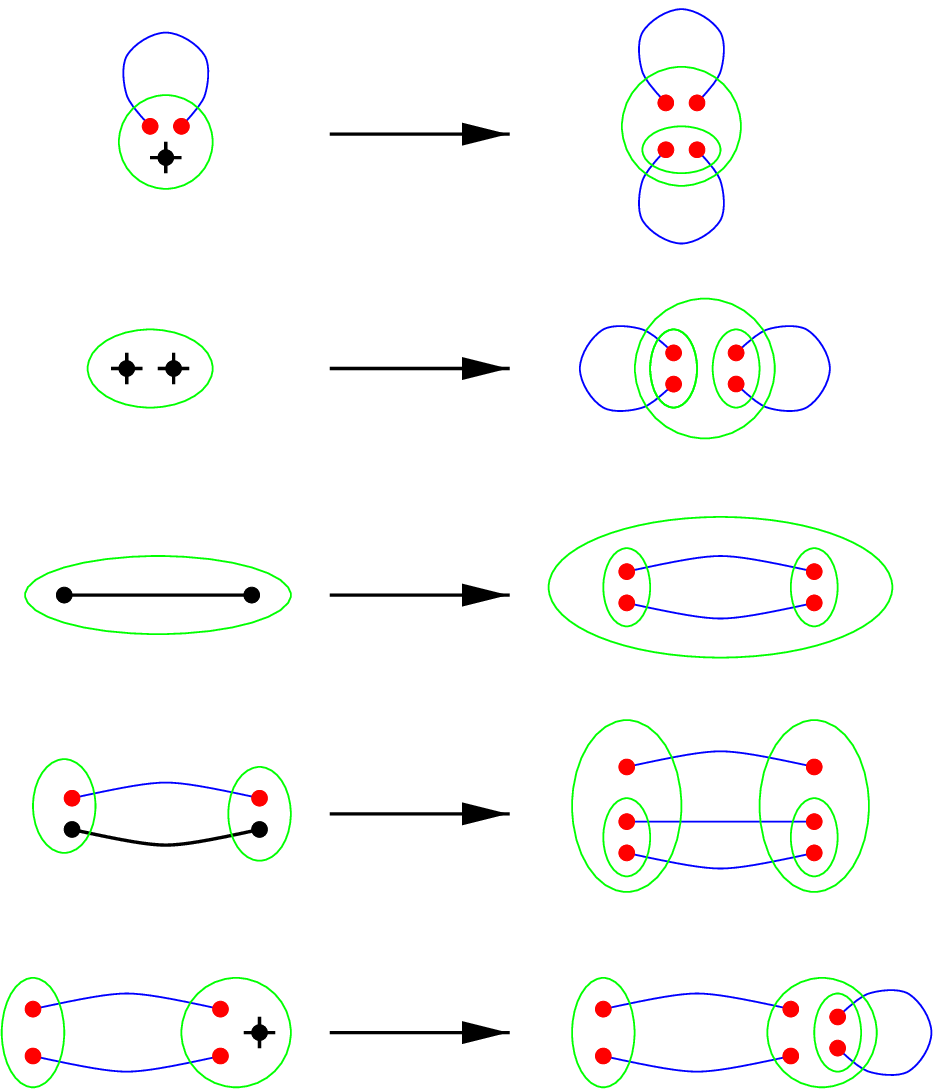}}
\centerline{\hyperdef\hOIdiag{userfigure}{uOldiag}{\fOIdiag}}
\endinsert
The corresponding subtractions can be written as products of 
$\CT_\CP$ operators, as in \eRIloop , for 1-and-2-loop divergent diagrams,
but now some 1-loop diagrams may be ``contained" into 2-loop diagrams
(the precise concept of inclusion for diagrams is given in \rIIM\ and
will be recalled below).
One thus starts to see the appearance of ``nested" subtractions.

This subtraction procedure can be iterated to 3-loops, etc$\ldots$
It produces a renormalization operation, that is expressed as a sum
of subtraction operations over all possible families of ``nested" divergent
diagrams, that we shall call ``forests", in analogy with the well known
Zimmermann's forests of renormalization theory.
We now give the general formula for the renormalization operation.

\subsec{Renormalization in terms of forests of divergent diagrams:}
In the above sketchy discussion, we also have not taken into account the
fact that the subtraction operators $\CT$ are properly defined by
\eIExp\ for rooted diagrams $\CP_w$, and this introduces some subtleties
when we iterate the subtractions.
To deal with this problem, let us recall some definitions of \rIIM .

\medskip
\item{$\bullet$}
A diagram $\CP$ is a family $\{\CC_1,\CC_2,\ldots\}$
of (possibly several) disjoint subsets $\CC_i$ of end-points,
called the connected components of the diagram.
\item{$\bullet$}
A diagram $\CP_1$ is {\it included} in a diagram $\CP_2$ if every connected
component of $\CP_1$ is included in a connected component of $\CP_2$
(not necessarily the same).
This is denoted $\CP_1\prec\CP_2$.
\item{$\bullet$}
Let us consider two rooted diagrams ${\CP_1}_{w_1}$ and ${\CP_2}_{w_2}$ such
that $\CP_1\prec\CP_2$. 
The rootings $w_1$ and $w_2$ are said to be {\it compatible} if the root
of every connected component of $\CP_2$ is either a root of a connected
component of $\CP_1$, or not in any connected component of $\CP_1$.
This is denoted ${\CP_1}_{w_1}\prec{\CP_2}_{w_2}$.
\item{$\bullet$}
The intersection $\CP_1\wedge\CP_2$ of two diagrams $\CP_1$ and $\CP_2$ is
the maximal diagram included in $\CP_1$ and $\CP_2$ (i.e. such that
$\CP\prec\CP_1$ and 
$\CP\prec\CP_2$ $\Rightarrow$ $\CP\prec\CP_1\wedge\CP_2$).
\item{$\bullet$}
Similarly, the union $\CP_1\vee\CP_2$ of two diagrams $\CP_1$ and $\CP_2$ 
is the minimal diagram containing $\CP_1$ and $\CP_2$ (i.e. such that
$\CP_1\prec\CP$ and $\CP_2\prec\CP\ \Rightarrow\ 
\CP_1\vee\CP_2\prec\CP$).
\item{$\bullet$}
A {\it forest} $\CF$ is a collection of diagrams such that every
pair $(\CP_1,\CP_2)$
of elements of $\CF$ are either disjoint ($\CP_1\wedge\CP_2=\emptyset$) or
included $(\CP_1\prec\CP_2$ or $\CP_2\prec\CP_1$).
The empty set $\emptyset$ is considered as a forest.
\item{$\bullet$}
A {\it compatibly rooted forest} $\CF_\oplus$ is a collection of rooted diagrams
such that the corresponding diagrams form a forest and the rootings are all
mutually compatible (if the diagrams are disjoint the rootings are by
definition compatible).
\item{$\bullet$}
To any compatibly rooted forest $\CF_\oplus$ we associate a {\it weight factor}
$W(\CF_\oplus)$, defined as the product over all roots $w_i$ of the inverse
of the number of points in the {\it largest connected component} $\CC_i$
of the diagrams of $\CF$ which has $w_i$ as its root.
\eqn\eWForest{
W(\CF_\oplus)\ =\ \prod_{w_i}{1\over|\CC_i|}
}
It is such that one has in particular 
\eqn\esumW{
\sum_{{\rm compatible}\atop{\rm rootings}\ \oplus}
W(\CF_\oplus)\ =\ 1
}

\medskip

With these notations, we shall define a renormalization operator ${\bf R}$
which generalizes \eRIloop\ and is expressed simply as a sum over all 
(compatibly rooted)
{\it forests of divergent diagrams}, i.e. of (rooted)
diagrams which, together with
the dipole which join the end points in the various connected components,
form a divergent molecule such as those depicted on \frootdiag .

The last subtle point in this construction is to define the product of
subtractions operators
$\CT_{{\CP_1}}$
and
$\CT_{{\CP_2}}$
for two diagrams such that ${\CP_1}\prec{\CP_2}$.
We omit for simplicity the rootings $w_1$ and $w_2$, but we have to remember
that we assume that these rooting are taken into account, and that they are
compatible for nested diagrams.
The subtraction operator for a single diagram $\CP$ is defined by 
\eChiP\ and \euvirSub , i.e. by
\eqn\eSubTe{
\CT_\CP I\ =\ \sum_{\sigma<\sigma_{\rm max}}\CT_\CP^{(\sigma)}I\,+
\,\CT_\CP^{(\sigma_{\rm max})} I\cdot
\Chi_\CP
}
with $\CT_\CP^{(\sigma)}I$ the term of degree $\sigma$ of the expansion
of $I$ with respect to the (rooted) diagram $\CP$.
The last term singles out the ``marginal" term, with degree
\eqn\eSigNot{\eqalign{
\sigma_{\rm max}\ &=\ -\,D\,\sum_{{\rm components}\atop\CC_i\in\CP}(|\CC_i|-1)
\cr
&=\ -\,D(\#\hbox{\ of points in\ }\CP-\#\hbox{\ of components of\ }\CP)\cr
}
}
and $\Chi_\CP$ is the characteristic function
\eqn\eCHIPP{
\Chi_\CP\ =\ \prod_{{\rm components}\atop\CC_i\in\CP}\Chi_{\CC_i}
}
which enforce the constraint that the term with degree $\sigma_{\rm max}$ is
subtracted only if all the distances within each connected component of
$\CP$ are smaller than the I.R. cut-off $L$ (as introduced in \eChiP).
If we now consider two distinct diagrams $\CP$ and $\CP'$ in a forest, there are
two cases:
\item{-} $\CP$ and $\CP'$ are disjoint ($\CP\wedge\CP'=\emptyset$).
In that case, the operators
$\CT_{\CP}$ and $\CT_{\CP'}$, as defined by \eSubTe , commutes and their
action on $I$ is given by
\eqn\eTTdis{
%\CT_{\CP} \CT_{\CP'}I \ =\ \CT_{\CP'} \CT_{\CP} I
%}
%\eqn\eTTprec{
\eqalign{  
\CT_{\CP} \CT_{\CP'}I \ &=\ \CT_{\CP'} \CT_{\CP} I\ =\ 
%\cr
%&
\sum_{\sigma<\sigma_{\rm max}}\sum_{\sigma'<\sigma'_{\rm max}}\CT^{(\sigma)}
{\CT'}^{(\sigma')}I\
\cr
&
+ \ \sum_{\sigma'<\sigma'_{\rm max}}\CT^{(\sigma_{\rm max})}{\CT'}^{(\sigma')}I\cdot\Chi_\CP\
%\cr
%&
+ \ \sum_{\sigma<\sigma_{\rm max}}\CT^{(\sigma)}{\CT'}^{(\sigma'_{\rm max})}I\cdot\Chi_{\CP'}\
\cr
&
+ \ \CT^{(\sigma_{\rm max})}{\CT'}^{(\sigma'_{\rm max})}I\cdot\Chi_\CP\Chi_{\CP'}\
\cr
}
}
\item{-} $\CP$ and $\CP'$ are nested.
Let us assume that $\CP$ is included in $\CP'$ ($\CP\prec \CP'$).
Then two operators $\CT^{(\sigma)}$ and ${\CT'}^{(\sigma')}$ which pick
the terms of degree $\sigma$ when expanding $\CP$ and the term with degree
$\sigma'$ when expanding $\CP'$ commute too (this is a consequence of the MOPE).
The only difficult point is to decide how the dilation of $\CP$ will act on the
I.R. cut-off $\Chi_{\CP'}$. 
It turns out that there are different possible choices, which result
in different renormalization procedures, which are equally consistent.
We shall choose the most simple: we state that the operators
$\CT^{(\sigma)}$ (which extract the homogeneous parts of the integrand)
act on the integrands $I$ but do not act on the I.R. cut-offs $\Chi$.
This means specifically that if $\CP\prec \CP'$, the two operators $\CT_\CP$ and
$\CT_{\CP'}$ still commute and their actions on the integrand $I$ is
{\it still given by} \eTTdis .
Let us note that if we had iterated stricto-sensu the subtraction procedure
that we have discussed at 1-loop and 2-loop orders, we would have obtained a
slightly different definition, which is more complicated to handle.
\medskip

This prescription extends trivially to the product of the $\CT_\CP$ operators
for an arbitrary number of diagrams $\CP$ within a given forest.
With this definition, we can now give the general form for the subtraction
operator ${\bf R}$, which acts on the integrand $I$ (of the form \eIntgrd\ 
and \eIntnc ) of the integrals over internal position space.

\medskip\noindent
{\bf Definition:}

The subtraction operator ${\bf R}$
is given by the sum over all possible compatibly rooted forests
$\CF_\oplus$ made of diagrams
{\it which correspond to U.V. divergent connected molecules}
\eqn\eRfull{
{\bf R}\ =\ \sum_{\CF_\oplus} W(\CF_\oplus) \prod_{\CP\in\CF_\oplus}\left( -\,
\CT_\CP\right)
}

\medskip
Let us stress that in \eRfull\ the empty forest $\emptyset$ (with no elements)
is included, and contributes by the identity operator 
\eqn\eTempty{
W(\emptyset)\prod_{\CP\in\emptyset}\left(-\,\CT_\CP)\right)\ =\ 1
}
which does not change the integrand $I$.

\medskip
We can now state the main result:
\medskip
\noindent{\bf Theorem}

\item{(1)} The renormalized manifold integral
\eqn\eRenInt{
\int\prod_{j=1}^{2N} d^Dx_j\, {\bf R}\,I(x_j)
}
with the integrand $I(x_i)$ defined by \eIntnc , is U.V. finite
for $\epsilon=0$.
\item{(2)} Similarly, the renormalized manifold integral
\eqn\eRenIntC{
O^{(N)}_{\bf R}\{z_i\}\ =\ 
\int\prod_{j=1}^{2N} d^Dx_j\, {\bf R}\,I^{\rm conn}(x_j)
\ ,
}
with the integrand $I^{\rm conn}(x_i)$ defined by \eIntgrd , is U.V. and
I.R finite for $\epsilon=0$.
\item{(3)} 
\eRenIntC\ defines in perturbation theory the renormalized expectation value
of the observable $O$ (given by \eNeutObs ), by 
\eqn\eRenObs{
\langle O\rangle_{\bf R}\ =\ \sum_{N=0}^\infty {(-\bR/2)^N\over N !} \,
O^{(N)}_{\bf R}\{z_i\}
}
$\bR$ is here the (new) renormalized coupling constant.
The renormalization procedure defined by the ${\bf R}$ subtraction
operator amounts to add to the bare Hamiltonian $\CH/\kBT$ counterterms
proportional to multi-local operators $\Phi$, with coefficient defined in
perturbation theory and containing U.V. divergences (i.e. poles in $1/\epsilon$)
of the form
\eqn\eGenCT{
\Delta\CH/\kBT\ =\ \sum_{\Phi}\Delta_\Phi(\bR,\epsilon)\int\!\cdots\int \Phi
}

\medskip
The forest structure of the ${\bf R}$ subtraction operator and the MOPE
ensures point (3), i.e. the fact that renormalization amounts to add 
multilocal counterterms to the Hamiltonian.
This ensures that the model of self-avoiding tethered manifolds is 
perturbatively renormalizable in the usual sense.

In the rest of this section we give the outline of the proof of points
(1) and (2), i.e. of the finiteness of the renormalized theory.

\subsec{Proof of U.V. convergence}
\medskip
\noindent{\it Subtraction in terms of nests}
\medskip

First we shall write the subtraction operation ${\bf R}$ in terms of
all possible diagrams, irrespective to the fact that they correspond or not
to connected or multiconnected molecules or not, and that these molecules
are U.V. divergent or not.
For that purpose, we recall the concept of {\it nest} already introduced
in \rIIM , and familiar in perturbative quantum field theory.
\item{$\bullet$} A {\it nest} $\CN$ is a forest with no disjoint elements.
In other words, it is a family of included diagrams
\eqn\eNestDef{
\CN\ =\ \{\CP_1\prec\CP_2\prec\cdots\prec\CP_M\}
}
Again, the empty set $\emptyset$ is considered as the nest with no elements.
\item{$\bullet$} Similarly, a {\it compatibly rooted nest} $\CN_\oplus$ is a 
compatibly rooted forest with no disjoint elements, and we associate to it
a weight factor $W(\CN_\oplus)$ defined by \eWForest .

\item{$\bullet$} Finally, there is a (purely technical) subtlety associated to
diagrams with a connected component made of one single point
$\CC=\{w\}$.
Indeed, subtracting the non-existing divergence associated  to a single point
amounts to perform no subtraction, i.e. to perform the subtraction associated with the empty set $\emptyset$, and this must not be counted twice, otherwise one gets zero.
Therefore, as explained in \rIIM, when expressing renormalization in terms of
nests, one must consider only the so-called {\it complete diagrams}.

\item{$\bullet$} We denote by $\CG$ the set of the $2N$ end-points in the
manifold integral \conmint .
A {\it complete diagram} of $\CG$ is a diagram $\CP$
which contains all the points in $\CG$.
Its connected components $\CC_i$ may be trivial, if they contain a single
point ($\CC=\{i\}$), or non-trivial, if they contains at least two points.
In other words, a complete diagram is nothing but a partition of $\CG$.
For instance, to the empty diagram $\emptyset$ we associate the complete
diagram  $\CP_\emptyset$ and the trivial subtraction operator $\CT_\emptyset$
\eqn\eemptydiag{
\emptyset\ \to\ \CP_\emptyset\,=\,\{\{i\},i\in\CG\}
\qquad;\qquad\CT_\emptyset\,=\ 1
}
while the maximal complete diagram is 
\eqn\eGdiag{
\CP_\CG\ =\ \{G\}
}
All the definitions of Subsec.~4.5 apply to complete diagrams, and
we define a complete nest simply as:

\item{$\bullet$} A {\it complete nest} is a nest of complete diagrams of $\CG$,
which {\it contains $\CP_\emptyset$ as its smallest element}
\eqn\ecomNest{
\CN\ =\ \{\CP_\emptyset\prec\CP_1\prec\cdots\prec\CP_M\}
}
Thus the smallest complete nest $\CN_\emptyset$ is not the empty set
$\emptyset$, but $\{\CP_\emptyset\}$.

With these definitions, and those of Subsec.~4.5, we have

\medskip
\noindent
{\bf Lemma}\par\nobreak
When acting on the integrand $I(x_j)$, the operator ${\bf R}$ defined by the 
forest formula \eRfull\ may be replaced by the equivalent formula, given now
by a sum over all possible {\it compatibly rooted complete nests} $\CN_\oplus$,
made of all possible complete diagrams
\eqn\eRnest{
{\bf R}\ =\ -\,\sum_{\CN_\oplus} W(\CF_\oplus) \prod_{\CP\in\CN_\oplus}\left( -\,
\CT_\CP\right)
}

\medskip
\noindent{\bf Proof}\par\nobreak
We first have to show that we can replace in \eRfull\ the sum over forests
made of U.V. divergent connected molecules by the sum over forest made of
connected molecules (U.V. convergent or divergent).
This is easy, since if in a forest
there is one diagram $\CP$  corresponding to a U.V. convergent molecule,
from power counting the action of the $\CT_\CP$ operator on $I$ gives
zero.

The second step is to show that we can replace the sum over forests of
connected molecules by the sum over nests.
To show this, let us consider the following simple situation:
Let $\CP_1$ and $\CP_2$ be two disjoint diagrams corresponding to two
disjoint connected molecules, and let us consider
the forest made of $\CP_1$ and $\CP_2$
\eqn\eTwoFor{
\CF\ =\ \{\CP_1,\CP_2\}
}
It gives in \eRfull\ the subtraction operator
\eqn\eRTwo{
\CT_{\CP_1}\CT_{\CP_2}
}
From $\CP_1$ and $\CP_2$ we can build the union diagram $\CP=\CP_1\vee\CP_2$
corresponding to the (disconnected) union of the two molecules.
We associate to this forest $\CF$ the three different nests
\eqn\eThreNe{
\{\CP\}
\quad,\quad
\{\CP_1,\CP\}
\quad,\quad
\{\CP_2,\CP\}
}
which give in \eRnest\ the subtraction operator
\eqn\eRthre{
-\CT_\CP+\CT_{\CP_1}\CT_{\CP} +\CT_{\CP_2}\CT_\CP
}
The key point is that these three nest are equivalent to the forest $\CF$.
Indeed, when applied to the integrand $I$, we have 
\eqn\eRequal{
\left(-\CT_\CP+\CT_{\CP_1}\CT_{\CP}+\CT_{\CP_2}\CT_\CP\right)I
\ =\ \left(\CT_{\CP_1}\CT_{\CP_2}\right) I
}
This is in fact quite simple.
We can rewrite \eRequal\ as
\eqn\eReqbis{
(1-\CT_{\CP_1})\CT_\CP+(1-\CT_{\CP_2})\CT_\CP+\CT_{\CP_1}\CT_{\CP_2}\ =\ 
\CT_\CP
}
and use the fact that all these operators probe the scaling behavior of the
integrand under the dilations of $\CP_1$ and $\CP_2$.
$\CT_{\CP_1}$ picks the terms with degrees
$\sigma_1\le{\sigma_1}_{\rm max}$ w.r.t. $\CP_1$,
$\CT_{\CP_2}$ picks the terms with degrees
$\sigma_2\le{\sigma_2}_{\rm max}$ w.r.t. $\CP_2$,
while $\CT_\CP$ picks the terms the terms with degree
$\sigma=\sigma_1+\sigma_2\le{\sigma_{\rm max}=
\sigma_1}_{\rm max}+{\sigma_2}_{\rm max}$ w.r.t. $\CP$.
Each of the three operators on the l.h.s. of \eReqbis\ picks one of the
{\it disjoint} subsets of terms, which are respectively
\eqn\eDisSubS{\eqalign{
&
\{
\sigma_1>{\sigma_1}_{\rm max},
\sigma_1+\sigma_2\le{\sigma_1}_{\rm max}+{\sigma_2}_{\rm max}
\}
\cr
&
\{
\sigma_2>{\sigma_2}_{\rm max},
\sigma_1+\sigma_2\le{\sigma_1}_{\rm max}+{\sigma_2}_{\rm max}
\}
\cr
&
\{
\sigma_1\le{\sigma_1}_{\rm max},
\sigma_2\le{\sigma_2}_{\rm max}
\}
\cr
}
}
and whose union is 
\eqn\eSubUn{
\{
\sigma_1+\sigma_2\le{\sigma_1}_{\rm max}+{\sigma_2}_{\rm max}
\}
}
that is the set of terms picked by $\CT_\CP$.
Moreover, the I.R. cut-off associated with the ``border terms" such that
$\sigma_1={\sigma_1}_{\rm max}$ or 
$\sigma_2={\sigma_2}_{\rm max}$ coincide in the l.h.s and the r.h.s. of
\eRequal , since by definition  (see \eCHIPP)
$\Chi_{\CP}=\Chi_{\CP_1}\Chi_{\CP_2}$.
This proves \eRequal .

This simple example can be easily generalized. 
To any compatibly rooted nest $\CN_\oplus$ in \eRnest\ one can associate a
unique forest $\CF_\oplus$ in \eRfull ,
made of the {\it connected molecules} which make the components of the
elements of the nest $\CN$.
Thus to any $\CF_\oplus$ we associate a class of nests
$\{\CN_\oplus\to\CF_\oplus\}$, and these classes form a partition of the set
of all nets.
Then one can show that the sum of the subtractions  within a given class
reproduces the subtraction of the corresponding forest
\eqn\eNetoFo{
\sum_{\CN_\oplus\to\CF_\oplus}\prod_{\CP\in\CN_\oplus}(-\CT_\CP)\ =\ 
\prod_{\CP\in\CF_\oplus}(-\CT_\CP)
}
Finally, the weight factors $W(\CN_\oplus)$ of all the nets within a class
are equal to the weight factor $W(\CF_\oplus)$ of the corresponding forest.

The third and final step is to complete all the diagrams in each nest by the
trivial connected components made of the points which are not in the diagram,
thus making all the diagrams and the nests complete.
One easily checks that this amounts to add the factor $(-1)$ in front of
the subtraction operation, to cancel the subtraction $(-\CT_\emptyset)=-1$
associated with the trivial complete diagram $\CP_\emptyset$.

This establishes the lemma.

\medskip
\noindent{\it Decomposition of the integration domain into Hepp sectors}
\medskip
The next step is to decompose the domain of integration $\CA$
(over the positions of the internal points $x_j$, or equivalently over
the mutual distances $|x_j-x_{j'}|$)
into  disjoint ``sectors" $\CH$,
which generalize the well known concept of Hepp sectors in perturbative
quantum field theory.
Each sector corresponds to a particular subdomain for the distance integration,
where the distances  may vanish in a particular, well defined order.
Thus each sector will isolate a family of potential U.V. divergences.

This is done exactly as in \rIIM.
\item{$\bullet$} A {\it saturated nest} is a maximal complete nest of $\CG$,
i.e. a complete nest $S$ made of $2N+1$ distinct complete diagrams.
\eqn\eSatNest{
\CS\ =\ \{\CP_\emptyset\prec\CP_1\prec\cdots\prec\CP_{2N-1}\prec
\CP_{2N}=\CP_\CG\}
}
One shows that one goes from a diagram $\CP_R$ of $\CS$ to the next one,
$\CP_{R+1}$ by fusing two distinct connected components of $\CP_R$ into a single
connected component of $\CP_{R+1}$.

\item{$\bullet$}
To each saturated nest $\CS$ is associated a {\it extended Hepp sector} 
$\CH^\CS$, defined  by the inequalities between distances inside each 
diagram $\CP_R$ of $\CS$.
%Defining the {\it diameter} of a diagram $\CP$ as
%\eqn\eDiamP{
%{\rm diam}(\CP)\ =\ {\rm Sup}(\hbox{distances inside each connected component
%$\CC_i$ of $\CP$})
%}
%$\CH^\CS$ is given by
%\eqn\edefHS{
%0\,=\,{\rm diam}(\CP_\emptyset)<{\rm diam}(\CP_1)<
%\cdots < {\rm diam}(\CP_R)<\cdots<{\rm diam}(\CP_\CG)
%}
Given a diagram $\CP$ composed of connected components $\{\CC_1,\CC_2,\cdots\}$,
we define the distance $d_{ij}$ between  two components $\CC_i$ and $\CC_j$
of $\CP$ as the minimal Euclidean distance between a point in $\CC_i$ and
a point in $\CC_j$.
\eqn\eDisCC{
d_{ij}\ =\ {\rm min}(|x-y|;\ x\in\CC_i,\ y\in\CC_j)
}
As in \rIIM\  we associate to $\CP$ the {\it minimal distance}
$d_{\rm min}(\CP)$ between the connected components of $\CP$
\eqn\edminCP{
d_{\rm min}(\CP)\ =\ \min_{\CC_i\ne\CC_j\in\CP} d_{ij}
}
If $\CP$ has only one connected component ($\CP=\{\CC\}$), by definition
$d_{\rm min}(\CP)=\infty$.
Then, given a saturated nest $\CS$ defined by \eSatNest , the sector $\CH^\CS$ 
is defined by the subdomain of the integration domain $\CA$ such that
\eqn\edefHS{
d_{\rm min}(\CP_\emptyset)<d_{\rm min}(\CP_1)<
\cdots < d_{\rm min}(\CP_R)<\cdots<d_{\rm min}(\CP_\CG)=\infty
}

Following \rIIM, these sectors are disconnected and form a partition of the
integration domain $\CA$
\eqn\esectprop{
\CH^\CS\cap\CH^{\CS'}=\emptyset\qquad\hbox{if}\ \CS\ne\CS'\qquad;
\qquad\bigcup_\CS\CH^\CS\,=\,\CA
}
Inside a given sector $\CH^\CS$, U.V. divergences may occur only if some
distances vanish in a hierarchical order compatible with the inequalities
\edefHS .

\medskip\noindent{\it Overlapping divergences}
\medskip
Then we study the short distance behavior of the renormalized integrand 
${\bf R} I$ separately within each sector $\CH^\CS$.
At that stage, we face the standard problem of overlapping divergences:
the subtraction operator ${\bf R}$ contains subtractions relative to
diagrams which do not correspond to the subsets of distances which are 
allowed to vanish in a given sector.

The solution to this problem turns out to be exactly the same as the one 
of \rIIM, itself inspired from the construction of
\rBerLam\ 
in perturbative quantum field theory.
For a given  Hepp sector $\CH^\CS$,
we shall classify the nests $\CN$ of diagrams which appear
in Eq.~\eRnest\ into disjoint equivalent classes $\CC$. 
Each equivalent class $\CC$ is uniquely characterized by a ``minimal nest"
$\CN_{\rm min}$ and a ``maximal nest" $\CN_{\rm max}$, and is made of all the
nests $\CN$ such that $\CN_{\rm min}\subset\CN\subset\CN_{\rm max}$,
that is of the nests $\CN$  made of all the diagrams of $\CN_{\rm min}$ and 
of some of the diagrams of $\CN_{\rm max}$.
Then we show that it is enough to apply the restricted subtraction operator
${\bf R}_\CC$, obtained by reducing \eRnest\ to the nests of a given class
$\CC$, to make the subtracted integrand ${\bf R}_\CC I$ U.V. convergent
in the sector $\CH^\CS$.
Summing over all different classes $\CC$ we get the subtracted integrand
${\bf R}I$ and its U.V. convergence in the sector $\CH^\CS$ is ensured.

This analysis can be repeated inside each sector $\CH^\CS$ (the decomposition in
classes $\CC$ is different), and thus ensures the U.V. finiteness of the
renormalized integral.

We now recall this general construction.

Let $\CS=\{\CR^0=\CP_\emptyset\prec\CR^1\prec\cdots\prec\CR^{2N}=\CP_g\}$
be a saturated nest, and 
$\CN=\{\CP_0=\CP_\emptyset\prec\CP_1\prec\cdots\prec\CP_T\}$ be a compatibly
rooted complete nest.
We denote $\{w\}_T$ the set of roots of $\CP_T$.
For convenience in notations we add to $\CN_\oplus$ the maximal element
$\CP_{T+1}=\CP_\CG$.
Out of the saturated nest $\CS$ which defines a sector $\CH^\CS$ and of the 
nest $\CN$, which defines a subtraction operation, we construct the
diagrams
\eqn\etableau{\eqalign{
%P^I_J\,=\,\Big(\big(R^I\backslash(P_J\backslash\{w\}_J)\big)\vee P_J\Big)\wedge P_{J+1}
\CP^I_J\,&=\,\big(\CR^I\vee_{\{w\}_J}\CP_J\big)\wedge\CP_{J+1}\cr
\CR^I\vee_{\{w\}_J}\CP_J\,&=\,
\big(\CR^I\backslash(\CP_J\backslash\{w\}_J)\big)\vee \CP_J\cr
}
}
One shows that
\eqn\eprotab{
\CP^0_J\,=\,\CP_J\qquad;\quad \CP^I_J\prec \CP^{I+1}_J
\ .}
As in \rIIM\ we shall call the set of $\CP^I_J$ a tableau.
Hence these diagrams form a nest $\CN_{\rm max}$, which depends on $\CS$ and on
$\CN$,  and which contains the original nest $\CN$.
Note that a large number of the elements $\CP^I_J$ of $\CN_{\rm max}$ 
coincide.
Moreover, one shows that there is a unique minimal nest $\CN_{\rm min}$ such
that the above construction gives the same $\CN_{\rm max}$, and that all the
nests which give $\CN_{\rm max}$ contain $\CN_{\rm min}$.
In other words
\eqn\eNminNmax{
\CN\ \to\ \CN_{\rm max}\qquad\Rightarrow\qquad \CN_{\rm min}\subset\CN\subset
\CN_{\rm max}
}
and all such nests form an equivalent class $\CC_\CS(\CN_{\rm min})$
\eqn\eEquClass{
\CC_\CS(\CN_{\rm min})\ =\ \{\CN :\ 
\CN_{\rm min}\subset\CN\subset\CN_{\rm max}
\}
}
(We do not discuss here the subtleties related to the fact that $\CN_{\rm min}$
and $\CN_{\rm max}$ are rooted nests, like $\CN$, and that the rootings of all
the nests in an equivalent class $\CC_\CS$ must be compatible rootings.
These points are discussed in details in \rIIM.)

The interest of this construction is that, {\it within the
Hepp sector $\CH^\CS$}, the diagrams of the minimal nest $\CN_{\rm min}$ are
{\it not U.V. divergent}, while the diagrams of the maximal nest 
$\CN_{\rm max}$ which are not in $\CN_{\rm min}$ are divergent.
If we consider the reduced subtraction operator ${\bf R}_{\CC}$ relative to
an equivalence class of nests
\eqn\eRCC{
{\bf R}_{\CC}\ =\ -\,\sum_{\CN\in\CC} W(\CN)\prod_{\CP\in\CN}(-\CT_\CP)
}
using \eEquClass\ we can rewrite it as
\eqn\eRCmin{
{\bf R}_{\CC}\ =\ -\,W(\CN_{\rm min})\prod_{\CP\in\CN_{\rm min}}(-\CT_\CP)
\prod_{\CP\in\CN_{\rm max}\backslash\CN_{\rm min}}(1-\CT_\CP)
}
This makes clear that applying ${\bf R}_{\CC}$ to the integrand $I$, we
retain the counterterms relative to the diagrams in $\CN_{\rm min}$ and we
subtract the singular part relative to the diagrams in
$\CN_{\rm max}\backslash\CN_{\rm min}$.

To show that this is enough to make the subtracted integrand U.V. finite in
the sector $\CH^\CS$, let us first give a simple example, inspired from the
Subsection 7.3 of \rIIM .

\medskip\noindent{\it A simple example}

We take as a ``sector" the domain when the distances in a single diagram 
$\CR$ become small.
In other words, we consider the nest $\CS$ with only one non-trivial element.
\eqn\eCSex{
\CS\ =\ \{\CR^0=\CP_\emptyset,\CR^1=\CR\}
}
This is not a saturated nest but this is enough for our example.
Similarly, we take for subtracting nest $\CN$ a nest with a single non-trivial
rooted diagram $\CP$, i.e. we take
\eqn\eCNex{
\CN\ =\ \{\CP_0=\CP_\emptyset,\CP_1=\CP\}
}
As a further simplification, we consider that both $\CR$ and $\CP$ have
only one non-trivial connected component (i.e. with more than one point), that
we denote also $\CR$ and $\CP$, and we denote by$w$ the root of $\CP$.

Generically, the integrand subtracted w.r.t. $\CP$, $(1-\CT_\CP)I$ ,
is not U.V.  convergent when the distances inside $\CR$ become small.
If we apply the above general construction to our example, two distinct cases
must be considered (see
\fig\fclassR{\hyperref\hclassR{Tableau construction in the simple case of
2 diagrams}}%
):
\midinsert
\centerline{\epsfxsize=15.truecm\epsfbox{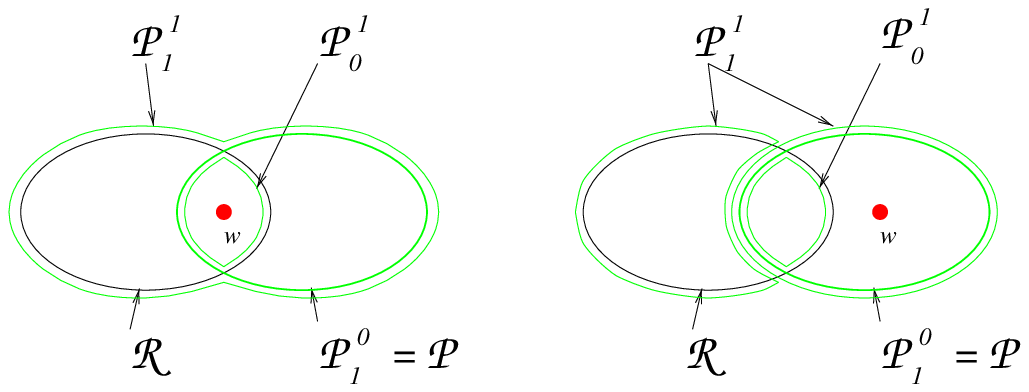}}
\centerline{\hyperdef\hclassR{userfigure}{uclassR}{\fclassR}}
\endinsert
\medskip\noindent If $w\in\CR$ then we find the tableau
\eqn\eTabexone{
\matrix{\CP^0_0=\{\emptyset\}&\CP^1_0=\{\CR\cap \CP\}\cr
        \CP^0_1=\{\CP\}&\CP^1_1=\{\CR\cup\CP\}\cr
       }
}

\medskip\noindent If $w\not\in\CR$ then we find the tableau
\eqn\eTabextwo{
\matrix{\CP^0_0=\{\emptyset\}&\CP^1_0=\{\CR\cap \CP\}\cr
        \CP^0_1=\{\CP\}&\CP^1_1=\{\CP,\CR\backslash\CP\}\cr
       }
}
In both cases the minimal and maximal nests are
\eqn\eNminex{
\CN_{\rm min}\ =\ \CN\ =\ \{\CP^0_0\prec\CP^0_1\}
\qquad;\qquad
\CN_{\rm max}\ =\ \{\CP^0_0\prec\CP^1_0\prec\CP^0_1\prec\CP^1_1\}
}
and the restricted subtraction operator ${\bf R}_\CC$ for the equivalence class
$\CC=\{\CN;\ \CN_{\rm min}\subset\CN\subset\CN_{\rm max}\}$ is
\eqn\eRCCex{
{\bf R}_\CC\ =\ (1-\CT_{\CP^1_1})(-\CT_\CP)(1-\CT_{\CP^1_0})
}

We want to show that when applying ${\bf R}_\CC$ to the integrand $I$,
the result is U.V. convergent when we contract the distances inside $\CR$.
Let us first consider the unsubtracted integrand $I$, and rescale
simultaneously the distances inside $\CP^1_0$, $\CP$ and $\CP^1_1$ by
the respective scaling factors $\lambda_1$, $\lambda$ and $\lambda_2$.
Since these diagrams are nested,
we can apply successively the MOPE for each rescaling to show that $I$ has
a small $\lambda$ expansion of the form
\eqn\eMExpI{
I\ \to\ \sum_{\sigma_1,\sigma,\sigma_2} \lambda_1^{\sigma_1}\lambda^{\sigma}
\lambda_2^{\sigma_2} \ I^{(\sigma_1,\sigma,\sigma_2)}
}
Now we apply the restricted subtraction operator ${\bf R}_\CC$ to $I$.
${\bf R}_\CC I$ has an expansion similar to \eMExpI , but from the definition
of the $\CT$ operators, the $\sigma$'s are restricted by the inequalities
\eqn\eSigInEx{
\sigma_1>\sigma_{\rm max}(\CP^1_0)
\ ,\ 
\sigma\le\sigma_{\rm max}(\CP)
\ ,\ 
\sigma_2>\sigma_{\rm max}(\CP^1_1)
}
With $\sigma_{\rm max}(\CP)$ defined by \eSigNot .

Let us now consider how ${\bf R}_\CC I$ behaves when the distances inside
$\CR$ become small.
For that purpose we now perform a similar rescaling of $\CR$ by a factor
$\lambda_\CR$.
Since from the MOPE under a rescaling of $\CP$ by $\lambda$, $I$ is decomposed
into terms relative to the diagram $\CP$ only (the coefficients of the MOPE),
times terms relative to the original diagram where the whole diagram $\CP$ has
been replaced by its root $w$, it is easy to see that under the rescaling
of $\CR$ by $\lambda_\CR$, each term $I^{(\sigma_1,\sigma,\sigma_2)}$ of the
expansion \eMExpI\ of $\CR_\CC I$ is homogeneous and scales as
\eqn\eScalR{
I^{(\sigma_1,\sigma,\sigma_2)}\ \to\ \lambda^{\sigma_\CR}
I^{(\sigma_1,\sigma,\sigma_2)}
\qquad{\rm with}\qquad
\sigma_\CR\ =\ \sigma_1+\sigma_2-\sigma_\CP
}
Hence ${\bf R}_\CC I$ has an expansion
\eqn\eMExpRCI{
{\bf R}_\CC I\ \to\ \sum_{\sigma_\CR} \lambda_\CR^{\sigma_\CR}\ I^{(\sigma_\CR)}
}
and from the inequalities \eSigInEx
\eqn\eSigRC{
\sigma_\CR\ > \sigma_{\rm max}(\CP^1_0)-\sigma_{\rm max}(\CP)
+\sigma_{\rm max}(\CP^1_1)
%= -D(|\CR_1|+|\CR_2|-|\CP|-1)
%=\ -D(|\CR|-1)\cr
%\sigma_{\rm max}(\CR)
}

In our example, $\CP^0_1$ and $\CP^0_1$ have one component, and $\CP^1_1$
has one component if $w\in\CR$, but two components if $w\not\in\CR$.
This implies that
\eqn\eSigRIn{
\sigma_{\rm max}(\CP^1_0)=-D(|\CR\cap\CP|-1)\ ,\ 
\sigma_{\rm max}(\CP^1_0)=-D(|\CP|-1)\ ,\ 
\sigma_{\rm max}(\CP^1_1)=-D(|\CR\cup\CP|-1\ \hbox{or}\ -2)
}
and, since $|\CR|=|\CR\cup\CP|+|\CR\cap\CP|-|\CP|$, that
\eqn\eSigRFi{
\sigma_\CR\ >\ -D(|\CR|-1)\ =\ \sigma_{\rm max}(\CR)
}
This last inequality ensures that the subtracted integrand ${\bf R}_\CC I$ is
not U.V. singular w.r.t. $\CR$.

\medskip\noindent{\it General Argument}

This can be generalized to get the general proof.
First let us consider the slightly more general case where we have a single
diagram $\CR$ and a nest with a single diagram $\CP$, as above, but now $\CR$
and $\CP$ are general complete diagrams with an arbitrary number of 
non-trivial components.
Out of the nests $\CS$ and $\CN$ defined by \eCSex\ and \eCNex\ we
construct the 4 complete diagrams 
\eqn\eTabexthree{
\matrix{\CP^0_0=\CP_\emptyset&\CP^1_0=\CR\wedge\CP\cr
        \CP^0_1=\CP&\CP^1_1=\CR\vee_{\{w\}}\CP\cr
       }
}
The minimal and maximal nests, and the subtraction operation ${\bf R}_\CC$ are
still defined by \eNminex\ and \eRCCex .
A term of respective degrees $\sigma^I_J$ w.r.t $\CP^I_J$ has still degree
\eqn\eSigRtwo{
\sigma_\CR\,=\ -\sigma^0_0+\sigma^1_0-\sigma^0_1+\sigma^1_1
}
w.r.t. $\CR$ (this generalizes \eScalR , note that $\sigma^0_0=0$).
Hence in ${\bf R}_\CC I$ we get only terms of degree
\eqn\eSigRbis{
\sigma_\CR \,>\,
-\sigma_{\rm max}(\CP^0_0)
+\sigma_{\rm max}(\CP^1_0)
-\sigma_{\rm max}(\CP^0_1)
+\sigma_{\rm max}(\CP^1_1)
}
Since we deal with complete diagrams, their number of points is always $2N$,
and  \eSigNot\ reduces to
\eqn\eSMCom{
\sigma_{\rm max}(\CP)\,=\,D(\#_{\rm com}(\CP)-2N)
\qquad;\qquad \#_{\rm com}(\CP)\,=\,\hbox{number of components of }\CP
}
U.V. convergence is ensured if
\eqn\eSRICV{
\sigma_\CR\, > \, \sigma_{\rm max}(\CR)
}
A sufficient condition for \eSRICV\ is (since $D>0$)
\eqn\eSRSuff{
 \#_{\rm com}(\CP^1_1)
-\#_{\rm com}(\CP^0_1)
+\#_{\rm com}(\CP^1_0)
\ge\#_{\rm com}(\CR)
}
(we used $\#_{\rm com}(\CP^0_0)=\#_{\rm com}(\CP_\emptyset)=2N$)
or equivalently
\eqn\eSRSbis{
\#_{\rm com}(\CR\vee_{\{w\}}\CP)+\#_{\rm com}(\CR\wedge\CP)\ge
\#_{\rm com}(\CP)+\#_{\rm com}(\CR)
}
\eSRSbis\ follows from the fact that by construction 
\eqn\eVeeWVee{
\CR\vee_{\{w\}}\CP\,\prec\,\CR\vee\CP\ \Rightarrow\ 
\#_{\rm com}(\CR\vee_{\{w\}}\CP)\ge\#_{\rm com}(\CR\vee\CP)
}
and from the general inequality valid for any partition $\CA$ and $\CB$ of
a finite set $\CG$ (remember that complete diagrams are nothing but partitions
of $\CG$)
\eqn\eIneqPar{
\#_{\rm com}(\CA\vee\CB)+
\#_{\rm com}(\CA\wedge\CB)\,\ge\,
\#_{\rm com}(\CA)+ \#_{\rm com}(\CB)
}
q.e.d.

\medskip
We now consider the general case.
Let
\eqn\eCSgen{
\CS=\{\CR^0=\CP_\emptyset\prec\CR^1\prec\cdots\prec\CR^{2N-1}\prec\CR^{2N}
=\CP_\CG\}
}
be a saturated nest and
\eqn\eCNmgen{
\CN_{\rm min}=\{\CP_0=\CP_\emptyset\prec\CP_1\prec\cdots\prec\CP_T\}
}
be a minimal nest (w.r.t. $\CS$).
To $\CN_{\rm min}$ is associated the maximal nest
\eqn\eCNMgen{
\CN_{\rm max}=\{\CP^I_J ;\ I=0,\cdots ,2N\ ,\ J=0,\cdots ,T\}
}
and the subtraction operator ${\bf R}_\CC$ defined by \eRCmin .
Using the factorization properties which follow from the MOPE, a term of 
respective degrees $\sigma^I_J$ w.r.t. $\CP^I_J$ has degree
\eqn\eSigRI{
\sigma_{\CR^I}\,=\,\sum_J(\sigma^I_J-\sigma^0_J)
}
w.r.t. $\CR^I$.
U.V. convergence in the Hepp sector $\CH^\CS$ will be ensured if we have
\eqn\eSRHepp{
\sigma_{\CR^I}\,>\,\sigma_{\rm max}(\CR^I)\quad,\qquad I=0,\cdots, 2N
}
Since in \eSigRI\ the $\CP^0_J=\CP_J\in \CN_{\rm min}$, and the
$\CP^I_J\in\CN_{\rm max}$,
in the expansion of ${\bf R}_\CC I$ w.r.t. $\CR^I$, we get only terms
with degree
\eqn\eSigRII{
\sigma_{\CR^I}\,>\,\sum_J\sigma_{\rm max}(\CP^I_J)-\sigma_{\rm max}(\CP^0_J)
}
Hence a sufficient condition for \eSRHepp , i.e. for U.V. convergence, is that
for all I we have
\eqn\eSMInI{
\sum_J\sigma_{\rm max}(\CP^I_J)-\sigma_{\rm max}(\CP_J)\,\ge\,
\sigma_{\rm max}(\CR^I)
}
or equivalently, since we deal with complete diagrams,
\eqn\eComInA{
\sum_{J=0}^T\#_{\rm com}(\CP^I_J)-\#_{\rm com}(\CP_J)\,\ge\,
\#_{\rm com}(\CR)-2N
}
We thus have reduced the problem to a given $I$, i.e. to a given
element of the saturated nest $\CS$, $\CR^I$,
but we still have an inequality involving all the $J$'s, i.e. all the elements
$\CP_J$ of $\CN_{\rm min}$.
We now show that we can reduce the problem to a simple one inside 
each $\CP_J$.
Let us call $\CR^I_{(J+1)}$ the restriction of $\CR^I$ inside $\CP_{J+1}$
\eqn\eCRIJplusI{
\CR^I_{(J+1)}\ =\ \CR^I\wedge\CP_{J+1}
}
The diagram $\CP^I_J$, defined by \etableau, depends on $\CR^I$ only through its
restriction $\CR^I_{(J+1)}$ to $\CP_{J+1}$, since one has
\eqn\eCRIJter{
\CP^I_J\,=\ \CR^I_{(J+1)}\vee_{\{w\}_J}\CP_J
}
And noting that from \eCRIJplusI\ 
\eqn\eCRzeroTplus{
\CR^I_{(0)}\,=\,\CP_\emptyset\qquad,\qquad\CR^I_{(T+1)}\,=\,\CR^I
}
we may rewrite the r.h.s. of \eComInA\ as
\eqn\eComEqB{
\#_{\rm com}(\CR)-2N=\,\sum_{J=0}^T
\#_{\rm com}(\CR^I_{(J+1)})-\#_{\rm com}(\CR^I_{(J)})
}
Hence, a sufficient condition for U.V. convergence w.r.t. $\CR^I$ is that
for each $J$, one has
\eqn\eComInC{
\#_{\rm com}(\CP^I_J)-\#_{\rm com}(\CP_J)
\,\ge\,
\#_{\rm com}(\CR^I_{(J+1)})-\#_{\rm com}(\CR^I_{(J)})
}
But we can now check this inequality {\it inside each component of $\CP_{J+1}$
separately}.
Indeed, considering each component of $\CP_{J+1}$ as a whole $\CG$,
the restriction of $\CP_J$ to this component is a complete rooted
diagram $\CP$ of $\CG$, with roots $\{w\}$,
the restriction of $\CR^I_{(J+1)}$ is another complete diagram $\CR$,
the restriction of $\CR^I_{(J)}$ is nothing but $\CR\wedge\CP$ and
the restriction of $\CP^I_J$ is nothing but $\CR\vee_{\{w\}}\CP$.
Thus the inequality \eComInC\ reduces to 
\eqn\eComInD{ 
\#_{\rm com}(\CR\vee_{\{w\}}\CP)-\#_{\rm com}(\CP) 
\,\ge\,  
\#_{\rm com}(\CR)-\#_{\rm com}(\CR\wedge\CP)
}
This is nothing but \eSRSbis\ that we proved above.
\medskip
We have thus proved that within a Hepp sector $\CH^\CS$, each restricted
subtraction operator ${\bf R}_\CC$ makes the integrand U.V. convergent.
Since 
\eqn\eRsumRC{
{\bf R}\ =\ \sum_\CC {\bf R}_\CC
}
the full subtraction operator ${\bf R}$ makes the integrand U.V.  convergent
in all Hepp sectors, hence in the whole integration domain $\CA$.
This ends the proof of U.V. convergence, that is of point (1) of the theorem
of Sec.~4.5.

\subsec{I.R. finiteness and other points:}

It remains to show point (2) of the Theorem, namely that applying the same
subtraction procedure to  the connected amplitudes defines renormalized
expectation values of invariant observables which are both U.V. and I.R.
finite at $\epsilon=0$.

It is not difficult to prove U.V. finiteness, along the same line as above.
The connected manifold integrand $I^{\rm conn}(x_j)$ in \eRenIntC\ 
, defined by the connected e.v. \eIntgrd , has similar factorization
properties under the short distance expansion as the manifold integrand
$I(x_j)$.
Indeed, they can be written as sum of products of $I(x_j)$'s, and applying
the MOPE to $I^{\rm conn}$ amounts to apply the MOPE to the $I$'s.
Then we can repeat step by step the arguments of Sec.~4.6, to prove
U.V. finiteness of the renormalized manifold integrals \eRenIntC .

Similarly we can show that renormalizing the connected manifold integrals 
amounts to add to the bare Hamiltonian the same counterterms as when
renormalizing the non-connected manifold integrals.

The most important point is to show that the connected manifold integrals
\eRenIntC\ are also I.R. finite, i.e. that no additional divergences occur
when some of the distances become very large ($x_j\to\infty$).
We shall not give a detailed proof of this result here, since it is
lengthy and not particularly illuminating.
We have to follow the procedure of Subsec.~4.2, which was used to show
I.R. finiteness of unrenormalized connected manifold integrals, checking that
the U.V. counterterms do not introduce additional I.R. divergent terms.
This is a consequence of our definition of the subtraction operation
$\CT_\CP$, given by \euvirSub\ 
\eqn\eTCbis{
\CT_\CP I\ =\ \sum_{\sigma<\sigma_{\rm max}}\CT^{(\sigma)}I\,+\,\CT^{(\sigma_{\rm max})} I\cdot
\Chi_\CP
}
When we dilate the distances inside the diagram $\CP$ by a global factor $\rho$,
the first terms $\CT^{(\sigma)}I$ scale as
$\rho^\sigma$, with $\sigma<\sigma_{\rm max}$, and are I.R. convergent when
$\rho\to\infty$ (since they are U.V. divergent when $\rho\to 0$).
The only dangerous term is the last one, with $\sigma=\sigma_{\rm max}$, which
is logarithmically divergent at $\epsilon=0$.
However, because of the I.R. cut-off $\chi_\CP$ (defined by \eChiP ), this
term vanishes if some of the distances inside $\CP$ becomes large, and is
not present in the large $\rho$ asymptotic expansion of $\CT_\CP I$.
This is the reason why this I.R. cut-off was introduced.
This implies that the U.V. counterterms are not I.R. divergent at $\epsilon=0$,
and since the bare connected manifold integrals are I.R. finite, that the
renormalized manifold integrals  are also I.R. finite, as long as the
external points $z_i$ (in the observables $O_{\{\qvec_i\}}(\{z_i\})$)
stay at a finite distance from each other.

\medskip
Another subtle point, that we have not discussed, is related to the
distributional character of the manifold integrals.
We have checked that the U.V. power counting of the subtracted integrands
${\bf R}I$ is good enough.
However, if we had dealt with functions, which had to be integrated with a
well defined measure $d^Dx_j$, it would be necessary to check that the
U.V. bounds are uniform over the integration domain $\CA$.
This point has been checked explicitly for the simpler model of \rIIM\ 
describing a membrane interacting with a single point.
It would be possible to study this problem for our model, although the
techniques of \rIIM\ become very heavy.
However, one has to remember that the measure $d^Dx_j$ is in fact defined as
a distribution over distance space.
This measure is singular on the boundaries of $\CA$, and these singularities
are treated by a finite part prescription (i.e. by integration by parts).
To have a complete proof of U.V. finiteness, one should check explicitly
that these integrations by parts do not interfere with the U.V. singularities
of the integrand (which occur on a much smaller subset of the boundary of
$\CA$).
Explicit calculations at first  and second orders show that this is indeed
the case, and
we do not see any reason why this might not be the case at higher orders,
but we do not have a full proof of this claim.

%\vfill\break
\newsec {Scaling Behaviour}
\subsec {RG Equation}
The outcome of the previous section is that the theory is renormalizable through
exactly two renormalizations, a wave function renormalization and a coupling
constant renormalization.
More precisely, let us define a renormalized Hamiltonian $\CHR[\rvecR]$,
function of a renormalized field $\rvecR(x)$, of a renormalized coupling
constant $\bR$ and of a renormalization momentum scale $\mu$, by
\eqn\eRenHam{
%\beta\,{\CHR}
{\raise.2ex\hbox{$\CHR[\rvecR]$}/\raise -.2ex\hbox{${\rm k}_{\rm B}T$}}
={Z(\bR)\over 2}\int\!d^Dx\,\big(\nabla_x\rvecR (x)\big)^2+
{\bR Z_b(\bR) \mu^\epsilon
\over 2}\int\! d^Dx\int\! d^Dx'\ \delta^{d}\big(\rvecR (x)-\rvecR (x')\big)
\ .}
One can build in perturbation theory two renormalization factors
$Z(\bR)$ and $Z_b(\bR)$, which are functions of $\bR$ and of $\epsilon$,
so that the theory, when expressed in terms of the renormalized quantities
$\rvecR$ and $\bR$, is UV finite when $\epsilon\to 0$.
This means that if we consider any physical observable $O(\rvecR)$,
such as the two point function
\eqn\eTwoPtFct{
F\ =\ (\rvecR(x)-\rvecR(y))^2
\ ,}
its average value 
\eqn\eRenAv{
F_{\sbf R}(x,y) =\ \langle(\rvecR(x)-\rvecR(y))^2\rangleR\ =\ 
{\int \CD[\rvecR(x)]\,F(\rvecR)\,\ee^{-
\CHR[\rvecR]/\kBT
}
\over \int \CD[\rvecR(x)]\,\ee^{-\CHR[\rvecR]/\kBT}}
}
is a UV finite function of $\bR$ as $\epsilon\to 0$,
when evaluated with the renormalized Hamiltonian $\CHR$.
As usual, the renormalization factors $Z$ and $Z_b$, when expanded in $\bR$,
have UV divergences, which are poles in $\epsilon$
\eqn\eZCT{
\eqalign{
Z(\bR)\ &=\ 1\,+\,\bR({A\over\epsilon}+\ldots)\,+\,\CO(\bR^2)\cr
%\ldots\cr
Z_b(\bR)\ &=\ 1\,+\,\bR({B\over\epsilon}+\ldots)\,+\,\CO(\bR^2)\cr
%\ldots\cr
}}

A major consequence of this renormalizability is the existence of
renormalization group equations, which as usual come from the freedom in the
choice of the renormalization scale $\mu$, and which can be derived as follows.
The renormalized Hamiltonian $\CHR[\rvecR]$ is equal to the bare Hamiltonian
$\CH[\rvec]$ given by \eEdwards\ by the change from renormalized to
bare quantities
\eqn\eRenBare{
\rvec\,=\,Z^{1/2}(\bR)\rvecR\quad;\quad b\,=\,\bR\,Z_b(\bR)\,Z^{d/2}(\bR)\,\mu^\epsilon
}
This means for instance that
\eqn\eEVRB{
\langle(\rvecR(x)-\rvecR(y))^2\rangleR\ =\ Z^{-1}\,
\langle(\rvec(x)-\rvec(y))^2\rangle
\ .}
Considering that $b$ is fixed and varying $\mu$, \eRenBare\ defines a
renormalization group flow for $\bR$, characterized by the Wilson $\beta$
function
\eqn\eBeta{
\beta(\bR)\ =\ \left.\mu{\partial\over\partial\mu}\right|_{b}\bR
}
Similarly, the anomalous dimension for the field $\rvecR$ is given by
\eqn\eDimAn{\eqalign{
\nu(\bR)\ &=\ \left.\mu{\partial\over\partial\mu}\right|_{b,\rvec}
\ln (\rvecR\mu^\nu) \cr
&=\ \nu\,-{1\over 2}\,\left.\mu{\partial\over\partial\mu}\right|_{b}
\ln Z\cr
}}
where $\nu=(2-D)/2$ is the canonical dimension of $\rvec$.
Since the functions $\beta$ and $\nu$ are dimensionless, they depend only
on $\bR$ (and on $\epsilon$ and $D$), and not on $\mu$.
A variation of the renormalization scale $\mu$ for the two-point function
\eRenAv\ is absorbed into a change of 
$\bR$ and a wave-function renormalization, through the renormalization
group equation
\eqn\eRGequ{
\left[\mu{\partial\over\partial\mu}\,+\,\beta(\bR){\partial\over\partial\bR}
\,+\, 2(\nu-\nu(\bR))\right]F_{\sbf R}(x-y)\ =\ 0
\ .}
From \eRGequ\ one derives the Callan-Symanzik equation, which gives the 
variation of the two-point function under global rescaling of the distance
$\ell=|x-y|$.
Thanks to the simple homogeneity relation 
\eqn\eHomogen{
F_{\sbf R}(\ell,\mu)\ =\ \mu^{-2\nu}\,f(\mu\ell)
}
the renormalization group equation gives
\eqn\eCSequ{
\left[\ell{\partial\over\partial\ell}\,+\,\beta(\bR){\partial\over\partial\bR}
\,-\, 2\nu(\bR)\right]F_{\sbf R}(\ell)\ =\ 0
\ .}
As usual, if there is an infra-red stable fixed point $\bR^\star>0$, such that
$\beta(\bR^\star)=0$, $\beta'(\bR^\star)>0$, the large $\ell$ behaviour of
$F_{\sbf R}(\ell)$ will be governed by this fixed point, leading to
\eqn\eScalBeh{
F_{\sbf R}(\ell)\ \sim\ \ell^{\,2\nu^\star}\qquad;\qquad\nu^\star\,=\,
\nu(\bR^\star)
}
i.e. to the fractal dimension $d_{\rm frac}$ of the membrane
\eqn\eFracDim{
d_{\rm frac}\ =\ D/\nu^\star
\ .}

%\vfill\break
\subsec {1-Loop Results}

We now apply the MOPE formalism to compute the renormalization group functions
for the model of self-avoiding tethered membranes at one loop, and show how
to recover the results of [\xref\rKaNe-\xref\rArLu-\xref\rBDone]
for the scaling indices at first order in $\epsilon$.
For simplicity let us denote by $\phi\{x,y\}$ the two-point operator
\eqn\eDefphi{
\phi\{x,y\}\ =\ \delta^d(\rvec(x)-\rvec(y))
\ .}
The U.V. divergent diagrams at first order are represented on 
\fig\fOldiagx{The diagrams giving \hyperref\hOldiagx{divergences at one loop}}
\midinsert
%\centerline{\epsfxsize=10.truecm\epsfbox{c1loopdiagx.eps}}
\centerline{\epsfbox{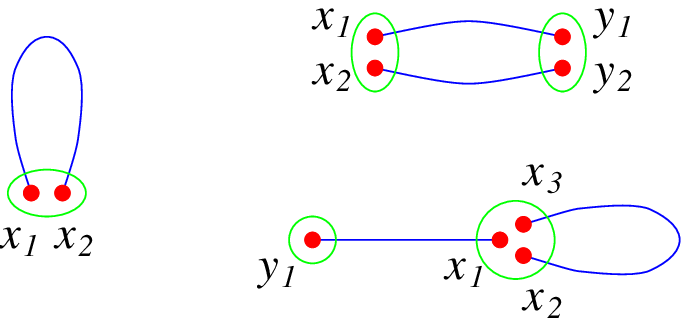}}
\centerline{\hyperdef\hOldiagx{userfigure}{uOldiagx}{\fOldiagx}}
\endinsert
We need the MOPE for the contraction
$\phi\{x_1,x_2\}\to\hbox{\sl local operators}$ when
$x_1\to x_2$, which is 
\eqn\ephimope{
\phi\{x_1,x_2\} {\buildlim x_1\to x_2\under=}
\Cbf_\phi^{\bf 1}\{x_1-x_2\}\,{\bf 1}(x)\,+\,
\Cbf_\phi^{\alpha\beta}\{x_1-x_2\}\,
\npr\nabla_\alpha\rvec(x)\nabla_\beta\rvec(x)\npr\,+\,\ldots
}
with $x=(x_1+x_2)/2$.
We start from the OPE 
\eqn\ephiope{\eqalign{
\ee^{\ii\kvec(\rvec(x_1)-\rvec(x_2))}\ =\ &\Big[1\,
+\,\ii(x_1^\alpha-x_2^\alpha)\npr\nabla_\alpha\rbf^\mu(x)\npr\kbf_\mu\,\cr
-\,{1\over 2}(x^\alpha_2-x^\alpha_1)&(x^\beta_2-x^\beta_1)
\npr\nabla_\alpha\rbf^\mu(x)\nabla_\beta\rbf^\nu(x)\npr
\kbf_\mu\kbf_\nu\,+\,\cdots\Big]\,\ee^{\kvec^2 G(x_1-x_2)}
\cr}
}
The MOPE \ephimope\ is obtained by integrating over $\kvec$ , leading to
\eqn\eCphione{
\Cbf_\phi^{\bf 1}\{x_1-x_2\}\ =\ (4\pi)^{-d/2}\,(-G(x_1-x_2))^{-d/2}
}
\eqn\eCphitwo{
\Cbf_\phi^{\alpha\beta}\{x_1-x_2\}\ =\ -\,{1\over 4}\,(4\pi)^{-d/2}\,
(-G(x_1-x_2))^{-1-d/2}(x_1^\alpha-x_2^\alpha)(x_1^\beta-x_2^\beta)
}
The MOPE for $\phi\phi\to\phi$ is of the form
\eqn\ephiphi{
\phi\{x_1,y_1\}\phi\{x_2,y_2\} {\buildlim{x_1\to x_2\atop y_1\to y_2}\under =}
\Cbf_{\phi\phi}^\phi\{x_1-x_2, y_1-y_2\}\,
\phi\{x,y\}\,+\,\cdots
}
where $x=(x_1+x_2)/2$, $y=(y_1+y_2)/2$.
We use the OPE
\eqn\epphiope{\eqalign{
\ee^{\ii\kvec_1(\rvec(x_1)-\rvec(y_1))}
&\ee^{\ii\kvec_2(\rvec(x_2)-\rvec(y_2))}
\ =\ \ee^{\ii(\kvec_1\rvec(x_1)+\kvec_2\rvec(x_2))}
\ee^{-\ii(\kvec_1\rvec(y_1)+\kvec_2\rvec(y_2))}\cr
=\ &\left(\ee^{\ii(\kvec_1+\kvec_2)\rvec(x)}+\cdots\right)
\ee^{-\kvec_1\kvec_2 G(x_1-x_2)}\,
\left(\ee^{-\ii(\kvec_1+\kvec_2)\rvec(y)}+\cdots\right)
\ee^{-\kvec_1\kvec_2 G(y_1-y_2)}\,\cr
}}
The coefficient $\Cbf_{\phi\phi}^\phi$ is obtained by integrating over
$\kvec_1$ and $\kvec_2$, after inserting the identity
$1=\int d^d\kvec\,\delta^d(\kvec-\kvec_1-\kvec_2)$
%\eqn\ekkonektwo{
%1\ =\ \int d^d\kvec\,\delta^d(\kvec-\kvec_1-\kvec_2)
%}
so that we get for $\phi\phi$
\eqn\epphiint{
\int\kern-1.ex\int\kern-1.ex\int
{d^d\kvec_1\over(2\pi)^d}\,{d^d\kvec_2\over(2\pi)^d}\,d^d\kvec
\,\delta^d(\kvec-\kvec_1-\kvec_2)\,\ee^{\ii\kvec(\rvec(x)-\rvec(y))}
\,\ee^{-\kvec_1\kvec_2(G(x_1-x_2)+G(y_1-y_2))}
}
After expanding
$\delta^d(\kvec-\kvec_1-\kvec_2)=\delta^d(\kvec_1+\kvec_2)+\cdots$,
the integral over $\kvec$ builds the operator $\phi\{x,y\}$, and the integral
over $\kvec_1$ gives the $\Cbf_{\phi\phi}^\phi$ coefficient
\eqn\eCphiphi{
\Cbf_{\phi\phi}^\phi\{x_1-x_2,y_1-y_2\}\ =\ (4\pi)^{-d/2}\,
\left[-G(x_1-x_2)-G(y_1-y_2)\right]^{-d/2}
}

Finally, the MOPE for the one-particle-reducible diagram is easily found to
factorize into
\eqn\ephiprime{
\phi\{x_1,y_1\}\phi\{x_2,x_3\}{\buildlim{x_2,x_3\to x_1}\under =}
\Cbf_\phi^{\bf 1}\{x_2,x_3\}\,\phi\{x_1,y_1\}\,+\,\cdots
}
This shows that the divergence naively associated to this diagram is in fact
not present, and is a reflect of the relevant divergence $\phi\to{\bf 1}$.

Let us now compute the renormalization factors $Z$ and $Z_b$ at first
order in $\bR$, i.e. the coefficients $A$ and $B$ in \eZCT .
One starts from the action \eRenHam\ without counterterms, that is
with $Z=Z_b=1$.
The perturbative expansion gives at order $\bR$ a term
\eqn\eFrstFact{
-{\bR\mu^\epsilon\over 2}\int\kern-1.ex\int d^D x_1 d^D x_2\,\phi\{x_1,x_2\}
}
When $x_1\to x_2$, it gives a singular contribution (a pole in $\epsilon$)
equal to
\eqn\eFrstSing{
-{\bR\mu^\epsilon\over 2}\int d^D x\,\npr(\nabla\rvecR(x))^2\npr\,
\int_{|x_1-x_2|\le L} 
\kern-15.pt d^D (x_1-x_2)\, {\delta_{\alpha\beta}\over D}\,
C_\phi^{\alpha\beta}\{x_1-x_2\}
}
$L$ is an arbitrary I.R. length scale, and since the residue of the pole does
not depend on $L$, we can set it to $L=\mu^{-1}$ (the inverse of the
renormalization mass scale).
The last integral over $(x_1-x_2)$ can be performed explicitly, and gives
\eqn\eFrstInt{
\int_{|y|\le\mu^{-1}}\kern-15pt d^Dy\,{1\over D}\,C^{\alpha\alpha}_\phi\{y\}\,=\,
{\mu^{-\epsilon}\over\epsilon}\,{-1\over 4D}\,(4\pi)^{-{d\over 2}}\,
(2-D)^{1+{d\over 2}}\,
\left({2\,\pi^{D/2}\over\Gamma(D/2)}\right)^{2+{d \over 2}}
}
This pole is therefore subtracted by a wave-function renormalization, by
choosing 
\eqn\eZonelp{
Z\ =\ 1\,+\, {\bR\over\epsilon}\,{1\over 4D}\,(4\pi)^{-{d\over 2}}\,
(2-D)^{1+{d\over 2}}\,
\left({2\,\pi^{D/2}\over\Gamma(D/2)}\right)^{2+{d \over 2}}
}
At order $\bR^2$ we get a factor
\eqn\eScndFact{
{\bR^2\mu^{2\epsilon}\over 8}\int\kern-1.ex\int\kern-1.ex\int\kern-1.ex\int
d^D x_1 d^D y_1 d^D x_2 d^D y_2 \,\phi\{x_1,y_1\}\,\phi\{x_2,y_2\}
}
When $x_1\to x_2$, $y_1\to y_2$ simultaneously, it gives a singular contribution
equal to
\eqn\eScndSing{
{\bR^2\mu^{2\epsilon}\over 8}\int\kern-1.ex\int d^D x\, d^D y\,\phi\{x,y\}\,
\int\kern-1.ex\int_{{|x_1-y_1|\le L\atop |y_1-y_2|\le L}}
\kern-15.pt d^D(x_1-x_2)
d^D (y_1-y_2)\, \Cbf_{\phi\phi}^\phi\{x_1-x_2, y_1-y_2\}
}
The singular part of this last double integral can be calculated
explicitly, leading to
\eqn\eScndInt{
\int\kern-1.ex\int_{{|u|\le\mu^{-1}\atop |v|\le\mu^{-1}}}
\kern-15.pt d^Du\, d^D v\, \Cbf_{\phi\phi}^\phi\{u,v\}\ =\ 
\mu^{-\epsilon}\,\left[{1\over\epsilon}\,%(4\pi)^{-{d\over 2}}\,
{(2-D)^{-1+{d\over 2}}\over(4\pi)^{d\over 2}}\,
\left({2\,\pi^{D\over 2}\over\Gamma(D/2)}\right)^{2+{d\over 2}}\,
{\Gamma\left({D\over 2-D}\right)^2\over\Gamma\left({2D\over 2-D}\right)}
\,+\,\cdots\right]
}
The same pole comes also from the situation where $x_1\to y_2$, $y_1\to x_2$.
These two poles are subtracted by a coupling constant renormalization, by
choosing
\eqn\eZbonelp{
Z_b\ =\ 1\,+\,{\bR\over\epsilon}\,{1\over 2}
\,(4\pi)^{-{d\over 2}}\, (2-D)^{-1+{d\over 2}}\,
\left({2\,\pi^{D\over 2}\over\Gamma(D/2)}\right)^{2+{d\over 2}}\,
{\Gamma\left({D\over 2-D}\right)^2\over\Gamma\left({2D\over 2-D}\right)}
}

From \eZonelp\ and \eZbonelp\ is is straightforward to compute the
renormalization group functions, using \eRenBare, \eBeta\ and \eDimAn .
We find 
\eqn\eBetaol{
\beta(\bR)\ =\ -\epsilon\bR\,+\,{\bR^2\over 2}\,
{(2-D)^{-1+{d\over 2}}\over(4\pi)^{d\over 2}}\,
\left({2\,\pi^{D\over 2}\over\Gamma(D/2)}\right)^{2+{d\over 2}}\,
\left[
{\Gamma\left({D\over 2-D}\right)^2\over\Gamma\left({2D\over 2-D}\right)}
+{d\over 2}{(2-D)^2\over 2D}\right]
}
\eqn\eNuol{
\nu(\bR)\ =\ {2-D\over 2}\,+\,{\bR\over 8D}\,
{(2-D)^{1+{d\over 2}}\over(4\pi)^{d\over 2}}\,
\left({2\,\pi^{D\over 2}\over\Gamma(D/2)}\right)^{2+{d\over 2}}
}
The $\bR^2$ coefficient of $\beta$ is positive for $0<D<2$. Therefore
for small positive $\epsilon>0$, the $\beta$ function has an I.R. attractive
fixed point $\bR^\star\propto\epsilon$, which governs the large distance
scaling behavior of the membrane.
Using \eScalBeh\ the anomalous dimension at order $\epsilon$ is
\eqn\eNustar{
\nu^\star\ =\ {2-D\over 2}\,+\,\epsilon\,\left[d+{4D\over (2-D)^2}
{\Gamma\left({D\over 2-D}\right)^2\over\Gamma\left({2D\over 2-D}\right)}
\right]^{-1}
}
This result agrees with that given in [\xref\rArLu,\xref\rKaNe,\xref\rBDone]
(it differs only by a term of order $\CO(\epsilon^2)$, and one has to remember that the expansion parameter ${\varepsilon}=4D-d(2-D)$ used in these papers is
not the expansion parameter $\epsilon$ used here, but is 
$\varepsilon=2\epsilon$).

%\vfill\break
\newsec {Finite Membranes}
\subsec{Extension of the model to finite membranes}
Up to now we have considered infinite membranes only: the model defined
by the action \eEdwards\ is defined on the infinite Euclidean space $\RR^D$.
It is important to extend our results to finite membranes.
In particular, the direct renormalization method used in
[\xref\rKaNe,\xref\rArLu,\xref\rBDone]
to perform one-loop calculations makes a crucial use of the finiteness of
the membrane to define renormalized quantities, such as the dimensionless
second virial coefficient $g$, and to study its flow with the size of the
membrane $z$ as $z\to\infty$.
In order to prove the consistency of the direct renormalization method, we need
to relate it to the field theoretical renormalization method developed here.
Moreover, several scaling indices characterize finite membranes.
This is the case of the configuration exponent $\gamma$, and of the contact
exponents $\theta$'s for membranes with boundaries.

Let us first consider a finite tethered membrane without boundary.
It corresponds for instance to an irregular lattice with coordination
defects and link varying elastic moduli.
Since on such a lattice there is no natural coordinate system, it is better
to describe such a system in the continuum limit in a coordinate independent
way.
This means that we choose arbitrarily some coordinate system $\{x\}$ on our
manifold $\CM$ (this is always possible locally, and we want to find a
continuum Hamiltonian which describes the elastic properties of such
a membrane.
If we first neglect self-avoidance and consider a phantom membrane, its elastic
properties are simply encoded by a fixed internal Riemannian metric
$g_{\alpha\beta}(x)$ on the manifold $\CM$.
The elastic Hamiltonian is still Gaussian, and of the form
\eqn\eElHaC{
{\raise.2ex\hbox{$\CH_{\rm el}[\rvec]$}/\raise -.2ex\hbox{${\rm k}_{\rm B}T$}}
\ =\ {1\over 2}\,\int_\CM d^Dx\,\sqrt{g(x)}\,
\partial_\alpha\rvec(x)\,g^{\alpha\beta}(x)\,\partial_\beta\rvec(x)
} 
As before, $\alpha=1,\cdots,D$ label the internal coordinate indices.
This Hamiltonian transforms covariantly under changes of coordinates:
the same membrane, with the same internal elastic properties, is described in
the coordinates $x'$ by the metric
$g'_{\alpha\beta}(x')={\partial x^\gamma\over\partial x'^\alpha}
{\partial x^\delta\over\partial x'^\beta} g_{\gamma\delta}(x)$.

%Its internal geometry is fixed (tethered membrane), this means that in
%a continuum description this geometry is given by a fixed (non-fluctuating)
%internal metric $g_{\alpha\beta}(x)$ on the manifold $\CM$.
%As before, $\alpha=1,\cdots,D$ label internal coordinates indices, and
%$x=\{x^\alpha\}$ are local coordinates on $M$.
%The Hamiltonian must depend on the configuration $\rvec(x)$, and on the
%fixed internal metric, but not on the choice of local coordinate $x$.
%This reparametrization invariance fixes the Hamiltonian \eEdwards\ to be
It is now easy to generalize the Hamiltonian \eEdwards\ for self-avoiding
membranes.
The covariance under a change of coordinates fixes it to be
\eqn\eEdCurv{\eqalign{
{\raise.2ex\hbox{$\CH[\rvec]$}/\raise -.2ex\hbox{${\rm k}_{\rm B}T$}}
\ =\ &{1\over 2}\,\int_\CM d^Dx\,\sqrt{g(x)}\,
D^\alpha\rvec(x)\,D_\alpha\rvec(x)\cr
+\,&
{b\over 2}\int_\CM d^Dx\,\sqrt{g(x)}\,\int_\CM d^Dx'\,\sqrt{g(x')}\ 
\delta^{d}\big(\rvec (x)-\rvec (x')\big)\cr
}
}
The $D_\alpha$ and $D^\alpha$ are covariant and contravariant derivatives for
the internal metric $g$.
Formally, the perturbative expansion can be constructed along the same lines
as for an infinite membrane. 
It involves the same diagrams and the same manifold integrals
\eZManInt\ and \eVManInt .
Simply, the two point function $G(x,y)$ on $\RR^D$ given by \eMlssProp\ has to
be replaced by the massless propagator on the surface $\CM$, $G_\CM(x,y)$,
solution of
\eqn\eProCurv{
-\,\Delta_x\,G_\CM(x,y)\ =\ {1\over\sqrt{g(y)}}\,\delta^D(x-y)\,-\,{1\over
{\rm Vol}(\CM)}
}
where $\Delta=D^\alpha D_\alpha$ is the scalar Laplacian on $\CM$ with respect to
$x$.
The additional source term, proportional to the inverse of the area of the
manifold ${\rm Vol}(\CM)=\int_\CM d^Dx\,\sqrt{g(x)}$, is here to take care of the
zero mode of the Laplacian.
This zero mode is a constant and is associated with the global
translation invariance $\rvec(x)\to\rvec(x)+\rvec_0$ of the model.
The propagator, solution of \eProCurv, is defined up to a constant.
However, this arbitrariness disappear in the partition function \eZPertExp\ and
in the translationally invariant observables \eCorrFirst .
Indeed, the determinant $\Delta\{x_i\}$ and the minors $N^{lm}\{x_i,z_n\}$
involve only differences of propagators $G_\CM(x,y)$.

For {\it open membranes}, i.e. membranes with a boundary, one must specify
the boundary conditions.
Translation invariance implies that the natural boundary conditions are 
{\it Neumann boundary conditions}: the normal derivative $D_\perp\rvec$ of the
field $\rvec$ must vanish on the $(D-1)$-dimensional boundary $\partial \CM$ of
$\CM$
\eqn\eNeumann{
D_\perp\rvec(x)\ =\ n^a(x)D_a\rvec(x)\ =\ 0\qquad\hbox{if\ }x\in\partial \CM
\ ,}
where $n^a$ is the unit vector normal to the boundary.

%\vfill\break
\subsec{Examples of $D$-dimensional manifolds}
The above described construction is somewhat formal.
In practice perturbative calculations have to be performed for
$\epsilon=2D-(2-D)d/2$ small and  non-integer $0<D<2$.
A general formulation of curved (Riemannian) manifolds with non-integer
dimensions is lacking, but one can device several explicit examples where
calculations are possible.

A simple case is discussed in \rIIM : the
constant curvature $d$-dimensional sphere $\CS_D$  with radius $L$.
By embedding the sphere in $\RR^{D+1}$, one can generalize the concept of
distance geometry in $\RR^D$ to $\CS_D$, and use SO($D+1$) invariance
(instead of Euclidean invariance E($D$)) to write the measure 
$\sprod\limits_{i=1}^Md^Dx_i\sqrt{g(x_i)}$ for
$M$ points over $\CS_D$ as a distribution 
$d\mu_\CM\{\ell_{ij};D,L\}$ over the $M(M-1)/2$ geodesic
distances $\ell_{ij}$ between these points.
This distribution has a simple form when expressed in terms of the so-called
chord distance on the sphere (see \rIIM\ for details).
This distribution is a meromorphic function of
$D$, and allows for an analytic continuation to $0<D<2$.
The massless propagator $G_{\CS_D}(x,y)$ can also be defined on $\CS_D$ for
non-integer $D$:
it is a function of the geodesic distance $\ell$ between $x$ and $y$ only
(thanks to SO($D+1$) invariance), and has for instance the following integral
representation
\eqn\eProSph{
G_{\CS_D}(\ell)\ =\ L^{2-D}\,{\Gamma\big({D-1\over 2}\big)
\over 4\,\pi^{{D+1\over 2}}}\,
\int_0^\infty {dr\over r}\,\left[|1+r^2-2r\cos(\ell/L)|^{{1-D\over 2}}
\,-|1-r|^{1-D}\right]
\ ,}
where $L$ is the radius of the sphere, which allows for an explicit
analytic continuation to $0<D<2$.
Eq. \eProSph\ can be derived simply by starting from the massless propagator in
infinite flat space in $D+1$ dimensions.
Details are given in appendix B.

A more general class of examples is provided by taking for the manifold $\CM$
the product of a ``physical" 2-dimensional manifold $\CM_2$ and of
a $(D-2)$-dimensional sphere $\CS_{D-2}$.
\eqn\eManPro{
\CM\ =\ \CM_2\otimes\CS_{D-2}
}
$\CM_2$ has a fixed metric $g^{(2)}_{\alpha\beta}$, and $\CS_{D-2}$ 
has the constant curvature metric of the sphere with radius $L$.
When $L\to\infty$ we get $\CM=\CM_2\otimes\RR^{D-2}$.
Points $\bar x$ on $\CM$ are labeled by their coordinates $x$ in $\CM_2$ and
$x'$ on $\CS_{d-2}$, $\bar x=(x,x')$.
The integration measure for $P$ points on $\CM$ is the product of the ordinary
measure $\sprod d^2x\sqrt{g^{(2)}(x)}$ on $\CM_2$ and of the distance measure
$d\mu_P(\ell'_{ij})$ on $\CS_{D-2}$.
The massless propagator $G_\CM(\bar x,\bar y)$ on $\CM$ is easily obtained from
the heat-kernel on $\CM_2$, $K_{\CM_2}(x,y;s)$ and that on $\CS_{D-2}$,
$K_{\CS_{D-2}}(\ell';s)$, where $\ell'$ is the geodesic distance between $x'$
and $y'$, by the relation
\eqn\eProdHK{
G_\CM(\bar x,\bar y)\ =\ \int_0^\infty ds \,K_{\CM_2}(x,y;s)
\,K_{\CS_{D-2}}(\ell';s)
\ .}
Similar tricks can be used to define the model on open manifolds, for instance
the $D$-dimensional ball $\CB_D$.
 
%\vfill\break
\subsec {MOPE and UV Singularities for finite Membranes}
Within this framework, the analysis of the short distance singularities
of the perturbative expansion of the model can be performed along similar
lines as for the infinite flat manifold.
In that case, the MOPE \eTheMOPE\ was derived from the OPE for the
product of exponentials of the $\rvec$ field inside the atoms $\CP$, given by
\eNorProd, \eAvVal\ and \eTayExp.
A similar OPE holds in curved space
\rBDW : when the points $x_i$ inside an atom
$\CP$ tend toward the point $x_\CP$, one has
\eqn\eOPEcsp{
\sprod_{i\in\CP}\ee^{\ii\kvec_i\cdot\rvec(x_i)}
 \ =\ \sum_{A}C^A\{y_i,\kvec_i\}\cdot \CN\!\left[
A(x_\CP)\ee^{\ii\kvec_\CP\cdot \rvec(x_\CP)}\right]
\ \ee^{-{1\over 2}\ssum\limits_{i,j\in\CP}\!\kvec_i\cdot\kvec_j
G_0(y_i-y_j)}
\ .
}
Here $y_i=x_i-x_\CP$ and
$\kvec_\CP=\ssum_{i\in\CP}\kvec_i$ is the total momentum of the atom $\CP$.
$G_0(y_i,y_j)$ is the free massless propagator in the infinite flat space
with the metric $g_{\alpha\beta}(x_\CP)$ at the point $x_\CP$:
\eqn\eGZcu{
G_0(y_i,y_j)\ =\ -{\|y_i-y_j\|_{x_\CP}^{2-D}\over(2-D)S_D}\qquad;\qquad
\|y_i-y_j\|_{x_\CP}^2\ =\ (y_i^\alpha-y_j^\alpha)(y_i^\beta-y_j^\beta)g_{\alpha\beta}
(x_\CP)
}
There are two crucial differences between \eOPEcsp\ and the OPE in flat
infinite space .
\item{(1)} The sum over the operators $A(x_\CP)$ is now performed on
operators which are not only monomials of the derivatives
of $\rvec$ ($\nabla\rvec(x_\CP)$, $\nabla^2\rvec(x_\CP)$, $\ldots$),
but depend also on the metric tensor $g$ and of its derivatives
%also of derivatives of the metric tensor $g$
($g(x_\CP)$, $\nabla g(x_\CP)$, $\nabla^2 g(x_\CP)$, $\ldots$) at the
point $x_\CP$ ($\nabla=\partial/\partial x$ is the ordinary derivative).
\par
\item{(2)} The standard normal product in flat space $\npr[\ ]\npr$ (defined by
factorizing out the expectation value of the single operator $[\ ]$, i.e. by
subtracting all tadpole diagrams in the perturbative expansion of the
expectation values of observables involving the operator $[\ ]$) is replaced
by the general normal ordered prescription in curved space, $\CN [\ ]$, which
is defined in Appendix C.
The basic idea is to make the operators U.V. finite (in curved space) by
subtractions which do not depend explicitly on the background curvature.
This prescription
coincides in flat infinite space with the usual $\npr[\ ]\npr$ prescription.

\par

The coefficients $C^A$ of the OPE are still monomials in the $\kvec_i$
and are homogeneous functions of the $y_i$'s.
They depend only on the local metric $g$ at the origin $x_\CP$, since
they are polynomials of the $\|y_i-y_j\|^{2-D}_{x_\CP}$, and in particular
they do not depend on the derivatives
of the metric at the point $x_\CP$.
A direct consequence is that for the operators $A$ which do not involve
derivatives of the metric, that is for the operators which are already present
in the OPE in flat Euclidean space, the coefficients $C^A\{y_i,\kvec_i\}$ are
the same as those in \eTayExp .

The outline of the derivation of \eOPEcsp\ is given in Appendix C.
It relies on the short distance behavior of the propagator $G(x,y)$ in
general curved space, which itself can be written as an OPE involving
local operators $A$ which depend on the metric $g$ and its derivatives.

Following the derivation of the MOPE \eTheMOPE\ given in Section~3.2,
we can now derive a general MOPE in curved space.
The multilocal operators are of the form (similar to \ePhiExpl )
\eqn\eMOPcsp{
\Phi_{\{A_\CP,\mvec_\CP\}}\{x_\CP\}\ =\ \int d^d\rvec\sprod_{\CP\in\CM}
\left\{\CN\!\left[A_\CP(x_\CP)\,(\ii\nabla_\rvec)^{\mvec_\CP}\,
\delta^d(\rvec-\rvec(x_\CP))\right]\right\}\ ,
}
where the $A_\CP$ are the general local operators depending not only on
derivatives of $\rvec$ but also on the metric and its derivatives.
Let us consider a product of $M$ such multilocal operators $\Phi_i\{x\}$,
let us partition the set of points $\{x\}$ into a set of atoms $\{\CP\}$
and let the points $x$ which belong to  each atom $\CP$ collapse toward
$x_\CP$.
The short distance behavior of the product is then given by a MOPE
\eqn\eMOPEcsp{
\sprod_{i=1}^M\,\Phi_i\{x\}\ =\ \sum_\Phi\,C_{\Phi_1,\cdots,\Phi_M}^\Phi
\{y\}\cdot\Phi\{x_\CP\}
}
where as usual the $y$'s are the relative distances inside each atom $\CP$
($y=x-x_\CP$).
The coefficients $C_{\Phi_1,\cdots,\Phi_M}^\Phi \{y\}$ depend on the metric
$g$ only through the $\|y_i-y_j\|_{x_\CP}$'s inside each atom $\CP$, and
therefore on the metric $g(x_\CP)$ but not on the derivatives of the metric
at the $x_\CP$'s.
These coefficients are homogeneous functions of the $y$'s, whose degree
can be extracted by power-counting analysis from the dimensions of the
operators $A$ and of the $\mvec$'s in the multilocal operators $\Phi$.
As for the OPE, is is important to notice that the coefficients of the
MOPE  $C_{\Phi_1,\cdots,\Phi_M}^\Phi \{y\}$ for operators $\Phi$'s which
do not involve derivatives of the metric are the same as those of the
MOPE in flat Euclidean space.

As a consequence, the short-distance UV divergences will be proportional to
the insertion of relevant multilocal operators of the form
\eMOPcsp , i.e. operators such that $\int\!\cdots\!\int \Phi$ has a
scaling dimension $\ge 0$ at the critical dimension (corresponding to
$\epsilon=0$).
Moreover, since the dimensional regularization prescription that we use
is known to preserve the covariance under changes
of coordinates $x\to x'(x)$ on the manifold $\CM$, the U.V. divergences
(i.e. poles at $\epsilon=0$) will be proportional to multilocal operators which
transform covariantly under such changes of coordinates.

%\vfill\break
\subsec {Renormalization for finite Membranes}
Following the arguments of section 4,
the theory on a curved space will be renormalized and made
U.V. finite by adding the adequate counterterms to the Hamiltonian
\eEdCurv , which are proportional to all relevant and marginal
multilocal operators.
Thus, the situation for the renormalization of the model of self-avoiding
membranes with a fixed internal curved metric is similar to that of
local field theories in curved space: new counterterms have to be added,
corresponding to local, metric dependent, operators.

For a closed manifold $\CM$ and for dimension $D<2$, the only marginally
relevant operators are those which are in the Hamiltonian \eEdCurv .
The renormalized Hamiltonian which makes the theory finite for $\epsilon=0$ is
therefore
\eqn\eRHamC{\eqalign{
{\raise.2ex\hbox{$\CH_{\bf R}$}/\raise -.2ex\hbox{${\rm k}_{\rm B}T$}}
\ =\ &{Z(\bR)\over 2}\,\int_\CM d^Dx\,\sqrt{g(x)}\,
D^\alpha\rvecR (x)\,D_\alpha\rvecR(x) \cr
&+\ 
{1\over 2}\,{\bR\mu^\epsilon Z_b(\bR)}
\int_\CM d^Dx\,\sqrt{g(x)}\int_\CM d^Dx'\,\sqrt{g(x')}\ 
\delta^{d}\big(\rvecR (x)-\rvecR (x')\big)
\ ,}}
where $Z(\bR)$ and $Z_b(\bR)$ are the same renormalization factors as those 
calculated for the infinite flat space model (see \eRenHam, \eZCT).
Operators which depend on the derivatives of the metric, that is in a
covariant form on the curvature tensor and its covariant derivatives, are
irrelevant at $\epsilon=0$ for $D<2$.
The less irrelevant one is just the integral of the scalar curvature $R(x)$,
\eqn\eIntR{
\int_\CM d^Dx\,\sqrt{g(x)}\,R(x)
\ ,}
which has scaling dimension $2-D$.
Other operators, such as $\int \sqrt{g}\,R^{\alpha\beta}D_\alpha\rvec D_\beta
\rvec$, $\int\sqrt{g}\,R\int\sqrt{g}\,\delta^d(\rvec-\rvec)$, etc$\ldots$ have
larger scaling dimensions.

\medskip
For open membranes with a boundary, additional boundary operators have to
be taken into account.
Indeed, as for local field theories, the short distance behaviour of
products of multilocal operators is modified when some subset of points
collapse toward a point on the boundary $\partial\CM$ of the manifold $\CM$.
The MOPE \eMOPEcsp\ now contains multilocal operators of the form
\eMOPcsp\ with special boundary local operators $B(z)$, $z\in\partial\CM$,
in addition to the bulk operators $A(x)$.
Additional short distance divergences may occur on the boundary and have to be
subtracted by additional counterterms proportional to the relevant multilocal
operators containing boundary local operators.

For $D<2$, the most relevant boundary operator is simply the boundary integral
\eqn\eBOne{
L\ =\ \int_{\partial\CM}\!\!d^{D-1}y\,\sqrt{h(y)}
\ ,}
($h(y)$ being the metric on the boundary induced by the metric $g$ in $\CM$).
At first order, it is generated when the two end-points of a dipole collapse
toward the boundary.
Its scaling dimension is equal to $1-D$, and since $L$ is a purely
geometrical term, its expectation value and its scaling dimension
are not modified by the fluctuations of $\rvec$.
Thus it is irrelevant for $D>1$, marginally relevant for $D=1$, and strongly
relevant for $D>1$, like the volume integral $\int_\CM\!d^Dx\,\sqrt{g(x)}$
for $D>0$.
A higher dimensional operator is 
$\int_{\partial\CM}\sqrt{h(y)}\,k(y)$, with $k(y)$ the extrinsic curvature of
the boundary in $\CM$, and has dimension $2-D$, it is irrelevant for $D<2$.
Other irrelevant operators are for instance
$\int_{\partial\CM}\sqrt{h(y)}\,\int_\CM\sqrt{g(x)}\,
\delta^d(\rvec(y)-\rvec(x))$, 
$\int_{\partial\CM}\sqrt{h(y)}\,\int_{\partial\CM}\sqrt{h(y')}\,
\delta^d(\rvec(y)-\rvec(y'))$, etc$\ldots$

This implies that for $D=1$, i.e. for the case of polymers, an additional
counterterm of the form
\eqn\eBCT{
%Z_{\rm boun.}(\bR)\int_{\partial\CM}\!\!d^{D-1}\!y \,\sqrt{h(y)}
Z_{\partial}(\bR)\int_{\partial\CM}\!\!d^{D-1}\!y \,\sqrt{h(y)}
}
has to be introduced to make the renormalized theory finite.
The renormalization factor
%$Z_{\rm boun.}$
$Z_\partial$
depends only on $\bR$, and can be
viewed as a ``boundary coupling-constant" which has to be added to the
Hamiltonian to cancel the (divergent) boundary term generated by the
fluctuations of the membrane.
This new term does not modify the already existing renormalizations of the
bulk operators, $Z(\bR)$ and $Z_b(\bR)$.
This renormalization factor is of course not new, and can be recovered
from the standard treatment of the Edwards model.
If $D\ne 1$, the boundary operator is not superficially relevant, and no
renormalization is required if we use dimensional regularization.
The renormalized Hamiltonian \eRHamC\ is sufficient to make the theory
UV finite for $\epsilon=0$, both for closed and for open membranes.

\medskip

%\vfill\break
\subsec {Finite Size Scaling and Hyperscaling}
The fact that the renormalization factors are the same for the theory in
infinite $D$-dimensional space and in a finite space is very similar
to what occurs for local field theories.
This property of local field theories allows to derive the so-called
finite size scaling laws for critical systems in  finite geometries.
In the same way this property of self-avoiding membranes allows to 
derive scaling laws for the behaviour of large finite (closed or open)
membranes.

As a first application, let us study the scaling behavior of the partition
function of a closed membrane $\CM$.
The partition function is defined in the standard way as the sum over all
configurations $\rvec$, with only one difference with respect to \ePartFunc :
in order to remove the translational zero mode, which gives an infinite
factor equal to the volume of the $d$-dimensional space in which the membrane
fluctuates, one must fix the position of the center-of-mass $\rveccm$
\eqn\eCenMas{
\rveccm\ =\ {1\over{\rm Vol}(\CM)}\,\int_\CM d^Dx\,\sqrt{g(x)}\,\rvec(x)
\qquad;\qquad{\rm Vol}(\CM)\ =\ \int_\CM d^Dx\,\sqrt{g(x)}
}
The partition function is then
\eqn\ePFFin{
\CZ(b)\ =\ \int \CD[\rvec(x)]\ \delta^d(\rveccm)\,\ee^{-\CH[\rvec]/\kBT}
}
with the Hamiltonian $\CH[\rvec]/\kBT$ given by \eEdCurv .

If we now scale the internal size of the manifold $\CM$ by a factor $X$,
for instance by rescaling the metric
$g_{\alpha\beta}(x)\to X^2 g_{\alpha\beta}(x)$,
we expect that in the large $X$ limit, the partition function $\CZ(b,X)$
of the new manifold scales as
\eqn\eScPF{
\CZ(b;X)\ =\ X^{\gamma-1}\,({\rm cst.})^{{\rm Vol}(\CM)\cdot X^D}
}
where ${\rm cst.}$ is a non universal constant (depending for instance on
the way the model is regularized), but independent of the shape and
the size of the manifold,  and $\gamma$ a universal (independent of the
regularization) configuration exponent.
As we show now, if the dimension $D$ of the membrane is not an integer
($D\ne1,2$) this configuration exponent is simply related to the $\nu^\star$ 
exponent by the hyperscaling relation \rBDone
\eqn\eHySc{
\gamma=\ 1\,-\,\nu^\star d
\ .}
This relation turns out to be still valid when $D=1$ for closed
``1-dimensional membranes" with no boundary (self-avoiding loop), but is
violated for open ``1-dimensional membranes" (free open self-avoiding walks).
It is also not valid for the physical case of 2-dimensional membranes.

Let us first recall standard results for the partition function of the free
membrane ($b=0$) (see for instance \rBDtwo ).
In this case, we simply deal with the free field, and the partition function
is given by
\eqn\eFreePT{
\CZ(0)\ 
=\ \int \CD[\rvec(x)]\ \delta^d(\rveccm)\,\ee^{-\CH_{\rm el}[\rvec]/\kBT}\  
=\ \left[{{\rm det}'(-\Delta)\over {\rm Vol}(\CM)}\right]^{-d/2}
}
The ${\rm det}'(-\Delta)$ is the product over the non-zero eigenvalues $\lambda_n$
of the Laplacian on $\CM$, and is defined by analytic continuation in terms 
of the $\zeta$-function $\zeta(s)$ of the Laplacian as 
\eqn\eZeta{
{\rm det}'(\Delta)\ =\ \ee^{-\zeta'(0)}\qquad,\qquad
\zeta(s)\ =\ \sum_{n>0}\lambda_n^{-s}
}
If one rescales the manifold by a factor $X$ (i.e. the metric by a factor
$X^2$), the volume is changed by a factor $X^D$, and
the $\zeta$ function is changed by $\zeta(s)\to\zeta(s)X^{2s}$, and
therefore
\eqn\eDetSc{
{\rm det}'(\Delta)\ \to\ {\rm det}'(\Delta)\,X^{-2\zeta(0)}
}
$\zeta(0)$ can be calculated explicitly, and is found to be:
\eqn\eZeZero{\eqalign{
\zeta(0)\ =\ &-1\hphantom{\,+\,{\chi\over 2}}\qquad
\vtop{\hbox{if $D$ not an integer}}\cr
&-1\,+\,{\chi\over 2}\qquad
\vtop{\hbox{if $D=1$, with $\chi=0$ for closed loop,
$\chi=1$ for open chain}}\cr
&-1\,+{\chi\over 6}\qquad
%\vtop{\hbox{if $D=2$ and $\CM$ is closed or open with a smooth boundary}
%\hbox{$\chi=$~Euler Characteristics of $\CM$}}
\vtop{\hbox{if $D=2$ with $\chi$ the Euler characteristics of $\CM$}}
}}
The result for $D=2$ is valid for smooth open or closed manifolds $\CM$.
If $\CM$ has conical singularities or if its boundary $\partial\CM$ has corners
the result is modified%
%\foot{or instance, for a smooth manifold with}
, and can be computed using conformal invariance.
Thus for the free membrane we get the configuration exponent
\eqn\eGamZe{\eqalign{
\gamma_0\,-\,1 \ =\ &-\,\nu d\hphantom{\,+\,{\chi\over 2}d}\qquad
\hbox{if $D$ not integer}\cr
&-\,\nu d\,+\,{\chi\over 2}d\qquad\hbox{if $D=1$}\cr
&-\,\nu d\,+\,{\chi\over 6}d\qquad\hbox{if $D=2$}
}}
with $\nu=(2-D)/2$ the scaling dimension of the free field $\rvec$.
This result has a simple interpretation:
the l.h.s of \eGamZe\ is the dimension of the partition function
$\CZ(0)$, $-\nu d$ is the scaling dimension of the constraint
$\delta^d(\rveccm)$ which removes the translational zero mode;
the last term, present for integer $D$, is the contribution of the measure
$\CD[\rvec]$ given by the trace anomaly.

\medskip
We now consider the interacting theory ($b\ne 0$).
It is sufficient to determine the scaling behavior of the
normalized partition function
\eqn\eNorPF{
\bar\CZ (b)\ =\ {\CZ(b)\over \CZ(0)}
%\ =\ {\int \CD[\rvec(x)]\ \delta^d(\rveccm)\,\ee^{-\CH[\rvec]/\kBT}
%\over\int \CD[\rvec(x)]\ \delta^d(\rveccm)\,\ee^{-\CH_{\rm el}[\rvec]/\kBT}}
}
which can be computed in perturbation theory, using the perturbative
rules defined above, with the massless propagator $G_\CM(x,y)$ \eProCurv .
The constraint $\rveccm=0$ fixes the arbitrary additive constant in the
definition of the propagator so that $\int_\CM d^Dx\sqrt{g(x)}G_\CM(x,y)=0$.

We first consider the case $D\ne 1,2$.
Since the renormalized Hamiltonian
$\CHR[\rvecR]$ \eRHamC\ makes the theory UV finite,
the renormalized partition function, defined as
\eqn\eRenPF{
\bar\ZR(\bR;\mu)\ =\ {1\over \CZ(0)}\,
\int \CD[\rvecR(x)]\ \delta^d(\rvecRcm)\,\ee^{-\CHR[\rvecR]/\kBT}
}
is UV finite
at $\epsilon=0$ as a series in the renormalized coupling constant $\bR$.
$\mu$ is the renormalization momentum scale.
Performing the change of variables \eRenBare\ from renormalized to bare
quantities, $\bar\ZR(\bR;\mu)$ is simply re expressed in term of the bare
partition function $\bar\CZ(b)$
(treating the measure $\CD[\rvec]$ as dimensionless, which is justified in
perturbation theory)
\eqn\eZRtoZ{
\bar\CZ(b)\ =\ Z(\bR)^{-d/2}\,\bar\ZR(\bR;\mu)
}
where $Z(\bR)$ is the wave-function renormalization factor, and
$b(\bR,\mu)$ the bare coupling constant.
If we now rescale the membrane by a factor $X$, simple dimensional analysis
shows that
\eqn\eZResc{
\bar\CZ(b;X)\ =\ Z(\bR(X))^{-d/2}\,\bar\ZR(\bR(X);X\mu)
}
with $\bR(X)$ given by $b=b(\bR(X),X\mu)$.
From the renormalization group flows \eBeta\ and \eDimAn , we know that
for $b$ and $\mu$ fixed, $\bR(X)$ and $Z(\bR(X))$ obey 
\eqn\eXflows{
X{\partial\over\partial X}\bR\ =\ 
-\left.\mu{\partial\over\partial\mu}\right|_b\bR\ =
\ -\,\beta(\bR)\quad;\quad
X{\partial\over\partial X}\ln Z\ =\ 
-\left.\mu{\partial\over\partial\mu}\right|_b\ln Z\ =\ 2\,(\nu(\bR)-\nu)
}
For $\epsilon>0$, $\bR(X)\to\bR^\star$ as $X\to\infty$ and
$Z\sim X^{2(\nu^\star-\nu)}$.
Combining all these results we see that the partition function scales as
\eqn\eZscaling{
Z(b,X)\ \propto\ X^{-\nu^\star d}
}
hence \eHySc .

For $D=1$, the same argument applies to the closed chain.
For the open chain, we must take into account the additional additive
counterterm \eBCT .
The result is
\eqn\eHSbout{
\gamma\,-\,1\ =\ -\,\nu^\star\,d\,+\,2\,\eta^\star
\ ,
}
with $\eta^\star$ the anomalous end-point dimension
\eqn\eAnEP{
\eta^\star\ =\ \eta(\bR^\star)\qquad;\qquad\eta(\bR)\ =\ 
\left.\mu{\partial\over\partial\mu}\right|_b %Z_{\rm boun.}
Z_\partial
\ .
}
\medskip
For $D=2$, unfortunately, the perturbative scheme developed here is useless.
Indeed, the line $D=2$ corresponds to $\epsilon=4$ and is always at a finite
distance from the line $\epsilon=0$.
It is natural to expect that the curvature operator \eIntR\ will become
relevant, since its scaling dimension is always $2-D$, but its mixing
with the other operators cannot be studied perturbatively by an
$\epsilon$-expansion.

%\vfill\break
\subsec {Relation with Direct Renormalization}
%\medskip\centerline{\sl A voir avec Bertrand ?!}\medskip

As a second application of this renormalization formalism, let us derive
simply the validity of the direct renormalization method used by 
previous authors.

\medskip
We first recall the principle of direct renormalization, along the lines of
e.g. \rBDtwo .
It relies on the following ideas:

\noindent (1)
One considers only {\it finite} membranes, with internal extent $L<\infty$,
at $\epsilon>0$.
In this case, as we have seen before, physical observables, expressed as
perturbative series in the ``bare'' coupling constant $b$, are UV {\it and}
I.R. finite.
Then one introduces a {\it physical coupling constant} $g=g(b,L)$, which is
dimensionless and depends on $b$ and on the size of the membrane $L$.
An example of such a coupling constant is provided by the second virial
coefficient, defined as follows.

Let us extend the SAM model to the case of
two interacting copies $\CM$ and $\CM'$ of the same membrane.
The Hamiltonian is now
\eqn\eHamTwo{\eqalign{
\CH[\rvec,\rvec']/\kBT\ =\ &{1\over 2}\int_\CM (\nabla\rvec)^2\,+\,
{1\over 2}\int_{\CM'} (\nabla\rvec')^2\cr
&\,+\,{b\over 2}\,\left[
\int_\CM\int_\CM\delta^d(\rvec-\rvec)+
\int_{\CM'}\int_{\CM'}\delta^d(\rvec'-\rvec')+
2\int_\CM\int_{\CM'}\delta^d(\rvec-\rvec')
\right]
\ .}}
$\rvec$ and $\rvec'$ describe the embedding of $\CM$ and $\CM'$ respectively.
The partition function $\CZ(b)$ for a single membrane is defined by \ePFFin,
\eqn\eParOne{
\CZ(b)\ =\ \int \CD[\rvec(x)]\ \delta^d(\rveccm)\,\ee^{-\CH[\rvec]/\kBT}
}
while the connected partition function for two membranes $\CZ_{2,c}$ is
\eqn\eParTwo{
\CZ_{2,c}(b)\ =\ \int\CD[\rvec]\CD[\rvec']\,
\delta^d(\rvec_{\rm cm})
\,\left[\ee^{-\CH[\rvec,\rvec']/\kBT}-\ee^{-\CH[\rvec]/\kBT}
\ee^{-\CH[\rvec']/\kBT}\right]
\ ,}
and the radius of gyration ${\bf R}$ of a single membrane $\CM$ is
\eqn\eMeanR{
{\bf R}^2\ =\ {1\over {\rm Vol}(\CM)^2}\,\int_\CM\int_\CM
\langle (\rvec-\rvec)^2 \rangle\strut_b
\ .}
The second virial coefficient is then defined by
\eqn\eVirCoef{
g\ =\ -\,{\CZ_{2,c}\over \CZ^2}\,{\bf R}^{-d}
\ .}
It is a dimensionless quantity, and has a perturbative expansion in
$b$ of the form
\eqn\egexp{
g(b,L)\ =\ c_1\,bL^\epsilon\,+\,c_2\,(bL^\epsilon)^2\,+\,\cdots
}
where $c_1$, $c_2$, $\ldots$ are dimensionless coefficients which do not depend 
on the size $L$ of the membrane, but depend on its shape.

\noindent (2)
It then appears that the scaling functions, i.e. the dimensionless observables
$F(b,L)$ of the model, which can be expressed as functions of the physical
coupling constant $g$ only
\eqn\eFbtog{
F(b,L)\ =\ \tilde F(g)
\ ,}
have a perturbative expansion in $g$ which is UV {\it finite} when
$\epsilon\to 0$, contrary to their perturbative expansion in $b$.
An example of a scaling function is the effective scaling exponent
\eqn\eEffNu{
\nu(b,L)\ =\ \tilde\nu(g)\ =\ L{\partial\over\partial L}\,\ln ({\bf R})
\ ,}
where ${\bf R}$ is the gyration radius \eMeanR .

\noindent (3)
In particular, the Wilson function $W(g)$, defined as
\eqn\eWilg{
W(g)\ =\ \left.L{\partial\over\partial L}g\right|_b
\ ,}
is a regular function of $\epsilon$ when $\epsilon\to 0$ order by order in
$g$.
Its expansion reads
\eqn\eWilExp{
W(g)\ =\ \epsilon\,g\,-\,a_2(\epsilon)\,g^2\,+\,O(g^3)\qquad,\qquad
0< a_2(\epsilon)\,=\,O(1)\ {\rm as}\ \epsilon\to 0
\ .}
It thus has a zero $0<g^\star=O(\epsilon)$ for $\epsilon>0$, and from the
definition of $W(g)$, one sees that in the limit of a large membrane,
the physical coupling constant $g$ tends towards a {\it finite} value:
\eqn\eglim{
g(b,L)\ \to\ g^\star\qquad;\qquad L\to\infty\ ,\ b\ {\rm fixed}
\ .}
Since from point (2) $W(g)$ has a regular $g$-expansion when $\epsilon\to 0$,
inverting $W(g)=0$ gives $g^\star$ as a finite series expansion in $\epsilon$.
Similarly, the scaling functions also have a finite large $L$ limit
\eqn\eFlim{
F(b,L)\ \to\ F^\star\,=\,\tilde F(g^\star)\qquad;\qquad L\to\infty
\ ,}
which correspond to scaling exponents for large self-avoiding membranes.
In particular,
\eqn\enuLim{
\nu^\star\ =\ \tilde\nu(g^\star)\ =\ \lim_{L\to\infty}\nu(b,L)
}
corresponds to the scaling exponent $\nu^\star$ obtained in Sec.~5.1.
Since $\tilde F(g)$ has a regular expansion in $g$ when $\epsilon\to 0$,
and since
$g^\star$ has a regular expansion in $\epsilon$, all the scaling exponents
$F^\star$ have also a finite expansion in $\epsilon$.

\medskip
We now show why direct renormalization is valid.
The key property that validates the direct renormalization method is
(2), namely the fact that the scaling functions have a regular expansion in the
physical coupling constant $g$ when $\epsilon\to 0$.
This is a simple consequence of our renormalization formalism applied to
finite membranes.
Indeed, one can show that {\it the same counterterms} as those
which make the theory of a single membrane UV finite when $\epsilon\to 0$
also make the model \eHamTwo\ of two interacting membranes UV finite.
In other words, re-expressing the bare quantities $\rvec$ and $b$ in terms of
renormalized ones $\bR$ and $\rvecR$ according to \eRenBare\ makes the
perturbation theory in $\bR$ finite order by order when $\epsilon\to 0$.
This property is not completely trivial, but before proving it, let us show that
it ensures property (2).
Therefore let us assume that if we perform the change of variable \eRenBare\ 
\eqn\ebBtoR{
b\ =\ \bR\,Z_b(\bR)\,(Z(\bR))^{d/2}\,\mu^\epsilon\qquad\longleftrightarrow\qquad
\bR\ =\ \bR(b\mu^{-\epsilon})
\ ,}
the dimensionless scaling functions $F$ have a finite perturbative
expansion in $\bR$ (when $\epsilon\to 0$). 
By simple dimensional arguments, they depend on the dimensionless coupling
$\bR$, on the size of the membrane $L$ and on the additional renormalization
scale $\mu$ only through $\bR$ and through the dimensionless product $\mu L$.
They can therefore be written as
\eqn\eFRen{
F(b,L)\ =\ \bar F(\bR,\mu L)
\ ,}
where $\bar F$ has an UV finite perturbative expansion in $\bR$.
Since $\mu$ can be chosen arbitrarily, we can take $\mu=L^{-1}$ and we 
obtain that by the change of variable
\eqn\ebBRL{
\bR(b,L)\ =\ \bR(bL^\epsilon)
}
the scaling functions can be written as an UV finite perturbative expansion in
$\bR(b,L)$
\eqn\eFRL{
F(b,L)\ =\ \bar F(\bR(b,L), 1)\ \equiv\ \bar F(\bR(b,L))
\ .}
This is true in particular for the second virial coefficient \eVirCoef\ 
$g(b,L)$, which is a scaling function, and can be written as
\eqn\egtobR{
g(b,L)\ =\ \bar g(\bR(b,L))\ =\ g_1\,\bR\,+\,O(\bR^2) \quad;\quad 0<g_1=O(1)\ 
{\rm when}\ \epsilon\to 0
\ .}
Inverting this relation we can write $\bR(b,L)$ as a series in $g$, which is
UV finite order by order when $\epsilon\to 0$.
\eqn\ebRtog{
\bR(b,L)\ =\ \tilde\bR(g(b,L))
\ .}
Using \eFRL\ and \ebRtog\ one sees that any scaling function $F$ is UV finite
when expressed in terms of $g$
\eqn\eFtoFti{
F(b,L)\ =\ \bar F(\tilde\bR(g))\ \equiv\ \tilde F(g)
\ .}
This is nothing but property (2).

The Wilson function \eWilg\ is also a scaling function and is therefore
UV finite as a function of $g$. 
Using \ebBRL\ ,\ebBtoR\ and \eBeta\ one sees that
${\scriptstyle L}{\partial\over\partial L}\bR(b,L)=-\beta(\bR)$, and therefore
the Wilson
function $W(g)$ is simply related to the renormalization group $\beta$-function
$\beta(\bR)$ by
\eqn\eWtoB{
W(g)\ =\ -\,\left. {\partial \bar g\over \partial \bR}\,\beta(\bR)\right|_
{\bR=\tilde\bR(g)}
\ .}
Thus the renormalization group $g$-flow driven by $W$ is just
the mirror of the $\bR$-flow
driven by $\beta$, and the effective coupling constant
$g^\star=\bar g(\bR^\star)$, where $\bR^\star$ is the I.R. fixed point.
Similarly, one can show that the scaling exponents $\nu^\star$ given by
\enuLim\ and by \eScalBeh\ coincide.

\medskip
It thus remains to show that the two membranes model \eHamTwo\ is renormalizable
by the same counterterms \eRenBare\ as the one membrane model.
This model is in fact a particular case of the general model describing
two -- a priori different -- membranes $\CM_1$ and $\CM_2$,
whose respective configurations are described by the fields $\rvec_1$ and
$\rvec_2$, interacting by the Hamiltonian
\eqn\eHamtwo{\eqalign{
{\raise.2ex\hbox{$\CH[\rvec_1,\rvec_2]$}/\raise -.2ex\hbox{${\rm k}_{\rm B}T$}}
%\ =\ &{1\over 2}\,\int_{\CM_1}\!\! d^Dx_1\,\sqrt{g(x_1)}\,
\ =\ &{1\over 2}\,\int_{\CM_1}\!\! d^Dx_1\,(\nabla\rvec_1)^2\,+\,
{1\over 2}\,\int_{\CM_2}\!\! d^Dx_2\,(\nabla\rvec_2)^2\cr
+\,&{b_{11}\over 2}\int_{\CM_1}\!\!d^Dx\int_{\CM_1}\!\!d^Dy\,
\delta^d(\rvec_1(x)-\rvec_1(y))\cr
+\,&{b_{22}\over 2}\int_{\CM_2}\!\!d^Dx\int_{\CM_2}\!\!d^Dy\,
\delta^d(\rvec_2(x)-\rvec_2(y))\cr
+\,&{b_{12}}\int_{\CM_1}\!\!d^Dx\int_{\CM_2}\!\!d^Dy\,
\delta^d(\rvec_1(x)-\rvec_2(y))\cr
}}
Indeed, this Hamiltonian is the most general allowed by power counting
which describes two interacting membranes.
Physical observables have a series expansion in the coupling constants
$b_{11}$, $b_{22}$ and $b_{12}$, which involves the free propagators
$G_{\CM_1}$ and $G_{\CM_2}$ only, since the free correlator between the
two membranes vanishes
\eqn\eProVan{
\langle\rbf_1(x)\rbf_1(y)\rangle\strut_0\ =\ G_{\CM_1}(x,y)\qquad
\langle\rbf_1(x)\rbf_2(y)\rangle\strut_0\ =\ G_{\CM_2}(x,y)\qquad
\langle\rbf_1(x)\rbf_2(y)\rangle\strut_0\ =\ 0
}
This perturbative expansion involves the same diagrams as for the one membrane
model, but the end-points of the dipoles may belong separately to $\CM_1$
or $\CM_2$.
It is a not very difficult task to repeat the analysis of the short
distance singularities of Sec.~3.
One finds that the UV singular configurations are given by the same
molecules as depicted in \fmolecule , but the points within each atoms,
i.e. the subset of collapsing end-points, must belong to the same membrane
(which depends of course of the atom).
One can then repeat the MOPE analysis of Subsec.~3.2.
Indeed, the coefficients of the MOPE are obtained from the short distance
OPE of the vertex operators \eNorProd\ within each atom $\CP$, which is
characteristic of the local properties of each manifold separately.
The final result is that the short distance behavior of products of multi-local
operators $\Phi\{x_1,\cdots,x_P\}$ involving the two membranes is given by the
MOPE, and that while the operators that appear in the MOPE do now depend also on
which particular membrane each point $x_i$ of the operators belongs to, the
coefficients
${\bf C}\{y\}$ of the MOPE {\it do not}, since all the
points in each atom belong to the same membrane.
The power counting discussed in subsec.~3.3 remains also valid, and therefore
one deduces immediately that only the molecules with
one or two atoms are divergent at $\epsilon=0$.
There are now two classes of divergent diagrams with 1 atom, and 3 classes
with 2 atoms, depicted on 
\fig\ftwoMdiag{Divergent diagrams for the \hyperref\htwoMdiag{2 membrane}
model}.
\topinsert
\centerline{\epsfxsize=13.truecm\epsfbox{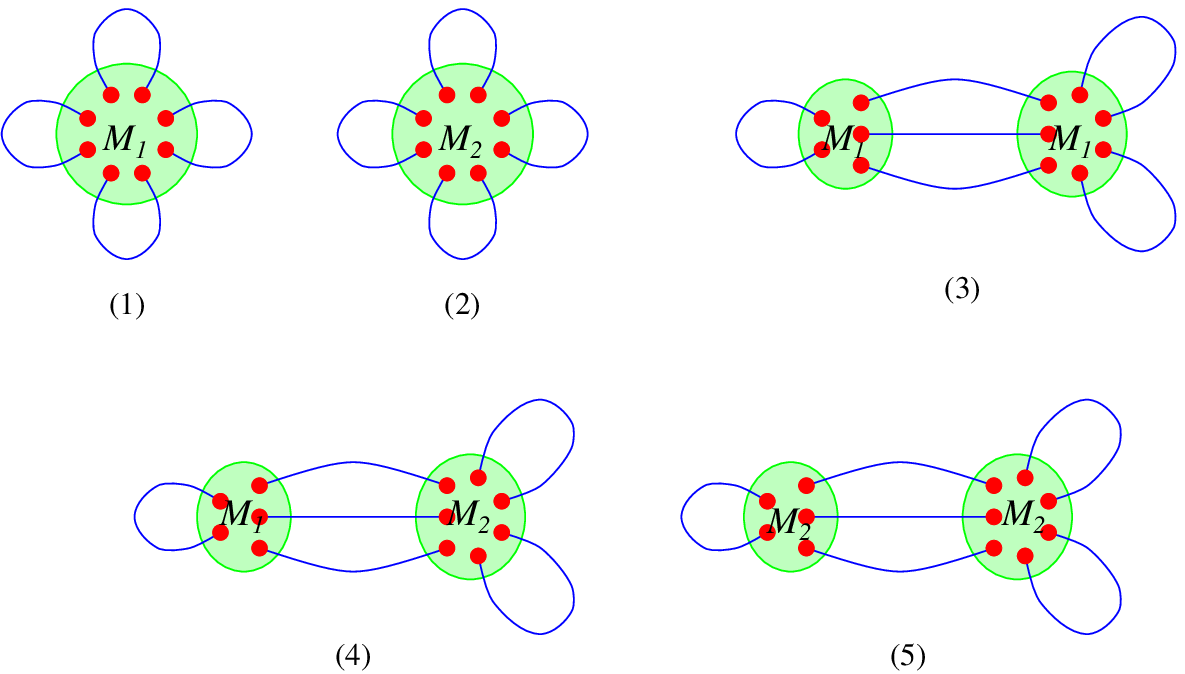}}
\centerline{\hyperdef\htwoMdiag{userfigure}{utwoMdiag}{\ftwoMdiag}}
\endinsert
These are exactly the same diagrams as those considered in sec.~3.
They contribute to the perturbative expansion just by different powers of 
$b_{11}$, $b_{22}$ and $b_{12}$.
As a consequence the renormalization analysis can also be performed along
the lines of sec.~4.
Diagrams (1) and (2) of \ftwoMdiag\ contribute to a multiplicative
renormalization of the first two Gaussian terms of the Hamiltonian
\eHamtwo , i.e. to a wave-function renormalization;
diagrams (3), (4) and (5) contribute to a multiplicative renormalization of
$b_{11}$, $b_{22}$ and $b_{12}$ respectively, i.e. to coupling constants
renormalization.
Moreover, simple inspection of the diagrams shows that in the perturbative
expansion, the divergences (1) and (3), associated only to the membrane $\CM_1$,
depend only on $b_{11}$, not on $b_{22}$ and $b_{12}$.
The same is true for divergences (2), (4) and $b_{22}$.
This implies that the general two membrane model \eHamtwo\ is
made UV finite at $\epsilon=0$ by the renormalized Hamiltonian
\eqn\eRHamtwo{\eqalign{
{\raise.2ex\hbox{$\CH^{\bf R}[{\rvecR}_1,{\rvecR}_2]$}/
\raise -.2ex\hbox{${\rm k}_{\rm B}T$}}
\ =\ &{Z({\bR}_{11})\over 2}\,\int_{\CM_1}\!\! d^Dx_1\,(\nabla{\rvecR}_1)^2\,+\,
{Z({\bR}_{22})\over 2}\,\int_{\CM_2}\!\! d^Dx_2\,(\nabla{\rvecR}_2)^2\cr
+\,&{{\bR}_{11}Z_b({\bR}_{11})\mu^\epsilon\over 2}\int_{\CM_1}\!\!d^Dx\int_{\CM_1}\!\!d^Dy\,
\delta^d({\rvecR}_1(x)-{\rvecR}_1(y))\cr
+\,&{{\bR}_{22}Z_b({\bR}_{22})\mu^\epsilon\over 2}\int_{\CM_2}\!\!d^Dx\int_{\CM_2}\!\!d^Dy\,
\delta^d({\rvecR}_2(x)-{\rvecR}_2(y))\cr
+\,&{{\bR}_{12}{\bar Z}_b({\bR}_{11},{\bR}_{22},\bR_{12})\mu^\epsilon}
\int_{\CM_1}\!\!d^Dx\int_{\CM_2}\!\!d^Dy\, \delta^d({\rvecR}_1(x)-{\rvecR}_2(y))\cr
}}
with the same counterterms $Z(\bR)$ and $Z_b(\bR)$ as those of the one membrane
theory \eRenHam .
Only the renormalization factor $\bar Z(b_{11},b_{22},b_{12})$ for $g_{12}$
is new.

A simple inspection of the diagrams shows that if the three coupling constants
are taken equal ${\bR}_{11}={\bR}_{22}={\bR}_{12}=\bR$, then the three classes
of 2-atom diagrams contribute with exactly the same factors, and therefore
the different $Z_b$ coincide
\eqn\eZequal{Z_b(\bR)\ =\ \bar Z_b(\bR,\bR,\bR)
\ .}
This completes the proof that the two membrane model \eHamTwo\ is renormalized
by the same counterterms than the one membrane model.

Finally, let us remark that a simpler argument (in fact equivalent) is to
consider again the proof of renormalizability for the one membrane model, and to
check that at all the steps, the fact that the membrane is connected or not is never important.

%\subsec {Calculations for the Sphere}
%\centerline{\sl Faut-il le faire?}

%\vfill\break
\subsec {Contact Exponents}

As a third application, let us discuss the contact exponents \rBDone\ for self-avoiding
membranes.
We refer to \rBDtwo\ for a general presentation of this problem.
They are defined as follows, as a generalization of the notion of contact
exponents for polymers \rDesCloJan .
Let us  consider the probability that two given points $x_1$ and $x_2$ on the membrane are separated by the vector $\rvec$ in external space.
This probability corresponds to the observable
\eqn\eProbR{
P(\rvec;x_1,x_2)\ =\ \langle\delta^d[\rvec-(\rvec(x_1)-\rvec(x_2)]\rangle
\ .}
For an infinite membrane it will have the scaling form
\eqn\eProbScal{
P(\rvec;x_1,x_2)\ =\ R^{-d}_{12}\,F\left(|\rvec|/R_{12}\right)\quad;\quad
R^2_{12}\ =\ \langle[\rvec(x_1)-\rvec(x_2)]^2\rangle
\ ,}
where $F$ is a scaling function.
The contact exponents $\theta$ are defined by the small $r$ behavior of
$F$
\eqn\eCoExDef{
F(r)\ \sim\ r^\theta\ ,\ r\to 0
\ .}
For open polymers there are three different contact exponents, depending on the position of the points
\item{$\theta_0$} if $x_1$ and $x_2$ are the two end points.
Then $\theta_0$ is given by the scaling relation $\theta_0=(\gamma-1)/\nu$.
\item{$\theta_1$} if $x_1$ is a end point and $x_2$ is on the chain
\item{$\theta_2$} if $x_1$ and $x_2$ are both on the chain. This exponent is the same for closed polymer chains.

Equivalently, they describe the behavior of $P(\rvec; x_1, x_2)$ for fixed
$\rvec$, $|x_1-x_2|\to\infty$
\eqn\ePLarX{
P(\rvec;x_1,x_2)\ \sim\ R_{12}^{-d-\theta}\ \sim\ |x_1-x_2|^{-\nu(d+\theta)}
\ .}
In this limit, we can take $\rvec=0$ and hence the scaling exponents $\theta$
are related to the scaling dimension $\Delta_\phi$ of the two point operator
\eqn\eScalPhi{
\phi(x_1,x_2)=\delta^d(\rvec(x_1)-\rvec(x_2))\quad,\quad
\langle\phi(x_1,x_2)\rangle\ \sim\ |x_1-x_2|^{\Delta_\phi}
\qquad\,\qquad |x_1-x_2|\to\infty
\, }
by the relation
\eqn\eDelTheta{
\Delta_\phi\ =\ -\nu(d+\theta)
\ .}

The scaling dimension $\Delta_\phi$ can be calculated in perturbation theory.
In our framework, perturbation theory is made UV finite when $\epsilon\to 0$ 
by the renormalizations \eRenBare .
However, in order to make the insertions of the bi-local operator $\phi(x_1,x_2)$ UV finite, an additional multiplicative renormalization is required, defining the renormalized operator $\phi_{\sbf R}$
\eqn\eRenPhi{
\phi_{\sbf R}(x_1,x_2)\ =
\ Z_\phi(\bR)\,\delta^d(\rvecR(x_1)-\rvecR(x_2))
\ =\
 Z_\phi(\bR)\,Z(\bR)^{d/2}\,\delta^d(\rvec(x_1)-\rvec(x_2))
\ .}
$Z$ the wave-function renormalization factor and $\bR$ the renormalized coupling constant.
The scaling dimension $\Delta_\phi(\bR)$ of $\phi_{\sbf R}$ is therefore given (using
standard arguments, similar to those of subsec.~5.1) by
\eqn\eDimPhi{
\Delta_\phi(\bR)\ =\ -\,d\,\nu(\bR)\,+\,\left. \mu{\partial\over\partial\mu}
\ln (Z_\phi)\right|_{b}\ =\ 
-d\,\nu(\bR)\,+\,\beta(\bR){\partial\over\partial\bR}\ln(Z_\phi(\bR))
\ ,}
where $\nu(\bR)$ is the scaling dimension of $\rvecR$, as defined in \eDimAn .
The second term in the r.h.s of \eDimPhi\ comes from the additional renormalization $Z_{\phi}$.
The scaling exponent $\Delta_\phi$ is obtained by taking \eDimPhi\ at the I.R. attractive fixed point $\bR^\star$
\eqn\eDimPhiSt{
\Delta_\phi\ =\ \Delta_\phi(\bR^\star)
\ .}

In order to compute $\Delta_\phi$, we must determine $Z_\phi$, that is the
short-distance singularities that are associated with the insertion of
$\phi(x_1,x_2)$.
These singularities are are obtained from the MOPE, and using power-counting
one can check that the divergent diagrams are those depicted on
\fig\fcontactdiag{\hyperref\hcontactdiag{Divergent diagrams} for the insertion
of $\Phi(x_1,x_2)$} , where interaction dipoles collapse toward the inserted dipole ($x_1,x_2)$.
These diagrams are very similar to the two atoms diagrams of \fdivdiag\ and \ftwoMdiag , but the divergence depends  whether the two points,
only one point,  or none of the points are on the boundary of the membrane. 
We now give a short derivation of the value for these contact exponents at
one loop.
\topinsert
\centerline{\epsfxsize=8.truecm\epsfbox{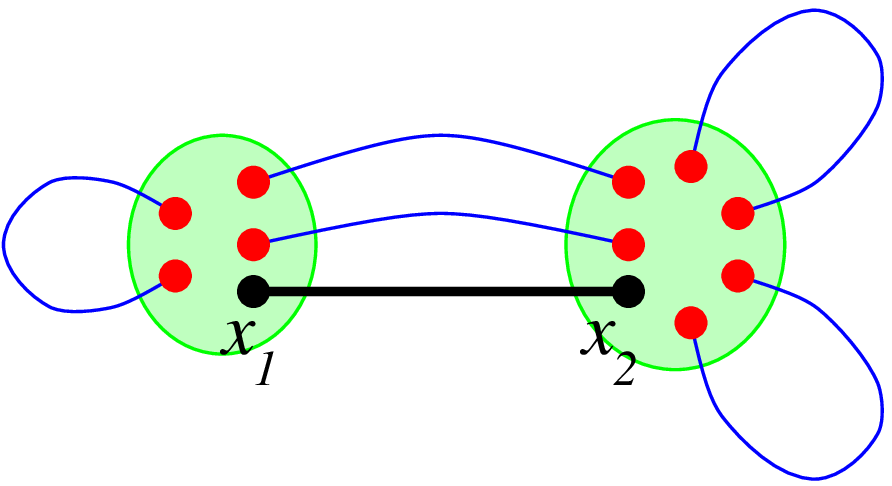}}
\centerline{\hyperdef\hcontactdiag{userfigure}{ucontactdiag}{\fcontactdiag}}
\endinsert

\medskip\noindent
{\bf Bulk contact exponent $\theta_2$:}
\medskip

This is the simplest situation.
At first order on perturbation theory the only divergence associated with
the insertion of $\phi(x_1,x_2)$ is when the two end-points of a dipole
tend respectively toward $x_1$ and $x_2$, as described on 
\fig\foloopcontact{\hyperref\holoopcontact{One loop divergent diagram} for the insertion
of $\Phi(x_1,x_2)$} . 
\topinsert
\centerline{\epsfxsize=5.truecm\epsfbox{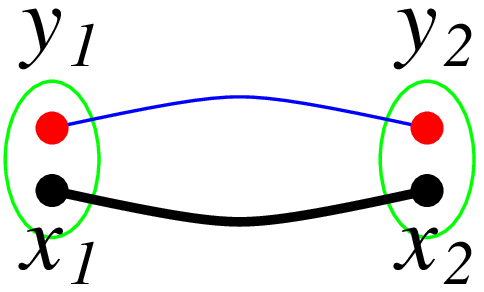}}
\centerline{\hyperdef\holoopcontact{userfigure}{uoloopcontact}{\foloopcontact}}
\endinsert
Therefore the singularity will be proportional to the MOPE coefficient for
the contraction $\phi\phi\to\phi$, calculated in sec.~5.2.
Taking into account the symmetry factors, we find that the counterterm $Z_\phi$ 
is at one loop
\eqn\eZcontone{
Z_\phi\ =\ 1\,+\,{\bR\over\epsilon}
\,(4\pi)^{-{d\over 2}}\, (2-D)^{-1+{d\over 2}}\,
\left({2\,\pi^{D\over 2}\over\Gamma(D/2)}\right)^{2+{d\over 2}}\,
{\Gamma\left({D\over 2-D}\right)^2\over\Gamma\left({2D\over 2-D}\right)}
\ .}
Using \eDelTheta\ and \eDimPhi\ we get that at one loop
\eqn\ethetwoone{
\Delta_\phi(\bR)\ =\ -d\,\nu(\bR)\,-\,
\bR\,(4\pi)^{-{d\over 2}}\, (2-D)^{-1+{d\over 2}}\,
\left({2\,\pi^{D\over 2}\over\Gamma(D/2)}\right)^{2+{d\over 2}}\,
{\Gamma\left({D\over 2-D}\right)^2\over\Gamma\left({2D\over 2-D}\right)}
\,}
and hence
\eqn\ethetwo{
%\theta_2\ =\ \epsilon\ {4\over (2-D)}\,\left[1+{d\over 2}{(2-D)^2\over 2D}
%{\Gamma\left({2D\over(2-D)}\right)\over\Gamma\left({D\over(2-D)}\right)^2}
%\right]^{-1}
\theta_2\ =\ \epsilon\ {4\over (2-D)}\,\left[1+(2-D)
{\Gamma\left({2D\over2-D}\right)\over\Gamma\left({D\over2-D}\right)^2}
\right]^{-1}
\,+\,{\cal O}(\epsilon^2)
\ .}
This agrees with the one-loop result of \rBDone .

$\theta_2$ can also be easily related to the two-membranes model
discussed in sec.~6.7.
Indeed, the divergent diagrams for the insertion of $\phi(x_1,x_2)$, as
described in \fcontactdiag , are exactly the diagrams (4) in \ftwoMdiag ,
with {\it one} of the lines between the two different membranes singled out.
As a consequence, the renormalization factor $Z_\phi(\bR)$ is related to
the counterterms for the two-membranes model by
\eqn\eZphiZtwoM{
Z_\phi(\bR)\ =\ {\partial\over\partial\bR'}\left[\bR'\,\bar Z_b(\bR,\bR,\bR')\right]_{\bR=\bR'}
\ .}
The derivative with respect to the inter-membrane coupling $\bR'$ allows to
single out one of the links between the two atoms in the diagram.

\medskip\noindent
{\bf Edge-bulk contact exponent $\theta_1$:}
\medskip
In one of the points (let us choose $x_1$) is on the boundary of an open membrane, the dimension $\Delta_\phi$ of the operator $\phi$ is modified.
Since we are interested only in local quantities, and since for $D<2$ no
additional boundary operators play a role, we may consider that locally around
$x_1$,
the membrane may be approximated by a half $D$-dimensional plane
${\cal H}=(x_\perp >0, x_\|\in \RR^{D-1})$.
We have seen that for a free membrane, we must choose Neumann boundary conditions for the field $\rvec$.
Hence, the propagator on ${\cal H}$ is not the massless propagator
$G(x,y)$ in $\RR^{D}$, given by \eMlssProp , but the Neumann propagator
\eqn\eNeuProp{
G_{\rm N}(x,y)\ =\ G(x,y)\,+\,G(x,\bar y)\qquad ,\qquad \bar y = y -2y_\perp
\hbox{\ is the mirror of \ }y\ .
}
The relevant singularity is now associated to the diagram of
\fig\fcontactBbou{
\hyperref\hcontactBbou{One loop divergent diagram} when one point is on the boundary.}.
\topinsert
\centerline{\epsfxsize=5.truecm\epsfbox{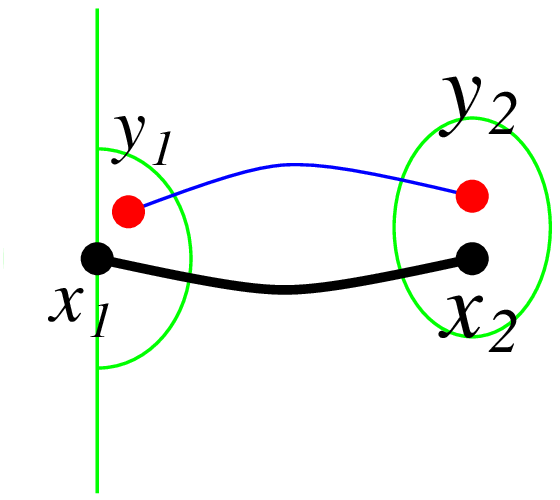}}
\centerline{\hyperdef\hcontactBbou{userfigure}{ucontactBbou}{\fcontactBbou}}
\endinsert
The corresponding MOPE coefficient is a simple modification of
\eCphiphi
\eqn\eCppBb{
{C_{\rm N}}_{\phi\phi}^\phi(x_1-y_1,x_2-y_2)\ =\ 
(4\pi)^{-d/2}\left[H_{\rm N}(x_1,y_1) -G(x_2-y_2)\right]^{-d/2}
\ ,}
where
\eqn\eHNeumann{\eqalign{
H_{\rm N}(x_1,y_1)\ =&\ {1\over 2}\big(G_{\rm N}(x_1,x_1)+G_{\rm N}(y_1,y_1)
-2 G_{\rm N}(x_1,y_1)\big)\cr
=&\ {1\over (2-D) S_D}\,\left( 2\,|x_1-y_1|^{2-D}-{1\over 2}|2 \,{y_1}_\perp |^{2-D}\right)
\ ,}
}
since $x_1$ is on the boundary.
The singular part of such a diagram is similar to that of \eScndInt\ 
and, denoting $u=y_1-x_1$ and  $v=y_2-x_2$, it is given by the small $u$ and $v$
singularity of the integral
\eqn\eDivBb{\eqalign{
\int_{u\in{\cal H}}\int_{v\in\RR} {C_{\rm N}}_{\phi\phi}^\phi(u,v)\ 
&=\ (2\pi)^{-d/2}\,\left[{\Gamma(D/2)\over 2(2-D)\pi^{D/2}}\right]^{-d/2}
\,\times\cr
&\hskip -3.truecm\times\,
\int_{u_\perp>0} \!\!\! d {u}_\perp \int d^{D-1}{u}_\| \int d^D v\,
\left[4\,\left[{u}_\perp^2+{u}_\|^2\right]^{2-D\over 2}-
\left|2\,{u}_\perp\right|^{2-D}+2\left|{v}\right|^{2-D}\right]^{-d/2}
}}
This singular part can be calculated explicitly.
It is equal to the singular part \eScndInt\ (which gives the counterterms
$Z_b$ and $Z_\phi$ in the bulk-bulk case), multiplied by the following factor,
$\omega(D)$,
which is simply given by the shift from the propagator $G$ to the Neumann
propagator $G_{\rm N}$
\eqn\eOmegaFac{\eqalign{
\omega(D)\ &=\ 
{\displaystyle\int_0^\infty d u_\perp\int d^{D-1} u_\|\  
\delta\Big[2\left(u_\perp^2+u_\|^2\right)^{2-D\over 2}-
{1\over 2}\left(2u_\perp\Big)^{2-D}-1\right]
\over
{\displaystyle\int d^D u\ \delta\left[|u|^{2-D}-1\right]}
}
\cr
&=\ 
{S_{D-1}\over 2\,S_D} \int_0^{u_\perp^{\rm max}}\!\!\!\!
d u_\perp \left[{1\over 2}+{1\over 4}(2u_\perp)^{2-D}\right]^{D\over 2-D}\,
\left[
\left[{1\over 2}+{1\over 4}(2u_\perp)^{2-D}\right]^{2\over 2-D}-u_\perp^2
\right]^{D-3\over 2}\cr
&{\rm with}\ u_\perp^{\rm max}\,=\,\left[2(1-2^{-D})\right]^{-1\over 2-D}
}
}
and $S_D={2\pi^{D/2}\over \Gamma(D/2)}$ is the volume of the unit sphere in
$\RR^D$.
This integral is convergent for $1<D<2$, and can be estimated numerically.
When $D\to 1$ $\omega(D)\to 1/2$, and it can be calculated with a finite part prescription at $u_\perp=u_\perp^{\rm max}$ for $0<D<1$.

Repeating the argument for $\theta_2$, we find that at order $\epsilon$, the
edge-bulk contact exponent $\theta_1$ is simply
\eqn\eThetaone{
\theta_1\ =\ \omega(D)\,\theta_2\,+\,{\cal O}(\epsilon^2)
}

\medskip\noindent
{\bf Edge-edge contact exponent $\theta_0$:}
\medskip
A similar reasoning allows to compute the edge-edge contact exponent 
$\theta_0$.
Now both points $x_1$ and $x_2$ are on the boundary, and the relevant
singularity is given by
\fig\fcontactbb{
\hyperref\hcontactbb{One loop divergent diagram} when both points are on the boundary.}
\topinsert
\centerline{\epsfxsize=5.truecm\epsfbox{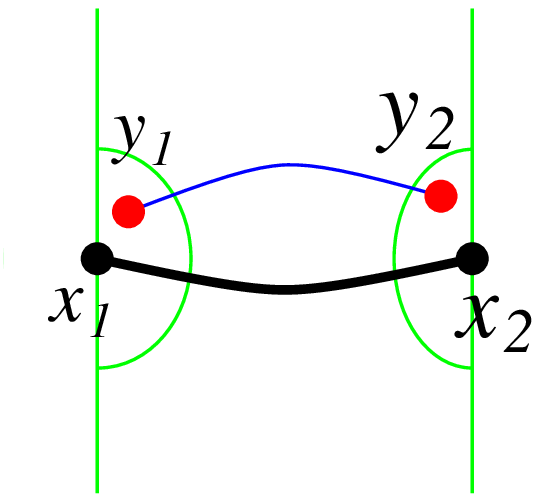}}
\centerline{\hyperdef\hcontactbb{userfigure}{ucontactbb}{\fcontactbb}}
\endinsert
with MOPE coefficient
\eqn\eCppbb{
{C_{\rm NN}}_{\phi\phi}^\phi(x_1-y_1,x_2-y_2)\ =\ (4\pi)^{-d/2}\left[H_{\rm N}(x_1,y_1)
+H_{\rm N}(x_2,y_2)\right]^{-d/2}
\ .}
The final result is at first order in $\epsilon$
\eqn\eThetazero{
\theta_0\ =\ \omega(D)^2\,\theta_2\,+\,{\cal O}(\epsilon^2)
}
\medskip\noindent
{\bf Comparison with previous results:}
\medskip
These results for $\theta_0$ and $\theta_1$ differ from those of \rBDone\ for
an open membrane with a smooth boundary.
Indeed in \rBDone\ one obtains at order $\CO(\epsilon)$ the result
$\theta2=2\theta_1=4\theta_0$, with $\theta_2$ given by \ethetwo ,
which differs from \eThetaone\ and \eThetazero ,
except for the case of open polymers $D=1$ (since $\omega(1)=1/2$).
This discrepancy in easily explained by the difference in the choice of
boundary conditions.
In \rBDone, an infinite membrane ${\cal M}\simeq\RR^D$ is considered,
with self-avoidance interaction acting only for points within a finite domain ${\cal D}\subset{\cal M}$ (with a smooth boundary).
For points outside ${\cal D}$ there is no self-avoidance, so the outside
membrane is phantom-like.
This boundary condition turns out to lead to different edge exponents than
our physical Neumann boundary conditions, which describe a finite membrane
${\cal M}$ (with a smooth boundary), with self-avoidance interactions.
This fact should not be considered as surprising.

%\vfill\break
\newsec {Other Models}

\subsec {Long Range Interactions}

The MOPE that we have derived for the multilocal operators
${\bf \Phi}_{\{A,\mvec\}}\{x\}$ given by \eMOPcsp , involving the
contact interactions $\delta^d(\rvec(x)-\rvec(y))$, can be generalized to
other operators, such as the long range Coulomb interaction (in external
$d$-dimensional space)
\eqn\eCoulomb{
\Psi_C(x,y)\ \propto\ |\rvec(x)-\rvec(y)|^{2-d}
\ .
}
It has a representation similar to \eExpRep
\eqn\eCouRep{
\Psi_C(x,y)\ =\ \int {d^D\kvec_1 d^d\kvec_2 \over (2\pi)^d}\,
\delta^d(\kvec_1+\kvec_2)\ |\kvec_1|^{-2}\,\ee^{\ii\kvec_1\cdot\rvec(x)}\,
\ee^{\ii\kvec_2\cdot\rvec(y)}
}
The only difference with the contact interaction is the appearance of a
new dipolar constraint replacing \eDipConst :
\eqn\eCouConst{
\CC_C\{\kvec_1,\kvec_2\}\ =\ (2\pi)^d\delta^d(\kvec_{1}+\kvec_{2})\,
|\kvec_1|^{-2}
\ .}

The short distance behavior of a product of such Coulomb operators can
be analyzed according to the analysis of Subsec.~3.2.
The corresponding MOPE will generate {\it the same contact operators}
${\bf\Phi}_{\{A,\mvec\}}\{x\}$ as in \eTheMOPE , with different
coefficients ${\bf C}_C^{\{A,\mvec\}}$.
Indeed, the original dipolar constraints $\CC$, now replaced by $\CC_C$, 
appear only in the coefficients \eCMOPE , not in the operators \ePhiMOPE.

If we now consider charged tethered membranes (with internal dimension $D$)
with non-screened repulsive Coulomb interactions in $d$ dimensional space,
which generalize polyelectrolytes, they are described by the Hamiltonian
\eEdwards , with $\delta^d(\rvec(x)-\rvec(x'))$ replaced by
$\Psi_C(x,x')$, and a new coupling constant $b>0$ corresponding to the
squared charge density.
The consequences of the new MOPE are twofold.
(1) The power counting is slightly modified.
The Coulomb interaction is relevant at large distance if
\eqn\eEpsCou{
\epsilon'\ =\ 2D-(d-2){2-D\over 2}\ \ge\ 0
\ .}
For $\epsilon'$ close to $0$, only the Gaussian operator $(\nabla\rvec)^2$
and the Coulomb interaction $\Psi_C$ need to be renormalized.
(2) Since it is not generated by the MOPE, the long range Coulomb interaction
$\Psi_C$ is in fact {\it not renormalized}.
This result holds to all orders in perturbation theory.
Physically it means that there are no screening effects for a uniformly
charged fluctuating object.

As a consequence the renormalized Coulomb Hamiltonian is of the form
\eqn\eRenCou{
{\raise.2ex\hbox{$\CH^{\bf R}$}/\raise -.2ex\hbox{${\rm k}_{\rm B}T$}}
\ =\ {Z(\bR)\over 2}\,\int d^Dx\,\big(\nabla_x\rvecR (x)\big)^2\, +\,
{\bR\mu^{\epsilon'}\over 2 }\,
\int d^Dx\int d^Dx'\ |\rvecR(x)-\rvecR(x')|^{2-d}
\ ,}
without coupling constant renormalization factor $Z_b$.
This renormalized Hamiltonian is equal to the bare one by the change from
renormalized to bare quantities as in \eRenBare
\eqn\eRenBCou{
\rvec\,=\,Z^{1/2}(\bR)\rvecR\quad;\quad b\,=\,\bR\,Z^{(d-2)/2}(\bR)\,
\mu^{\epsilon'}
\ .}
Therefore the RG function $\beta(\bR)$ and the anomalous dimension $\nu(\bR)$,
given by \eBeta\ and \eDimAn , are not independent, but related by
\eqn\eBNuCou{
\beta(\bR)\ =\ -\,2\,D\,\bR\,+\,(d-2)\,\bR\,\nu(\bR)
}
If, as can be checked at one loop, an UV fixed point $\bR^\star>0$ exists,
such that $\beta(\bR^\star)=0$, then the anomalous size  exponent
$\nu^\star=\nu(\bR^\star)$ is exactly given by
\eqn\enuCou{
\nu^\star\ =\ {2D\over d-2}
\ .}

This result coincides with the result obtained using a Gaussian
variational method
[\xref\rGuiPal-\xref\rLeDous],
an approximation which in this case turns out to
be exact. 
This result is of course meaningful only if $\nu^\star<1$.
If the r.h.s of \enuCou \ is larger than 1, this means that the long range
Coulomb interaction makes the object rigid and completely flat, with
$\nu=1$.
This occurs when $d\le 4$ for polyelectrolytes ($D=1$) and when $d\le 6$ for
charged membranes ($D=2$).

%\vfill\break
\subsec {Tricritical Behavior: $\Theta$ versus $\Theta ''$ Points}

Another application concerns interacting tethered membranes at the tricritical
point.
One might expect on physical grounds that for ``real'' tethered membranes,
there is a competition between two-body attractive interactions (for instance
long-range Van der Walls forces) and hard core repulsive interactions (for
instance hydration forces).
This is known to occur for polymers, and as a consequence there is
a low temperature collapsed phase, where attractive forces dominate, and
the high temperature swollen phase, where repulsive forces dominate.
These two phases are separated by a critical $\Theta$ point.
In the framework of the Edwards model, the effective
2-body coupling $b$ vanishes at this point, and one must introduce the next
most relevant operator.
For polymers it is well known that this operator is simply the 3-body
repulsive contact interaction
[\xref\rDeGeTri,\xref\rDesCloJan].

For membranes with arbitrary $1\le D\le 2$, the situation is more subtle.
Among all multilocal operators \eMOPcsp\ one has to consider two potentially
relevant operators: the 3-body operator
\eqn\eThrBod{
\Phi_3\{x,y,z\}\ =\ \delta^d(\rvec(x)-\rvec(y))\,\delta^d(\rvec(y)-\rvec(z))
\ ,}
with engineering dimension $\epsilon_3$ for the corresponding coupling 
\eqn\eepsthr{
\epsilon_3\ =\ 3D-2d{(2-D)\over 2}
\ ;}
and the modified 2-body operator
\eqn\eTwoBod{
\Phi''_2\{x,y\}\ =\ -\,\Delta_{\vec r}\delta^d(\rvec(x)-\rvec(y))
}
where $\Delta_\rvec$ is the Laplacian in $d$-dimensional space,
with corresponding dimension $\epsilon''_2$
\eqn\eEpstwo{
\epsilon''_2\ =\ 2D-(d+2){(2-D)\over 2}
\ .}
This last operator describes a 2-body interaction repulsive at short range
{\it and} attractive at larger range.

A simple analysis shows that the critical $\Theta$-line $\epsilon_3=0$ and the
$\Theta''$-line $\epsilon''_2=0$ intersect at $(D=4/3, d=6)$.
If $D<4/3$, and in particular for polymers, the 3-body operator
\eThrBod\ is the most relevant,
while for $D>4/3$ the modified 2-body operator \eTwoBod\ dominates.
Considering only this last operator,
the one loop calculation of the renormalization group flow and of the
scaling exponent $\nu^\star$ at order $\epsilon''_2$ is presented in the next
subsection.
The corresponding one loop results for the 3-body operator require numerical
calculations, and the analysis
of the cross-over between the two regimes at $D=4/3$, $d=6$ is more
complicated to study.
They can be found in
\ref\rDaWiIII{K.~J.~Wiese and F.~David, Nucl. Phys. B 450 [FS] (1995) 495.}%
.

%\vfill\break
\subsec {1-Loop Results for the $\Theta ''$ Point}

We first apply the MOPE formalism to compute at one loop the renormalization
group functions for the $\Theta''$ point.
We need the MOPE for $\Phi''_2\{x_1,x_2\}\to\hbox{\sl local operators}$ when
$x_1\to x_2$, which is 
\eqn\eMOmoone{
\Phi''_2\{x_1,x_2\} {\buildlim x_1\to x_2\under=}
\Cbf_{\Phi''_2}^{\bf 1}\{x_1-x_2\}\,{\bf 1}(x)\,+\,
\Cbf_{\Phi''_2}^{\alpha\beta}\{x_1-x_2\}\,
\npr\nabla_\alpha\rvec(x)\nabla_\beta\rvec(x)\npr\,+\,\ldots
}
with $x=(x_1+x_2)/2$.
Starting from the OPE \ephiope\ and integrating over the $\kvec$, now with an
extra kernel $\kvec^2$ to account for the $-\Delta_\rvec$, we obtain
\eqn\eCmoone{
\Cbf_{\Phi''_2}^{\alpha\beta}\{x_1-x_2\}\ =\ -\,{1\over 4}\,({d\over 2}+1)\,(4\pi)^{-d/2}\,
(-G(x_1-x_2))^{-2-d/2}(x_1^\alpha-x_2^\alpha)(x_1^\beta-x_2^\beta)
}
The MOPE for $\Phi''_2\Phi''_2\to\Phi''_2$ is of the form
\eqn\eMomotwo{
\Phi''_2\{x_1,y_1\}\Phi''_2\{x_2,y_2\} {\buildlim{x_1\to x_2\atop y_1\to y_2}\under =}
\Cbf_{\Phi''_2\Phi''_2}^{\Phi''_2}\{x_1-x_2, y_1-y_2\}\,
\Phi''_2\{x,y\}\,+\,\cdots
}
where $x=(x_1+x_2)/2$, $y=(y_1+y_2)/2$.
We use the OPE \epphiope\ and obtain, along the same line as in sec.~5.2
\eqn\eCmotwo{
\Cbf_{\Phi''_2\Phi''_2}^{\Phi''_2}\{x_1-x_2,y_1-y_2\}\ =\ (4\pi)^{-d/2}\,
\left({d^2\over 16}+{3d\over 8}-{1\over 2}\right)\,
\left[-G(x_1-x_2)-G(y_1-y_2)\right]^{-1-d/2}
}

We are now in position to compute the renormalization factors.
The renormalized action is
\eqn\eRACmod{
{Z''(\bR)\over 2}\,\int d^Dx (\nabla\rvecR)^2\,+\,
{\bR\,Z''_b(\bR)\mu^{\epsilon''_2}\over 2}\int\!\!\int  d^Dx\,d^Dx'\,
(-\Delta_\rvec)\delta^d(\rvecR(x)-\rvecR(x'))
\ .}

At one loop, one obtains $Z''$ from the singular part of the integral of
$\Cbf_{\Phi''_2}^{\alpha\beta}\{x_1-x_2\}$ as $\epsilon''_2\to 0$.
\eqn\eZmod{
Z''\ =\ 1\,+\, {\bR\over\epsilon''_2}\,{1\over 4D}\,(4\pi)^{-{d\over 2}}\,
(2-D)^{2+{d\over 2}}\,(1+{d\over 2})\,
\left({2\,\pi^{D/2}\over\Gamma(D/2)}\right)^{3+{d \over 2}}
}
Similarly we obtain $Z''_b$ from $\Cbf_{\Phi''_2\Phi''_2}^{\Phi''_2}$
\eqn\eZbmod{
Z''_b\ =\ 1\,+\,{\bR\over\epsilon''_2}\,{1\over 2}
\,(4\pi)^{-{d\over 2}}\, (2-D)^{{d\over 2}}\,
\left({d^2\over 16}+{3d\over 8}-{1\over 2}\right)\,
\left({2\,\pi^{D\over 2}\over\Gamma(D/2)}\right)^{3+{d\over 2}}\,
{\Gamma\left({D\over 2-D}\right)^2\over\Gamma\left({2D\over 2-D}\right)}
}

The renormalization group functions $\beta$ and $\nu$ are computed using
\eRenBare, \eBeta\ and \eDimAn .
We find for the anomalous exponent $\nu^\star$ at the $\Theta''$ point
(at order $\epsilon''_2$)
\eqn\eNustar{
\nu^\star\ =\ {2-D\over 2}\,+\,\epsilon''_2\,\left[d+2+{4D\over (2-D)^2}
{\Gamma\left({D\over 2-D}\right)^2\over\Gamma\left({2D\over 2-D}\right)}
\,{\left({d^2\over 16}+{3d\over 8}-{1\over 2}\right)\over\left({d\over 2}+1\right)}\right]^{-1}
}

%\subsec {Membrane interacting with one Impurity}
%\medskip{\sl Faut-il le faire?}

\newsec {Conclusions}

In this paper we have given a general formalism to study in perturbation theory
the generalized Edwards model, which describes the continuum limit for flexible
self-avoiding tethered membranes and polymers.
We have shown that the short distance behavior of this model in encoded into
a multilocal operator product expansion (MOPE), which generalizes Wilson's
operator product expansion.
This MOPE allows us to study in a systematic way  the short distance behavior
of the model, and to prove that it is renormalizable at all orders in
perturbation theory.
The formalism constructed here allows in particular to prove the validity of
the direct renormalization method, which was used previously to derive
at first order in perturbation scaling laws and scaling exponents for
large self-avoiding tethered membranes.
It also provides a direct proof of the consistency of direct renormalization for
polymers.
We have also presented a few applications of this formalism, including 
membranes with long range interactions and membranes at the $\Theta$-point.

Let us stress that our formalism is quite general, and should apply to a large
variety of interactions, provided that they are multi-local (with respect to
the ``internal" membrane space), singular at short distances in the external
embedding space, and invariant under global translations.
The method applies also naturally to the case of several interacting membranes
and polymers, and might be applicable also to membranes in disordered media,
via the replica method (as already done for polymers).

Let us also stress that the method is essentially perturbative.
Several important issues still have to be addressed in this context.
Is the perturbation series asymptotic, and in which sense is it summable?
In Quantum Field Theory it is expected, for instance, that the
$\varepsilon=4-d$ expansion is Borel summable; this conjecture appears to be
satisfied numerically, and is the basis of very efficient ressummation
methods which lead to very accurate predictions for critical exponents.
Almost nothing is known, even at the heuristic level, for the generalized
Edwards model.
This is a very important point, since to reach the point of
physical interest $D=2$ one must extrapolate up to $\epsilon=4$.
In this respect, it would be suitable to control the large orders of the
perturbative expansion.

One thing which is known is that at the first and second orders, the estimate
for the size exponent $\nu$ coincides at large $d$ with the result
$\nu_{\rm var}=2D/d$ of the Gaussian variational method
[\xref\rDaWiI,\xref\rDaWiII].
Arguments have been given that this is true to all orders in a $1/d$ expansion
\rDaWiII , with a difference $\nu-\nu_{\rm var}=\CO\big(\exp(-d)\big)$.
One should notice however that this property seems to be true only for the
size exponent $\nu$.
Also one should stress that no systematic $1/d$ expansion has been constructed
so far for this model.

Another important issue is the relevance of curvature terms.
It is known for two dimensional objects both the intrinsic (or Gaussian)
curvature and the extrinsic (or mean) curvature might become relevant.
For polymerized membranes, the intrinsic curvature does not fluctuate (see 
the discussion of section~6). So it should not play an important role,
except for the configuration exponent $\gamma$, which is expected to get an
extra dependence on the integrated curvature, i.e. of the Euler characteristics,
for $D=2$ exactly. On the contrary, the size exponent $\nu$ is insensitive to
the intrinsic curvature.
More drastic effects are expected from the extrinsic curvature, i.e. from the
bending rigidity of the membrane.
For phantom membranes (without self-avoidance) but with bending rigidity
it is known that a crumpling transition occurs, separating a flat phase (at high
rigidity) from a crumpled phase (at low rigidity).
In the flat phase self-avoidance is clearly irrelevant.
The generalized Edwards model considered in this paper is applicable to the
crumpled phase, and concentrates on the modification of its properties due to
self-avoidance.
It implicitly assumes that bending rigidity terms are irrelevant, which is
true close to $\epsilon=0$ (by power counting).
The important point is their relevance at the infra-red stable fixed point
(which describes the crumpled phase) for finite $\epsilon$, in particular
for $\epsilon=4$ ($D=2$).
A heuristic analysis can be obtained by simply comparing the size exponent
$\nu$ in the crumpled phase (with self-avoidance) with the size exponent
$\nu_{\rm cr}$ of phantom membranes at the crumpling transition.
It is expected that bending rigidity is relevant whenever $\nu>\nu_{\rm cr}$,
and that this will drive the membrane into the flat phase
[\xref\rGuiPal,\xref\rDaWiII].
The different estimates of \rGuiPal\ and of \rDaWiII\ lead to a lower critical
dimension $d_{\rm lc}$, below which two dimensional self-avoiding membranes
should be flat, lying in the range $3\le d_{\rm lc}< 4$.
Clearly it is a major issue to have more refined estimates of $d_{\rm lc}$,
in order to establish whether physical two dimensional membranes in three
dimensional space may be in a crumpled state or are always flat, a behavior
which would be very different from that of ordinary polymer chains.
\bigskip
\noindent{\bf Acknowledgments:}
\medskip
F.D. thanks K. Wiese for his interest and useful remarks.
We thank M. Berg\`ere for a careful reading of the manuscript.

%\vfill\break
\appendix {A}{General form of the MOPE}

It is not difficult to generalize the multilocal operator product expansion
\eTheMOPE , which describes the short distance behavior of a product
of bilocal operators $\delta^d(\rvec(x)-\rvec(y)$, to describe the short
distance behavior of a product of the most general multilocal operators.
These operators $\Phi$ are given by \ePhiExpl , they are characterized by a set
$\CI$ of ``end-points" $i$, by local operators $A_i$ and by $d$-uples
$\mvec_i\in\NN^d$:  
\eqn\ePhibis{
\Phi\ =\ \Phi_{\{A_i,\mvec_i\}}\{x_i\}\ =\ 
\int d^d\rvec\,\sprod_{i}
\Big\{\npr A_i(x_i)\,
(\ii\nabla_{\!\rvec})^{\mvec_i}\delta^d(\rvec-\rvec(x_i)) \npr\Big\}
\ .
}
We depict graphically such an operator by the starfish-like diagram of
\fig\fStarfish{Starfish representation of
\hyperref\hStarfish{a multilocal operator}}.
\topinsert
\centerline{\epsfxsize=5.truecm\epsfbox{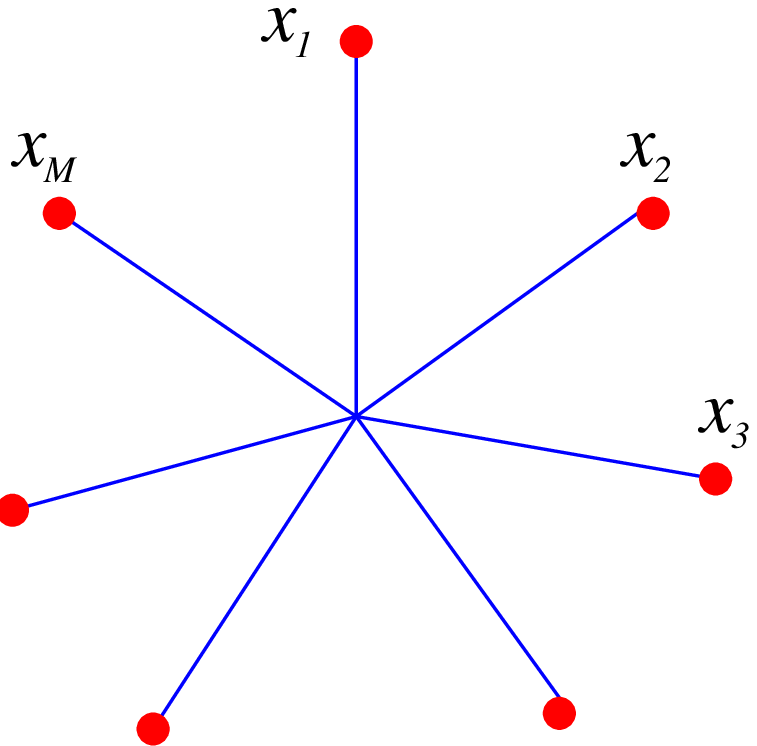}}
%\vskip-3.truecm\hskip2.truecm$\mvec_1$
\centerline{\hyperdef\hStarfish{userfigure}{uStarfish}{\fStarfish}}
%\centerline{\fStarfish}
\endinsert

We now consider $P$ such operators $\Phi_1,\cdots\ \Phi_P$, with different
end-points, and we are interested
in the behavior of the product of these operators when we partition the
set of all end-points $\CI=\CI_1\cup\cdots\cup\CI_P$ into a set of atoms
$\CP$, assign a given position $x_\CP$ to each atom $\CP$, and let the
end-points $x_i$, $i\in\CP$ inside each atom $\CP$ tend toward $x_\CP$.
In this limit, the operators form a molecule $\CM$, that we assume to be
connected.

The general MOPE takes the form of a sum over all multilocal operators
$\Phi\{x_P\}$ relative to the set of atoms $\CP$:
\eqn\eGMOPE{
\Phi_1\{x_i\}\cdots\Phi_P\{x_i\}\ =\ \sum_\Phi
\Cbf^\Phi_{\Phi_1\cdots\Phi_P}\{y_i\}\cdot\Phi\{x_\CP\}
}
where the coefficients $\Cbf^\Phi_{\Phi_1\cdots\Phi_P}$ 
depend on the relative coordinates $\{y_i=x_i-x_\CP\}_{i\in\CP}$ inside each
atom $\CP$, and not on the coordinates $x_\CP$ of the atoms themselves.
They are homogeneous functions of the $y_i$'s, with degree ${\rm deg}\Cbf$
which depend on the operators $\Phi_1,\cdots\Phi_P$, of $\Phi$ and of the
molecule $\CM$.

The derivation of \eGMOPE\ is parallel to that of \eTheMOPE .
One starts by rewriting each $\Phi_I$ $I=1,\cdots ,P$ as
\eqn\ePhiI{
\Phi_I\{x_i\}\ =\ \int d^d \rvec_I\int \sprod_{i\in\CI_I}\,
{d^d\kvec_i\over(2\pi)^d}\, [\kvec_i]^{\mvec_i}\,
\ee^{-\ii(\ssum\limits_{i\in\CI_I}\kvec_i)\rvec_I}
\,\sprod_{i\in\CI_I}\,\npr A_i(x_i)\ee^{\ii\kvec_i\rvec(x_i)}\npr
}
The integration over $r_I$ gives a neutrality constraint for each starfish
\eqn\eConsSF{
\CC_I\{\kvec_i\}\ =\ (2\pi)^d\,\delta^d(\ssum_{i\in\CI_I}\kvec_i)
}
Inside each atom $\CP$, the product of vertex operators dressed by the $A_i$'s
has an operator product expansion (which generalizes \eNorProd\eAvVal\eTayExp)
of the form
\eqn\eGenOPE{
\sprod_{i\in\CI_I}\,\npr A_i(x_i)\ee^{\ii\kvec_i\rvec(x_i)}\npr\ =
\ \sum_A\,C_{\{A_i\}%_{i\in\CP}
}%
^A\{y_i,\kvec_i\}\,
\ee^{-{1\over 2}\sum\limits_{i,j\in\CP}\kvec_i\kvec_j G(y_i,y_j)}
\,\npr A(x_\CP)\ee^{\ii(\ssum\limits_{i\in\CP}\kvec_i)\rvec(x_\CP)}\npr
}
where the coefficients $C_{\{A_i\}}^A$ are monomials in the $\kvec_i$'s, and
depend only algebraically of the $y_i$'s, i.e. contain only (possibly
non-integer and negative) powers of the relative coordinates $y_i-y_j$.
Inserting a momentum $\kvec_\CP$ for each atom via the identity
$1=\int d^d\kvec_\CP\delta^d((\ssum_{i\in\CP}\kvec_i)-\kvec_\CP)$, using the
connexity of the molecule $\CM$ to rewrite one of the starfish neutrality
constraints $\CI_I$ as a global neutrality constraint for the molecule
$\delta^d(\ssum_{\CP}\kvec_\CP)$, and now Taylor expanding in the
$\kvec_\CP$ one ends with \eGMOPE , with the explicit expression for
the coefficient 
\eqn\eCGenExp{\eqalign{
\Cbf^\Phi_{\Phi_1\cdots\Phi_P}\{y_i\}\ &=\ 
\ \int\sprod_{i\in\CI}
{d^d\kvec_i\over (2\pi)^d}\,
\mathop{{\sprod}'}_{I}\CC_I\{\kvec_i\}\, \sprod_{i\in\CI}[\kvec_i]^{\mvec_i}
\,\times\cr
&\hskip -3.em\times\sprod_{\CP\in\CM}
\Bigg\{C^{A_\CP}_{\{A_i\}_{i\in\CP}}\{y_i,\kvec_i\}%_{i\in\CP}
\,{1\over \mvec_\CP!}\,
(\nabla_\kvec)^{\mvec_\CP}\delta^d(-\!\ssum_{i\in\CP}\kvec_i)\,
\ee^{-{1\over 2}\!\ssum\limits_{i,j\in\CP}\!\kvec_i\cdot\kvec_j G(y_i,y_j)}
\Bigg\}\cr}
}
with $\sprod'$ being the product over the $I$'s but one.

%\vfill\break
\appendix {B}{The (massless) propagator on the sphere $\CS_D$.}
The propagator for a free field on the $D$-dimensional sphere $\CS_D$ with
constant curvature can be defined in various ways which allow for a proper
analytic continuation in $D$.
Let us consider the unit sphere with radius $L=1$ and volume
\eqn\eVolSD{{\rm Vol}(\CS_D)\ =\ S_{D+1}\ =\ {2\,\pi^{D+1\over 2}\over
\Gamma\left({D+1\over 2}\right)}
}
The massive propagator $G(x,y;t)$, where $t$ is the squared mass $t=m^2$,
is the solution of the equation
\eqn\eMassG{
(-\Delta_x\,+\,t)G(x,y;t)\ =\ {1\over\sqrt{g(y)}}\delta(x,y)
}
where $\Delta_x$ is the covariant Laplacian.
By O($D$) invariance it depends only of the geodesic distance $\ell$ between
$x$ and $y$, $G=G(\ell;t)$, and in radial coordinates \eMassG\ becomes
\eqn\eMassEq{
\left[{\partial^2\over\partial\ell^2}+(D-1)\cot(\ell)
{\partial\over\partial\ell}-t\right]G(\ell;t)\ =\ 0\qquad\hbox{for\ }\ell\neq
0\ \hbox{mod}(2\pi)
}
supplemented by the condition that at the origin the singular part
of $G(\ell;t)$ behaves as the massless propagator in plat space
\eqn\eLapBC{
G_{\CS_D}(\ell)\ {\buildlim\ell\to 0\under\simeq}\ 
\hbox{cst}\,-\,\ell^{2-D} {\Gamma(D/2)\over 2(2-D)\pi^{D/2}}\ +\ \cdots
}

By the change of variable $\cos(\ell)=z$ \eMassEq\ reduces to a Gegenbauer 
equation 
\eqn\eGegen{
(1-z^2)G''\,-\,DzG'\,-\,tG\ =\ 0
}
and $G(\ell;t)$ is expressible in terms of known special functions
(Gegenbauer functions or hypergeometric functions).

In the massless limit $t\to 0$ the propagator becomes I.R. singular.
Subtracting the I.R. singular part due to the zero mode of $\Delta$, we obtain
the massless propagator
\eqn\eMsslss{
G(\ell)\ =\ \lim_{t\to 0}\left( G(\ell;t)\ -\,{t^{-1}\over\hbox{Vol}(\CS_D)}
\right)
}
which is the solution of \eProCurv\ on $\CS_D$.

Explicit integral representations for the propagators on the unit sphere can
be obtained for instance by the following argument.
Let us embed the unit sphere $\CS_D$ in $\RR^{D+1}$ and use the spherical
coordinates $\vec X\to(r,\vec\phi)=(r,\phi,\varphi_1,\cdots\varphi_{D-2})$
in $\RR^{D+1}$.
If $G(\ell;t)$ is the massive propagator on $\CS_D$, it is easy to see that
the function on $\RR^{D+1}$
\eqn\eFtDef{
F(\vec X;t)\ =\ H(r;t)G(\phi;t)
}
satisfies the equation
\eqn\eDelF{
-\Delta_{\vec X} F\ =\ H(r;t) r^{-2}\,\delta(\vec\phi)
}
where $\delta(\vec\phi)$ is the delta function at the origin on the unit sphere,
provided that $H(r;t)$ satisfies
\eqn\eEquH{
\left[{\partial^2\over\partial r}\,+\,{D\over r}{\partial\over\partial r}\,+\,
{t\over r^2} \right]H(r;t)\ =\ 0
\ .}
Hence, from the knowledge of the massless propagator on $\RR^{D+1}$,
\eqn\eMPrDpO{
G(\vec X,\vec Y)\ =\ -\,{|\vec X-\vec Y|^{1-D}\over (1-D)S_{D+1}}
}
we can obtain the function $F(\vec X)$ as the Coulomb potential created by
a linear distribution of charge $\rho(r)=r^{D-2}H(r;t)$ along the positive
unit axis $\vec Y(r)=(r,0\cdots 0)$
\eqn\eFexpl{
F(\vec X)\ =\ \int_0^\infty dr\, H(r;t)\,r^{D-2}\,G(\vec X,\vec Y(r))
}
Solving \eEquH\ leads to
\eqn\eHexpl{
H(r;t)\ =\ r^{{1-D\over 2}\pm
\sqrt{\left({1-D\over 2}\right)^2-t}}
%H(r;t)\ =\ r^\alpha\quad;\quad \alpha\,=\,{1-D\over 2}\pm
%\sqrt{\left({1-D\over 2}\right)^2-t}
}
and from $F(\vec X;t)$ we obtain finally the integral representation
for the massive propagator on the sphere
\eqn\ePrSpEx{
G(\phi;t)\ =\ {\Gamma({\scriptstyle D-1\over2})\over 4\,\pi^{D+1\over 2}}\,
\int_0^\infty {dr\over r}\,
r^{ {{D-1\over 2}}\pm {\sqrt{\left({D-1\over2}\right)^2-t}}}
\,[r^2+1-2r\cos(\phi)]^{1-D/2}
%\beta(t)\ &=\ {{D-1\over 2}}\pm {\sqrt{\left({D-1\over2}\right)^2-t}}
}
The $\pm$ representations are equivalent via the change of variable 
$x\leftrightarrow 1/x$.
The integral \ePrSpEx\ is convergent for $1<D$, and allows for an explicit
analytic continuation in $D$.
For $D\le 1$ it is defined by a finite part prescription.

The massless propagator $G(\phi)$ is defined by \eMsslss\ , and has the integral
representation.
\eqn\eMsPr{
G(\phi)\ =\ 
{\Gamma({\scriptstyle D-1\over 2})\over 4\,\pi^{D+1\over 2}}\,
\int_0^\infty {dr\over r}
\,\left([r^2+1-2r\cos(\phi)]^{1-D\over 2}-\theta(1-r)\right)
\ ,}
with $\theta$ the Heaviside step function.
If we shift $G$ it by a constant so that $G(0)=0$, we get  
\eqn\eMsPrLi{\eqalign{
G(\phi)\ &\to\ G(\phi)-G(0)\cr
&=\ 
{\Gamma({\scriptstyle D-1\over 2})\over 4\,\pi^{D+1\over 2}}\,
\int_0^\infty {dr\over r}\, r^{D-1}
\,\left([r^2+1-2r\cos(\phi)]^{1-D\over 2}-[(r-1)^2]^{1-D\over 2}\right)
}
}

Finally, from \ePrSpEx\ one can construct the heat kernel $K(\phi;s)$ on the
$D$-dimensional sphere through the inverse Laplace transform with respect to $t$
\eqn\eHeKeSp{
K(\phi;s)\ =\ \int_{-\ii\infty}^{+\ii\infty}\hskip-1ex {dt\over 2\ii\pi}\,
\ee^{st}\,G(\phi;t)
}
%{\it We have not yet found a simple integral representation for $K(\phi;s)$.}

%\vfill\break
\appendix {C} {The short distance OPE in curved space}

In this appendix we give the outline of the derivation of the
OPE  for vertex operators \eOPEcsp\ in curved space.
It relies on the short distance behavior of the free massless propagator
$G(x,x')=\langle{\bf r}(x){\bf r}(x')\rangle_0$
on a general curved manifold $\CM$, when $x,x'\to z$.
This behavior can be inferred from the small proper-time behavior of the 
heat-kernel
\eqn\eHKDef{
K(x,x';s)\ =\ \langle x|\ee^{s\Delta}|x'\rangle
}
using the relation
\eqn\ePrHK{
G(x,x')\ =\ \int_0^\infty ds\,\left[K(x,x';s)-{1\over {\rm Vol}(\CM)}\right]
\ .}
The last term in \ePrHK\ is required in order to subtract the contribution of
the zero mode of the Laplacian $\Delta=D^\alpha D_\alpha$, which contributes
to the large $s$ limit of the heat-kernel:
$\lim_{s\to\infty}K(x,x';s)=1/{\rm Vol}(\CM)$.

The small $s$ behavior of $K$ is known to be given by the following
asymptotic expansion 
(see in particular the chapter 9 of \rBDW)
\eqn\eHKExp{
K(x,x';s)\ =\ (4\pi\,s)^{-D/2}\,%{\mit\Delta}(x,x')^{1/2}\,
\ee^{-\ell^2(x,x')/4s}\,
\sum_{r=0}^\infty\,a_r(x,x')\,s^r
\ .}
$\ell(x,x')$ is the geodesic distance between $x$ and $x'$.
$D$ is the dimension of the manifold $\CM$.
%${\mit\Delta}(x,x')$ is (related to) the Van Vleck-Morette determinant.
A crucial property of the functions $\ell^2(x,x')$
%, ${\mit\Delta(x,x')}$
and of the coefficients $a_r(x,x')$ is that, for a smooth metric, they are smooth
functions of $x$ and $x'$, at least as long as these points are close enough.
Their Taylor expansion at $x,x'=z$ depends only on the local properties
of the metric at $z$.
This means that at a given order $m$, the coefficients of their Taylor expansion
are a differential polynomial in the metric, that is
involve only the metric tensor $g(z)$ and its derivatives
$\partial g(z)$, $\partial\partial g(z)$, etc$\ldots$ at $z$ up to some
order (which depends on $m$).

%For instance, we have explicitly (omitting contracted indices)
%\eqn\eDelExp{
%{\mit\Delta}(x,x')\ =\ 1\,+\, (x-z)\,C\,g^{-1}(z)\,\partial g(z)\,+\,
%(x,-z)\,C'\,g^{-1}(z)\,\partial g(z)\,+\,\cdots
%}
%and more specifically, each coefficient of the term of degree $m$ of the
%expansion of $\mit\Delta(x,x')$
%(that is the sum of the terms of order $(x-z)^\mvec(x'-z)^{\mvec '}$, with
%$m=|\mvec|+|\mvec '|$)
%is a polynomial in $g^{-1}$, $g$ and its derivatives $\partial g$,
%$\partial\partial g$, $\partial\partial\partial g$, etc$\ldots$ at the point
%$z$, which (by homogeneity) contains exactly $m$ derivatives $\partial$.

For instance, the expansion of the coefficients $a_r(x,x')$ starts as 
\eqn\eACoef{\eqalign{
a_0(z,z)\ &=\ 1\qquad;\qquad a_1(z,z)\ =\ {1\over 6}\,R(z)\cr
a_2(z,z)\ &=\ {1\over 30}\,D^\alpha D_\alpha R(z)\,+\,{1\over 12}\,R^2(z)\,
-{1\over 180}\,R^{\alpha\beta}\,R_{\alpha\beta}\,+\,{1\over 180}\,
R_{\alpha\beta\gamma\delta}\,R_{\alpha\beta\gamma\delta}\cr
}}
which means that the term of degree 0 of the Taylor expansion of $a_r$
involves exactly $2r$ derivatives $\partial$ of the metric.
Then, the terms of degree $m$ of the Taylor expansion of $a_r$
(that is the sum of the terms of order $(x-z)^\mvec(x'-z)^{\mvec '}$, with
$m=|\mvec|+|\mvec '|$)
are differential polynomials of the metric which contains exactly
$2r+m$ derivatives $\partial$
(that is polynomial in $g^{-1}$, $g$ and its derivatives $\partial g$,
$\partial\partial g$, $\partial\partial\partial g$, etc$\ldots$ at the point
$z$).
Similarly for the squared geodesic distance we have
\eqn\eLSqr{\eqalign{
\ell^2(x,x')\ &=\ (x-x')^\alpha\,(x-x')\beta\,g_{\alpha\beta}(z)\,+\,\cdots\ \cr
&=\ \|x-x'\|^2_z\,+\,\cdots\cr
}
}
and the terms of degree $m>2$ contain exactly $m-2$ derivatives.

The $(x,x')\to z$ behavior of $G(x,x')$ is obtained by using
\ePrHK\ and \eHKExp .
For $s\sim\|x-x'\|_z$ in the integral \ePrHK\ the small $s$ behavior
of the heat-kernel dominates.
This generates a singular contribution for the
small distance expansion of $G$, which is proportional to $\|x-x'\|_z^{2-D}$,
times  powers of $(x-z)$ and $(x'-z)$.
\eqn\eGsing{\eqalign{
G_{\rm sing}(x,x')\ 
&=\ \sum_{r=0}^\infty\, (4\pi)^{-D/2}\,\Gamma\left({D-2\over 2}-r\right)\,
\left({\ell(x,x')\over 2}\right)^{2-D-2r}\,a_r(x,x')\cr
&=\ \|x-x'\|^{2-D}\,C(x,x')
\ ,}}
where $C(x,x')$ is an asymptotic series in $x-z$ and $x'-z$.
The coefficients of order $m$ of $B$ are differential polynomials
of the metric $g$ at $z$, with exactly $m$ derivatives $\partial$.
Therefore \eGsing\ can be written as a sum over differential monomials 
$B[g](z)$ of the metric at $z$, of the form
\eqn\eGsMo{
G_{\rm sing}(x,x')\ =\ \sum_{B[g]} C^B(x-z,x'-z))\,B[g](z)
}
where the coefficients $C^B$ are homogeneous function of $(x-z)$ and $(x'-z)$
of degree $2-D+m$, where $m$ is the total number of derivatives $\partial$ in
$B[g]$.
The leading contribution is given by the $r=0$ term, and is associated to the
unity operator $B_0={\bf 1}$, it is nothing
but the flat space propagator $G_0(x,x')= -\|x-x'\|^{2-D}/(2-D)S_D$ at $z$.

The rest of the $s$-integral in \ePrHK\ generates a regular contribution
$G_{\rm reg}(x,x')$, which is an asymptotic series in integer powers of
$x-z$ and $x'-z$.
By a formal Taylor series expansion the coefficients of this series define
the expectation values of the normal ordered operators quadratic in ${\bf r}$
\eqn\eGreg{
G_{\rm reg}(x,x')\ =\ \sum_{\mvec,\mvec'} {(x-z)^\mvec\over\mvec !}
{(x'-z)^{\mvec'}\over\mvec '}\,\langle\CN[\nabla^\mvec{\bf r}\nabla^{\mvec '}
{\bf r}](z)\rangle_0
}
The total asymptotic expansion of the propagator is
\eqn\eGAsEx{
G(x,x')\ =\ G_{\rm reg}(x,x')\,+\,G_{\rm sing}(x,x')
}
This decomposition of the short distance expansion of the propagator in terms
of the non-analytic part $G_{\rm sing}$ and the analytic part $G_{\rm reg}$
is unambiguously defined if $D$ is not an even integer.
This short distance expansion is nothing but the Operator Product Expansion for
two free field operators in curved space.
The operators that are generated are the normal ordered operators
$\CN[\nabla^\mvec{\bf r}\nabla^{\mvec '}{\bf r}]$ and the purely geometrical
local operators $B[g]$.

\medskip
Now we can derive the general OPE \eOPEcsp.
The product of exponential operators can be separated as in \eNorProd\ into
its normal product $\npr[\ ]\npr$ times its expectation value
\eqn\eNPSep{
\prod_i\ee^{\ii\kvec_i\cdot\rvec(x_i)}
\ =\ \langle
\prod_i\ee^{\ii\kvec_i\cdot\rvec(x_i)}
\rangle_0\,
\npr
\prod_i\ee^{\ii\kvec_i\cdot\rvec(x_i)}
\npr
}
The normal product $\npr[\ \ ]\npr$ is defined in the usual way
by subtracting all propagators between
pairs of $x_i$'s, and it is therefore analytic in the $x_i$'s when they
tend toward the same point $z$.
The singularities when $x_i\to z$ are contained into 
\eqn\eGaAv{
\langle \prod_i\ee^{\ii\kvec_i\cdot\rvec(x_i)} \rangle_0\ =\ 
\ee^{-{1\over 2}\sum\limits_{i,j}\kvec_i\cdot\kvec_j\,G(x_i,x_j)}\ 
=\ 
\ee^{-{1\over 2}\sum\limits_{i,j}\kvec_i\cdot\kvec_j\,G_{\rm reg}(x_i,x_j)}\ 
\ee^{-{1\over 2}\sum\limits_{i,j}\kvec_i\cdot\kvec_j\,G_{\rm sing}(x_i,x_j)} 
}
Let us {\it define} the normal ordered product
\eqn\eNOPdef{
\CN\left[\prod_i\ee^{\ii\kvec_i\cdot\rvec(x_i)}\right]\ =\ 
\npr \prod_i\ee^{\ii\kvec_i\cdot\rvec(x_i)} \npr\ 
\ee^{-{1\over 2}\sum\limits_{i,j}\kvec_i\cdot\kvec_j\,G_{\rm reg}(x_i,x_j)} 
}
Since $G_{\rm sing}(x_i,x_j)$ has an asymptotic Taylor expansion,
we can Taylor expand this normal ordered product around $z$ in powers of the
$y_i=x_i-z$.
We thus obtain an OPE similar to \eTayExp
\eqn\eTEOP{
\CN\left[\prod_i\ee^{\ii\kvec_i\cdot\rvec(x_i)}\right]\ =\ 
\sum_{A[\rvec]} C^A\{y_i,\kvec_i\}\ \CN\left[ A(z)\,
\ee^{\ii\kvec\cdot\rvec(z)}\right]
\ ,
}
($\kvec=\sum_i\kvec_i$)
where the sum runs over the local operators $A[\rvec](z)$ which are
monomials in the derivatives
of $\rvec$, that is on the same operators than in \eTayExp .
Since they are obtained from a formal Taylor expansion,
the coefficients $C^A$ are also exactly the same than those in \eTayExp .

Now we separate the singular part of the propagator into
\eqn\eGsSep{
G_{\rm sing}(x_i,x_j)\ =\ G_0(x_i,x_j)\ +\ \sum_{B[g]\ne B_0}C^B(y_i,y_j)\,
B[g](z)
}
and expand the last, subdominant sum in the exponential \eGaAv.
We obtain also an OPE, which involves only geometrical local operators
$B[g]$
\eqn\eGOPE{
\ee^{-{1\over 2}\sum\limits_{i,j}\kvec_i\cdot\kvec_j\,G_{\rm sing}(x_i,x_j)} 
\ =\ \sum_{B[g]}C^B\{y_i,\kvec_i\}\,B[g](z)\,
\ee^{-{1\over 2}\sum\limits_{i,j}\kvec_i\cdot\kvec_j\,G_0(x_i,x_j)} 
}

Putting together the OPE \eTEOP\ and \eGOPE\ we obtain an OPE of the form
\eOPEcsp :
\eqn\eTheOPE{
\prod_i\ee^{\ii\kvec_i\cdot\rvec(x_i)}\ =\ 
\sum_A C^A\{y_i,\kvec_i\}\ \CN\left[ A(z)\,\ee^{\ii\kvec\cdot\rvec(z)}\right]
\ 
\ee^{-{1\over 2}\sum\limits_{i,j}\kvec_i\cdot\kvec_j\,G_0(x_i,x_j)} 
\ ,}
where now the sum runs over all products $A=A[\rvec]B[g]$, that is over all
local operators $A$ which are monomials in
derivatives of $\rvec$ and depend on the metric $g$ and its derivatives at
$z$.
The  coefficients $C^A$ are nothing but $C^A=C^{A[\rvec]}C^{B[g]}$.
This ends the derivation of the OPE in curved space.

%\appendix {D--Z}{}
\listrefs
\listfigs
\vfill\eject
\end